\documentclass[prc,twocolumn,showpacs,showkeys,superscriptaddress,nofootinbib]{revtex4}
\usepackage{graphicx,amsmath,amssymb}

\def\be{\begin{eqnarray}}
\def\ee{\end{eqnarray}}
\def\bc{\begin{center}}
\def\ec{\end{center}}

\def\rmd{{\rm d}}
\def\om{\omega}

\def\prt{\partial}

\newcommand{\lsim}{\stackrel{\scriptstyle <}{\phantom{}_{\sim}}}
\newcommand{\gsim}{\stackrel{\scriptstyle >}{\phantom{}_{\sim}}}
\begin{document}
\title{Charge and isospin fluctuations in a non-ideal pion gas\\ with dynamically fixed particle number}
\author{E.E. Kolomeitsev}
\affiliation{Matej Bel University, SK-97401 Banska Bystrica, Slovakia}
\affiliation{BLTP, Joint Institute for Nuclear Research, RU-141980 Dubna, Russia}
\author{D.N. Voskresensky}
\affiliation{National Research Nuclear University (MEPhI), 115409 Moscow, Russia}
\affiliation{BLTP, Joint Institute for Nuclear Research, RU-141980 Dubna, Russia}
\author{M.E. Borisov}
\affiliation{Prokhorov General Physics Institute of the Russian Academy of Sciences}
\begin{abstract}
We study the behavior of the non-ideal pion gas with the dynamically fixed number of particles, formed on an intermediate stage in ultra-relativistic heavy-ion collisions. The pion spectrum is calculated within the self-consistent Hartree approximation. General expressions are derived for cross-covariances of the number of various particle species in the pion gas of an arbitrary isospin composition.  The behavior of the cross-variances is analyzed for the temperature approaching from above the maximal critical temperature of the Bose-Einstein condensation for the pion species $a=\pm,0$, i.e. for $T>\max T_{\rm cr}^a$. It is shown that in case of the system with  equal averaged numbers of isospin species, the variance of the charge, $Q=N_{+}-N_{-}$, diverges at $T\to T_{\rm cr}=T_{\rm cr}^a$, whereas variances of the total particle number, $N=N_{+} + N_{-} + N_{0}$, and of a relative abundance of charged and neutral pions, $G=(N_{+}+N_{-})/2 - N_{0}$, remain finite in the critical point. Then  fluctuations are studied in the pion gas with small isospin imbalance $0<|G|\ll N$ and $0<|Q|\ll N$\,  and shifts of the effective masses, chemical potentials and values of critical temperatures are calculated  for various pion species, and the highest critical temperature, $\mbox{max}T_{\rm cr}^{{a}}$ is found, above which the pion system exists in the non-condensed phase.
Various pion cross variances are calculated for  $T>\mbox{max}T_{\rm cr}^{{a}}$, which prove to be strongly dependent on the isospin composition of the system, whereas the variances of $N$ and $G$ are found to be independent on the isospin imbalance up to the term linear in $G/N$ and $Q/N$.
\end{abstract}


\maketitle
\section{Introduction}

The first hadrochemical calculations for heavy-ion collisions at energies $\gsim A$\,GeV~\cite{Montvay-Zimanyi-79,ZFJ} inspired considerations of a possibility for the Bose-Einstein condensation (BEC) of pions in baryon enriched matter. However, subsequent more detailed studies, see Ref.~\cite{Migdal:1990vm,Voskresensky:1993ud} for review, excluded such a possibility for heavy-ion collisions at such energies, focusing attention on a possibility of a liquid phase of the inhomogeneous ($k\neq 0$) pion condensation in dense warm nuclear matter resulting in a significant enhancement of the in-medium pion distributions with $k\neq 0$ at non-zero temperature $T\neq 0$ and baryon density $n\gsim n_0$, where $n_0=0.16\,{\rm fm}^{-3}$ is the nuclear saturation density.  Moreover, the pion BEC in baryon enriched matter did not manifest itself in experiments at GSI energies~\cite{Reisdorf:2006ie}.

Experimental evidence for a formation of the baryon-poor medium at midrapidity at SPS, RHIC and LHC energies~\cite{Afanasiev2002,Alt2008,Nayak:2012np,Abelev:2012wca,Adamczyk:2017iwn}
invited for investigations of the properties of a dense and hot purely pion gas. Spectra of produced pions proved to be approximately exponential at intermediate transverse momenta, $m_\pi\lsim p_T\lsim 7\,m_\pi$, but show an enhancement at low transverse momenta, here and below $m_\pi$ is the pion mass, $m_\pi=139$\,MeV, and we imply $\hbar=c=1$.  Already first attempts~\cite{Kataja:1990tp,mishust} to fit the $p_T$ pion distributions in heavy-ion collisions at 200\,$A$GeV by ideal-gas expressions required the pion chemical potential $\sim 120-130$\,MeV. The magnitude of the chemical potential depends on how medium flow~\cite{Lee-Heinz-Schnee-90} and resonance contributions~\cite{Schnedermann:1993ws} are included. Subsequent more detailed analyzes of the SPS data~\cite{Ferenc,Tomasik-Heinz02}, using the method proposed in~\cite{Bertsch94} of extraction of the pion freeze-out space density from the mid-rapidity particle densities and the femtoscopic radii, showed its significant enhancement at small transverse momenta.

Estimates in~\cite{Goity:1989gs} showed that at temperatures $T\lsim 130-140$\,MeV the rate of pion absorption becomes smaller than the rate of re-scattering. This implies that the pion number can be considered as approximately fixed during subsequent pion fireball expansion from the chemical freeze-out temperature up to a lower temperature of kinetic (thermal) freeze-out~\cite{Gerber:1990yb}.

The proximity of the pion chemical potential to the critical value of the BEC in the ideal gas initiated speculations that a Bose-Einstein condensate might be formed in the pion fireball prepared in ultrarelativistic nucleus-nucleus collisions. This idea motivated a study of the properties of a pion BEC possibly formed in collisions at highest SPS energy~\cite{Voskresensky:1994uz}, and then at RHIC and LHC energies. Using the results of the first pion femtoscopy experiments, the value of the density of the pion system at the kinetic freeze-out was estimated as $n\sim (1-6)n_0$. An ideal pion gas and an interacting relativistic pion gas within $\lambda \varphi^4$ interaction model were studied under the assumption that the number of pions of each species is dynamically fixed within the time interval between the chemical and kinetic freeze-outs. The most central collisions with high pion multiplicity were proposed as the most preferable for observation of effects of the pion BEC. Utilizing the Weinberg Lagrangian, the investigation of the pion BEC was continued in~\cite{KV95,Kolomeitsev:1996tv}, a role of inelastic reactions $\pi^+\pi^-\to 2\pi^0$, which however conserve the particle number, was also discussed.

We assume that pions freeze out from a hot fireball, which can be described by a temperature $T(\vec{r},t)$, a density $n (\vec{r},t)$, and chemical potentials $\mu_a (\vec{r},t)$ for each pion species ($a=\pm,0$) within time interval $0<t<\tau_{\rm exp}$ of the fireball expansion between chemical and kinetic freeze-outs, cf. analysis \cite{Hung:1997du} performed for SPS energies. Formation of hadrons at RHIC and LHC conditions  occurs  at temperature $T_{\rm had}$ after a cooling of expanding quark-gluon fireball. Fitting of the particle yields measured at RHIC and LHC energies showed that   $T_{\rm had}$ should approximately coincide with the temperature of chemical equilibration $T_{\rm chem} \simeq 155$\,MeV~\cite{Stachel}. The chemical non-equilibrium analysis of the LHC data on mean particle multiplicities~\cite{Petran:2013lja} shows that $T_{\rm had}\simeq T_{\rm chem}$, might be even lower, $140-145$\,MeV. The kinetic freeze-out temperature for hadrons is expected to be still lower, $T_{\rm kin}\sim 100-120$\,MeV as was evaluated in~\cite{Teaney:2002aj,Pratt1999,Melo:2015wpa,Prorok:2015vxa} for energies under consideration. The time scale of chemical equilibration for $T\simeq 100-120$\,MeV, was estimated as $\tau_{\rm abs}\sim 100$\,fm for SPS energies~\cite{Song1997,Pratt1999}, being thereby much longer than the typical time of the thermal equilibration in the system,  $\tau_{\rm term}\sim $\,few fm, and than the time of the fireball expansion, $\tau_{\rm exp}\sim 10-20$\,fm up to the kinetic freeze-out~\cite{Prakash1993,Prorok:2015vxa}. Here the absorbtion time  $\tau_{\rm abs}$ is a typical time for $1\leftrightarrow 3$ processes and the thermalization time $\tau_{\rm term}\sim \tau_{\rm elast}$ is characterized mainly by elastic $2\leftrightarrow 2$  processes.

Although it was shown in~\cite{GreinerGong} that in the course of the quasi-equilibrium isentropic expansion of the initially equilibrated ideal pion gas the chemical potential cannot reach the critical value $m_\pi$, the non-equilibrium overcooling effects may drive pions to the BEC~\cite{Voskresensky:1994uz,Voskresensky:1995tx,Voskresensky:1996ur}.
The BEC can also occur because of an additional injection of non-equilibrium pions from resonance decays~\cite{ornik}, decomposition of a blurred phase of hot baryon-poor and pion enriched matter existing before the chemical freeze-out  \cite{Voskresensky:2004ux}, sudden hadronization of supercooled quark-gluon plasma~\cite{CC94}, and a decay of the transient Bose-Einstein condensate of gluons or glueballs pre-formed at an initial stage in a heavy-ion collision, cf.~\cite{Blaizot:2011xf,Xu:2014ega,Kochelev16,Peshier2016,Tanji:2017suk}. Reference~\cite{Voskresensky:1996ur} demonstrated that before the formation of the Bose-Einstein condensate the initially non-equilibrium interacting pion gas in ultrarelativistic heavy-ion collisions should pass several stages including a wave-turbulence stage.

Pion spectra obtained at LHC in collisions with $\sqrt{s_{NN}}=2.76$\,TeV can be fitted~\cite{Begun:2014rsa} with the help of the pion ideal gas distribution and the chemical potential $\mu\simeq 134.9$\,MeV, being very close to the critical value of the BEC in the ideal pion gas. The existing estimates for the typical density of the pion fireball are contradictory~\cite{Begun:2014rsa,Teaney:2002aj} yielding values varying in a broad range,  from $n\sim 0.8\,n_0$ to $2.5\,n_0$ for LHC energies. Recently, experimentalists in the ALICE Collaboration observed a significant suppression of three- and four-pion Bose-Einstein correlations in Pb-Pb collisions at $\sqrt{s_{NN}}=2.76$\,TeV at the Large Hadron Collider (LHC)~\cite{Abelev-coherent,Adam-coherent}. This may indicate that there is a considerable degree of coherent pion emission in relativistic heavy-ion collisions~\cite{Akkelin-02,Wong-Zhang07}. Analysis~\cite{Begun:2015ifa} indicated that about 5$\%$ of pions could stem from the BEC. Further discussions of a BEC in heavy-ion collisions at LHC energies can be found in the review~\cite{Shuryak:2014zxa}.

The higher is the pion multiplicity the more probable is to observe effects of the pion BEC~\cite{Voskresensky:1994uz,Voskresensky:1996ur}. The event-by-event analysis is preferable thereby. The fluctuation effects ordinary are increased in the vicinity of the critical point of any phase transition. In particular, second-order phase transitions are accompanied by
fluctuations of the order parameter, observed in various critical opalescence phenomena in equilibrium systems~\cite{LLP8}. If the system undergoes a first order phase transition, fluctuations grow, provided the system crosses the spinodal instability border, cf.~\cite{Maslov:2019dep}.

References~\cite{Begun:2006gj,Begun:2008hq}  argued for the divergence of the normalized variance in the critical point of the BEC for the ideal pion gas. On the experimental side an enhancement of the normalized variance was observed in the high pion multiplicity events in $pp$ collisions in the energy range 50--70\,GeV~\cite{Kokoulina:2011ed,Ryadovikov12}. Here, care should be taken, when one compares theoretical expectations for the thermal fluctuation characteristics with results of actual measurements, which incorporate background contributions, the dependence on center-of-mass energy, other dynamical effects, collision centrality, kinematic cuts, etc., cf.~\cite{Asakawa-Kitazawa2015}. The most simple and still relevant description of fluctuations in a quasi-equilibrium system formed in heavy-ion collisions can be performed employing the grand-canonical ensemble formulation, since usually only a part of the system, typically around mid-rapidity, is considered. Thus energy and conserved quantum numbers may be exchanged with the rest of the system, which serves as a heat bath~\cite{Jeon:2003gk,Heiselberg:2000ti}.

Self-consistent account for a pion-pion interaction in the Hartree approximation demonstrated that variance of the pion number in the system with an equal averaged number of pion species remains finite at the critical temperature~\cite{KV18} as well as the skewness and kurtosis~\cite{BKV18}. A suppression of fluctuation effects occurs also due to the finiteness of the system~\cite{Begun-Gor-2008-FS}. Nevertheless, they remain to be enhanced near the critical point of BEC. Therefore, the appearance of a significant increase of particle number fluctuations could be considered as a signal that the pion system formed in heavy-ion collisions is approaching the BEC at some conditions.

In this work we  calculate characteristics of particle number fluctuations in the non-ideal hot pion gas with the dynamically fixed number of particles considered in the grand-canonical formulation.
In Sect.~\ref{sec:hartree} we remind a formalism for the description of the pion non-ideal gas with the dynamically fixed particle number ~\cite{KV18}. In Sect.~\ref{sec:covariance} we apply the self-consistent Hartree approximation for an  arbitrary relation between  various pion fractions. The results are applicable for temperatures $T>\max T_{\rm cr}^a$, where $T_{\rm cr}^a$ is the critical temperature of the BEC for the pion species $a=\pm,0$. We find general expressions for cross-variances of various pion species. Then the  behavior of the cross-variances is analyzed for the temperature approaching the critical temperature of the BEC  $\max T_{\rm cr}^a$. In Sect.~\ref{sec:fluc} we study fluctuations of the  charge and the relative number of charged and neutral pions in a system with equal averaged number of pions for each isospin species. In Sect.~\ref{sec:imbalance} we consider properties of a system with a small isospin imbalance, either with a small net charge or with a small difference between the number of charged and neutral pions at zero net charge. For this, in Sect.~\ref{subsec:prop} we calculate pion characteristics and in Sect.~\ref{subsec:fluct-imbal} we apply the results for estimations of the cross-variances in such systems. Conclusions are drawn in Section~\ref{sec:conclusion}.
Some  details of calculations are collected in Appendices~\ref{app:deriv}, \ref{app:dFdN}, and~\ref{app:Iexpan}.

\section{Non-ideal pion gas with dynamically fixed particle number. Formalism}\label{sec:hartree}

We use the simplest model for the description of a non-ideal pion gas with the interaction  $\lambda \vec{\phi}^4/4$, where $\vec{\phi}$ is the isospin vector in cartesian representation $\vec{\phi}=(\phi_1,\phi_2,\phi_3)$. Following arguments of Ref.~\cite{KV18} applying this model to a pion-enriched system created on an intermediate stage of a heavy-ion collision, we  keep in the Lagrangian density only the terms containing equal number of creation and annihilation operators. By this we explicitly take into account that processes with a change of the number of particles, generated by the dropped terms, do not occur within the time window $0<t<\tau_{\rm exp}$, cf.~\cite{Goity:1989gs,Gerber:1990yb}. The resulting Lagrangian density  reads~\cite{KV18}
\begin{align}
\mathcal{L}
&=\sum_{a=\pm}\big(\prt_\mu \varphi_a \prt^\mu \varphi_a^\dag - m_{\pi}^2 \varphi_a\varphi_a^\dag-\lambda (\varphi_a\varphi_a^\dag)^2\big)
\nonumber\\
&+\prt_\mu \varphi_0\prt^\mu \varphi_0^\dag- m_{\pi}^2 \varphi_0\varphi_0^\dag
- \frac32\lambda (\varphi_0\varphi_0^\dag)^2
\nonumber\\
&-4\lambda (\varphi_-\varphi_-^\dag) (\varphi_+\varphi_+^\dag) -
2\lambda\big[(\varphi_+\varphi_+^\dag) + (\varphi_-\varphi_-^\dag)\big](\varphi_0\varphi_0^\dag)
\nonumber\\
&-\lambda\big[ \varphi_+ \varphi_- \varphi_0^{\dagger 2}
+ \varphi_0^2 \varphi_+^{\dagger} \varphi_-^{\dagger}\big]
\,,
\label{ourlag}
\end{align}
where  $\varphi_a$ and $\varphi_a^\dag$ stand for annihilation and creation operators of a pion of type $a=+,-,0$, and
$\varphi_\pm+\varphi_\mp^\dag=(\phi_1\pm i\phi_2)/\sqrt{2}$ and $\varphi_0+\varphi^\dag_0=\phi_3$\,.
From the comparison with the leading terms of the effective Weinberg Lagrangian~\cite{Weinberg68}, the coupling constant can be estimated as $\lambda=m_\pi^2/2f_\pi^2\simeq 1.13$, where $f_\pi=93$\,MeV is the weak pion decay constant, and $m_\pi=139$\,MeV is the free pion mass (we neglect small explicit isospin symmetry breaking).

In the Lagrangian density~(\ref{ourlag}) we dropped the terms $\mathcal{L}_{3\leftrightarrow 1}$ containing non-equal number of creation and annihilation operators, responsible for the absorption and production processes, which are assumed to be not operative for $t<\tau_{\rm exp} \ll \tau_{\rm abs}$. Due to this, we  deal with three complex fields, whereas  initial Lagrangian for pions is formulated  in terms of three real fields $\phi_1,\phi_2,\phi_3$. The doubling of the degrees of freedom is related to the fact that we consider the system at time scale $\tau_{\rm elast}$ much less than  $\tau_{\rm abs}$.  So, e.g., the positive and negative pions are not treated anymore as particles and antiparticles for $t\ll \tau_{\rm abs}$, recall nonrelativistic Schr\"odinder description performed using the complex wave functions.

Interaction terms in two first lines in (\ref{ourlag}) allow for $\pi^-\pi^-$, $\pi^+\pi^+$ and $\pi^0\pi^0$ elastic re-scattering processes, which permit to equilibrate the energy-momentum in the corresponding $+,-$ and $0$ pion subsystems. The corresponding terms in the third line allow an exchange of the energy-momentum  in the $\pi^-\pi^+\to  \pi^-\pi^+$, and $\pi^+\pi^0\to \pi^+\pi^0$ and $\pi^-\pi^0\to \pi^-\pi^0$ reactions.
Presence of the last term in (\ref{ourlag}) allows to conserve the total number of pions permitting change between isospin fractions in the system in reactions $\pi^+\pi^-\leftrightarrow \pi^0 \pi^0$, not changing the total charge and isospin projection. The probabilities of all these processes are of the one and the same order. Within the self-consistent Hartree approximation ~\cite{KV18}, which we employ below to calculate the pion spectrum these terms do not contribute but they are assumed to be operative, establishing the local equilibrium in the time interval between chemical and kinetic freeze-outs. As the result, one can use relation between chemical potentials of the species, $2\mu_0 =\mu_+ +\mu_{-}$.

In the self-consistent Hartree approximation the modification of the spectrum of pions of type $a$ is reduced to the replacement of the vacuum pion mass by an effective mass~\cite{KV18}, which may depend on the species under consideration, $\om_a(p)=\sqrt{m_a^{*2}+p^2}$,
\begin{align}
m_a^{*2}&=m_\pi^2+\Pi_a\,,\nonumber\\
\Pi_a &=2\lambda
\sum_{b=+,-,0}[\mathcal{K}]_{ab}d_b m^{*2}_b \,,
\quad \mathcal{K}=\left[\begin{array}{ccc}2&2&1\\2&2&1\\1&1&3\end{array}\right]\,,
\label{pi-op}
\end{align}
where $a,b=\{+,-,0\}$,
and $d_b=d(m_b^{*2},\mu_b,T)$ are dimensionless functions of  the effective mass $m_b^{*2}$, chemical potential $\mu_b$ and temperature $T$,
\begin{align}
d(m^2,\mu,T) &= \intop \frac{\rmd^3 p}{(2\pi)^3}\frac{1}{2m^{2}\sqrt{m^{2}+p^2}}
\nonumber\\
&\times \frac{1}{e^{(\sqrt{m^{2}+p^2}-\mu)/T}-1}\,.
\label{d-def}
\end{align}

The ensemble-averaged pion densities are expressed through the chemical potentials as
\begin{align}
n_a (\mu_a,T) =\frac{\langle \hat{N}_a\rangle}{V}= h_a (m_a^{*2}(\mu_+,\mu_-,\mu_0,T),\mu_a,T),
\label{na-def}
\end{align}
where $V$ is the volume of the system and $m_a^{*2}(\mu_+,\mu_-,\mu_0,T)$ is the solution of Eqs. (\ref{pi-op}) and the function $h$ is defined as
\begin{align}
h(m^2,\mu,T)=\intop \frac{\rmd^3 p}{(2\pi)^3}\frac{1}{e^{(\sqrt{m^2+p^2}-\mu)/T}-1}\,.
\label{h-def}
\end{align}
Notice that relation (\ref{na-def}) makes sense only, if $m_a\ge \mu_a$, otherwise there appears a pole in the particle momentum distribution. The critical temperature of such an instability (we call it the Bose-Einstein instability), $T_{\rm cr}^a$ is determined by equation $m_a^{*}(T_{\rm cr}^a)=\mu_a(T_{\rm cr}^a)$. Thus, our consideration here and below is valid only for temperatures $T>\max_a T_{\rm cr}^a$. For $T<\max_a T_{\rm cr}^a$, Eq.~(\ref{pi-op}) becomes invalid, since it does not take into account a contribution of the Bose-Eistein condensate appeared  for the pions of that sort $\tilde{a}$, for which $T_{\rm cr}^{\tilde{a}}=\max_a T_{\rm cr}^a$.

\section{Variances and cross-variances of pion numbers for system of arbitrary composition }\label{sec:covariance}

In~\cite{KV18} it was argued that the minimum of free energy is realized in an interacting pion gas for a system with equal values of chemical potentials, i.e. with the equal average numbers of pions of different species. Variance of the pion number was studied for the symmetrical system, when averaged densities for all three pion fractions are equal. In heavy-ion collisions in some events  pionic subsystems are produced with compositions deviating from that corresponding to the most energetically favorable case. Therefore, it would be useful to extend  analysis of~\cite{KV18,BKV18} for such systems. For that we introduce three independent chemical potentials, $\mu_a$. Then the partition function in the grand-canonical ensemble is equal to
\begin{align}
Z(\mu_+,\mu_-,\mu_0,T)={\rm Tr}\,e^{-(\hat{H} -\sum_a\mu_a \hat{N}_a)/T},
\end{align}
where $\hat{H}$ is the Hamiltonian describing the system with conserved number of pions of each sort and which yields the pion spectrum with the effective mass (\ref{pi-op}) in the Hartree approximation. $\hat{N}_a$ is the operator of the number of pions of type $a$, ${\rm Tr}$ means  ensemble averaging and integration over the phase volume, $\langle g_a \int \frac{ V \rmd^3 p}{(2\pi)^3}(...)\rangle$, for a system of a large volume $V$;  $g_a$ is the isotopic degeneracy factor.

Besides the ensemble-averaged number of particles for the given species
\begin{align}
\langle \hat{N}_a\rangle=T\frac{\partial}{\partial \mu_a}\log Z\Big|_{T,V}\,,
\end{align}
one can calculate the higher order commulants of the particle number operators as derivatives of $Z$ with respect to the chemical potentials. The cross-variances of the number of pions of various sorts are determined as \cite{Sawyer:1989nu,Roepke:2017bad}
\begin{align}
\langle \hat{N}_a \hat{N}_b\rangle -\langle \hat{N}_a\rangle\langle \hat{N}_b\rangle=T^2\frac{\partial}{\partial \mu_a}\frac{\partial}{\partial \mu_b}\log Z\Big|_{T,V}
\label{aver-Z}
\end{align}
and normalized cross-variances are given by
\begin{align}
\varpi_{ab}=
\frac{\langle \hat{N}_{a} \hat{N}_{b}\rangle - \langle \hat{N}_{a}\rangle \langle \hat{N}_{b}\rangle}
{\sqrt{\langle \hat{N}_{a}\rangle \langle \hat{N}_{b}\rangle}}
= \frac{T}{\sqrt{n_a n_b}}\frac{\prt n_a}{\prt \mu_b}\Big|_{T,V}\,,
\label{cross-var}
\end{align}
with the obvious symmetrical relation $\varpi_{ab}=\varpi_{ba}$.
These general expressions allow to express  the normalized variance of the total number of particles, $N=N_{+} + N_{-} + N_{0}$ with $N_a$ standing for the number of pions of species $a$ in the given event, as
\begin{align}
\varpi_N &=\frac{\langle \hat{N}^2\rangle -\langle \hat{N}\rangle^2}{\langle \hat{N}\rangle}
=\sum_{a,b=\pm,0}\sqrt{c_a \, c_b}\,\varpi_{ab}\,,
\label{var-N-gen}\end{align}
where we introduced a relative fraction of pions of sort $a$:
\begin{align}
c_a &=\frac{\langle \hat{N}_{a}\rangle}{\langle \hat{N}\rangle}
=\frac{n_a}{n_+ + n_- + n_0 }\,.
\end{align}
Fluctuations of the  charge in the system, $Q=N_{+} - N_{-}$, are characterized by the quantity
\begin{align}
\varpi_Q
&=2 \frac{\langle \hat{Q}^2\rangle -\langle \hat{Q}\rangle^2}
{\langle \hat{N}_{+}\rangle + \langle \hat{N}_{-}\rangle}
\nonumber\\
&= \sum_{a,b=\pm}\frac{\sqrt{c_a \, c_b}}{\frac12(c_+ + c_-)}\varpi_{ab}(2\delta_{ab}-1)\,.
\label{var-Q-gen}
\end{align}
Imbalance of charged versus neutral pions,  $G=N_{\rm ch}-N_{0}$, where $N_{\rm ch}=\frac12(N_{+} + N_{-})$, is characterized by
\begin{align}
\varpi_G &=2\frac{\langle \hat{G}^2\rangle -\langle \hat{G}\rangle^2}
{\langle \hat{N}_{\rm ch}\rangle +\langle \hat{N}_{0}\rangle }
\nonumber\\
&= \sum_{a,b=\pm,0}\frac{\sqrt{c_ac_b}}
{1+c_0}\varpi_{ab}(1-3\delta_{a0})(1-3\delta_{0b})\,.
\label{var-G-gen}
\end{align}

As follows from Eqs.~(\ref{aver-Z}) and (\ref{cross-var})  the normalized cross-variance $\varpi_{ab}$ can be associated with fluctuations of particle densities
\begin{align}
\varpi_{ab} &=V
\frac{\langle \hat{n}_{a} \hat{n}_{b}\rangle - \langle \hat{n}_{a}\rangle
\langle \hat{n}_{b}\rangle}
{\sqrt{\langle \hat{n}_{a}\rangle \langle \hat{n}_{b}\rangle}}
= \frac{T}{\sqrt{n_a n_b}}\frac{\prt n_a}{\prt \mu_b}\Big|_{T,V}\,,
\label{varpi-def}
\end{align}
given by Eq.~(\ref{na-def}) and, through the effective pion masses, being functions of all three chemical potentials. Using relations $\delta n_a =-N_a \delta V/V^2$ one may introduce variance of the volume for fixed $N_a$ and $T$.

Speaking about fluctuations of intensive (not depending on $V$) and extensive (depending on $V$) variables one has to take into account that measuring characteristics of fluctuations at different experimental conditions may reflect different moments of the fireball evolution, like the chemical freeze-out, $(n_{\rm chem},T_{\rm chem})$, and the kinetic one, $(n_{\rm kin},T_{\rm kin})$. Right after the chemical freeze-out (for $t>0$) the pion annihilation and creation processes cease and the total number of pions $N$ does not change, therefore. Thus, if we consider an ideal detector with full $4\pi$ geometry, fluctuations of the total pion number reflect the state of the system at the chemical freeze-out. So, expression (\ref{var-N-gen}) describes fluctuations of the total pion number at $T(0)=T_{\rm chem}$, $V(0)=V(T_{\rm chem})$. However, the same quantity, $\varpi_N$, taken for $T=T_{\rm kin}$ and $V(T_{\rm kin})$ also characterizes fluctuations of the volume of the pion fireball at the kinetic freeze-out at measurements done in the $4\pi$ geometry.

The chemical potentials of pions for all three species evolve until the kinetic freeze-out occurring at $t=\tau_{\rm kin}$. Within this time window the system may reach the BEC point, at which various fluctuation characteristics may significantly grow. Although in the time interval  between chemical and kinetic freeze-outs the total number of pions remains not changed an exchange of particles between pion species continues owing to the $2\leftrightarrow 2$ reactions. Thus, if pions are measured in experiments with incomplete geometry and/or in a restricted momentum range, then the elastic pion-pion reactions and processes  $\pi^0\pi^0\leftrightarrow\pi^+\pi^-$ change populations of pions of different isospin species and in different momentum bins.  Therefore, there exists a kind of thermodynamic reservoir for the subsystem  of pions, which later reach detector, and the grand-canonical formulation can be relevant in such a situation. If one measures correlations  between pions emitted  at different angles and in various momentum bins, one may get an information about the state of the pion fireball at the kinetic freeze-out. Moreover, the quantities $\varpi_Q$, $\varpi_G$ characterize fluctuations in the system at the kinetic freeze-out.

However, any case one should bear in mind that comparison of the results of idealized calculations and real measurements is very uncertain without a detailed study of experimental conditions.
Thus we may say that, only if indeed a significant growth of fluctuation characteristics were observed, it could be associated with a closeness to the pion BEC either at the chemical freeze-out or at the thermal freeze-out, depending on the specifics of the measurement.

To calculate cross-variances (\ref{cross-var}) we need the derivatives of the densities
\begin{align}
\frac{\partial n_a}{\partial \mu_b}=
\frac{\partial h_a}{\partial \mu_b}\delta_{ab}
+\frac{\partial h_a}{\partial m_a^{*2}}\frac{\partial m_a^{*2}}{\partial \mu_b}\,,
\label{n-deriv}
\end{align}
where enter derivatives of the effective pion masses with respect to chemical potentials.  From Eq.~(\ref{pi-op}) taking into account the dependence of $d_a$ on $m_a^{*2}$ and $\mu_a$ we get
\begin{align}
\frac{\partial m_a^{*2}}{\partial \mu_b}
&= 2\lambda\!\!\!\sum_{c=\pm,0}\!\! \left[
\frac{\partial (d_cm_c^{*2})}{\partial \mu_b}+ \frac{\partial (d_cm_c^{*2})}{\partial m_c^{*2}} \frac{\partial m_c^{*2}}{\partial\mu_b}\right]\mathcal{K}_{ca}
 \,,
\label{dmdmu-eq}
\end{align}
\begin{align}
&\frac{\prt h_a}{\prt \mu_a}=m_a^{*2} I_1^{a}\,,\,\,
\frac{\prt h_a}{\prt m_a^*}=-m_a^{*2} I_3^{a}\,,
\nonumber\\
&m_a^{*}\frac{\partial d_a}{\partial \mu_a}=\frac{1}{2}I^a_3 \,,\,\,m_a^{*2}\frac{\partial d_a}{\partial m_a^{*2}}=-\frac14I^a_2 \,.
\end{align}
through auxiliary dimensionless quantities, $I^{(a)}_n$, being functions of $m_a^{*2}$, $\mu_a$ and $T$,
\begin{align}
\{I^{a}_1,I^{a}_2,I^{a}_3 \} &=\!\!\intop\!\! \frac{\rmd^3 p}{(2\pi)^3p^2}\Big\{\frac{\om_a^2(p)+p^2}{m_a^{*2}\om_a(p)},
\frac{1}{\om_a(p)}, \frac{1}{m_a^{*}}\Big\}
\nonumber\\
&\times \frac{1}{e^{(\om_a(p)-\mu_a)/T}-1}\,,
\label{I-def}
\end{align}
with $\om_a(p)=\sqrt{m_a^{*2}+p^2}$. All integrals (\ref{I-def}) diverge at the critical point of the induced BEC of pions of sort $a$, $T_{\rm cr}^{a}$, determined by the equation
$$m_a^*(T_{\rm cr}^{a})=\mu_a(T_{\rm cr}^{a}).$$
Indeed, for $\mu_a\to m_a^{*}-0$, we get
\begin{align}
I_n^{a}\Big|_{\mu_a\to m_a^*-0}\to
\frac{\,T}{2^{3/2}\pi\sqrt{m_a^{*}}\sqrt{m_a^*-\mu_a}}\,.
\label{I3-limit}
\end{align}
Finally solving (\ref{dmdmu-eq}) we find the  cross-variances:
\begin{align}
&\varpi_{\pm\pm}=\frac{m_{\pm}^{*2}T}{n_{\pm}}\left[
I_1^{\pm}
- \frac{\lambda \big[I_3^{\pm}\big]^2}
{D } ( 4 + 5\lambda  I_2^{0} )\right]\,,\,\,\,
\nonumber\\
& \varpi_{\pm\mp}=-\frac{m_{+}^{*}m_{-}^{*}T}{\sqrt{n_{\pm}n_{\mp}}}
\frac{\lambda I_3^{+} I_3^{-} }
{D} (4 + 5\lambda  I_2^{0})\,,
\nonumber\\
&\varpi_{0 0}=\frac{m_0^{*2}T}{n_{0}}\left[ I_1^{0}
-
\frac{\lambda [I_3^{0}]^2 }
{D}(6 +  5\lambda (I_2^{{+}} + I_2^{{-}}))\right]\,,
\nonumber\\
&\varpi_{\pm 0}=\varpi_{0\pm}=-\frac{m_{\pm}^{*}m_0^{*}T}{\sqrt{n_{\pm} n_{0}}}
\frac{ 2\lambda I_3^{\pm} I_3^{0}}
{D}\,,
\nonumber\\
& D=\lambda  I_2^{0} +  (4 + 5\lambda I_2^{0})
[1+ \lambda (I_2^{+}+I_2^{-})]\,.
\label{mixed-derivtives}
\end{align}
For an ideal pion gas ($\lambda=0$)  fluctuations of pions of different species are independent, $\varpi_{ab}=\varpi_{aa}\delta_{ab}$, and the normalized variances of the particle number are given by one simple expression $\varpi_{aa}=T\,I_1^{a}/n_a$. Thus, the particle number fluctuations in an ideal Bose gas diverge at the critical point of the BEC. The presence of this divergence in the variance of the particle number put in doubt~\cite{terHaar52,Fierz56,ZUK77} the applicability of the grand canonical description of an ideal Bose gas at temperatures close to the critical one. However, if the interaction is self-consistently taken into account, as it is done here within the Hartree approximation,   the divergence disappears  ~\cite{KV18,BKV18}. For example, in a system with only one sort of particles we obtain the following expression for the normalized variances
\begin{align}
\varpi_{aa}=\frac{m_a^{*2}T}{n_a}\Big[ I_1^{a}- \frac{\xi_a\lambda [I_3^{a}]^2}
{1+\xi_a\lambda I_2^{a}}\Big] \,,\quad \xi_{\pm}=1\,,\,\, \xi_0=\frac32\,.
\label{waa-single}
\end{align}
This expression can be rewritten as
\begin{align}
\frac{n_a}{m_a^{*2}T}\varpi_{aa} =  I_1^{a}- \frac{[I_3^{a}]^2}{I_2^{a}} + \frac{[{I_3^{a}}/{I_2^{a}}]^2}{\xi_a\lambda+{1}/{I_2^a}}\,.
\label{fin-comb}
\end{align}
Taking into account that  integrals $I_n^{a}$ diverge in a correlated way, see Eq.~(\ref{I3-limit}), we show that the divergent parts in the first two terms in Eq.~(\ref{fin-comb}) cancel each other exactly. The ratio of two divergent integrals in the numerator of the third term  also proves to be finite. To identify the finite remainder one can use the following identities among quantities  $I_n^{a}$, $n=1,2,3$,
\begin{align}
 I_1^{a} -  I_3^{a} &= 2\,(\tilde{d}_a + d_a)\,,
\nonumber\\
  I_2^{a} - I_3^{a} &= 2\,(\tilde{d}_a - d_a)\,,
\label{I123-relat}
\end{align}
where $d$ is defined in Eq.~(\ref{d-def}) and $\tilde{d}$ is given by
\begin{align}
\tilde{d}= \!\!\intop\!\! \frac{\rmd^3 p}{(2\pi)^3}
\frac{1}{2m^{*2} (\om (p)+m^*)} \frac{1}{e^{(\om (p)-\mu )/T}-1}\,.
\label{def-d-dt}
\end{align}
Both $d_a$ and $\tilde d_a$ remain finite at the critical temperature $T_{\rm cr}^{a}$. Now,
expressing, e.g., $I_1^a$ and $I_2^a$ through $I_3^a$ and substituting in the first two terms in (\ref{fin-comb}) we obtain the finite limiting value at $T\to T_{\rm cr}^a$ equal to
\begin{align}
\Big( I_1^{a}- \frac{[I_3^{a}]^2}
{I_2^{a}} \Big)\Big|_{T=T_{\rm cr}^{a}}=4\tilde{d}_{{\rm cr},a}\,,
\label{fin-comb-d}
\end{align}
where $\tilde{d}_{{\rm cr},a}\equiv \tilde{d}_{a}|_{T=T_{\rm cr}^{a}}$.
The final expression for the normalized variance $\varpi_{aa}$ at $T_{\rm cr}^a$ reads
\begin{align}
\varpi_{aa}(T_{\rm cr}^a)=\frac{\mu_{{\rm cr},a}^2\,T_{\rm cr}^a}{\xi_a \lambda n_a}\Big(1+
4\xi_a\lambda \tilde{d}_{{\rm cr},a}\Big)\,.
\label{waa-fin}
\end{align}
Note that although $\tilde{d}_a$, ${d}_a$ and $h_a$ are functions of three variables $m_a^*, \mu_a$ and $T$, the
quantities $\tilde{d}_{{\rm cr},a}$,
${d}_{{\rm cr},a}\equiv {d}_{a}|_{T=T_{\rm cr}^{a}}$ and $h_a|_{T=T_{\rm cr}^{a}}$
are already functions of only one variable $t_{{\rm cr},a}=T_{\rm cr}^{a}/\mu_{{\rm cr},a}$, where  $\mu_{{\rm cr},a}=\mu_a(T_{\rm cr}^{a})=m_a^* (T_{\rm cr}^{a})$.

Although the integrals in $d_a$ and $\tilde{d}_a$ cannot be evaluated analytically at arbitrary $T$, at $T=T_{\rm cr}^{a}$ we can write expansions in terms of
\begin{align}
&d_{{\rm cr},a} = \frac{ t_{{\rm cr},a}^{3/2}}{4 \sqrt{2}\pi^{3/2}}
\nonumber\\
&\times \Big(\zeta\big({\textstyle\frac32}\big)
+ \frac{3}{8}\, t_{{\rm cr},a} \zeta\big({\textstyle\frac52}\big) - \frac{15}{128} t_{{\rm cr},a}^2 \zeta\big({\textstyle\frac72}\big)+... \Big)\,,
\nonumber\\
&\tilde{d}_{{\rm cr},a}  = \frac{ t_{{\rm cr},a}^{3/2}}{8\sqrt{2}\pi^{3/2}}
\nonumber\\
&\times \Big(\zeta\big({\textstyle\frac32}\big)
+ \frac{9}{8}\, t_{{\rm cr},a} \zeta\big({\textstyle\frac52}\big)
- \frac{75}{128} t_{{\rm cr},a}^2 \zeta\big({\textstyle\frac72}\big)+...\Big) \,,
\label{ddt-exp}
\end{align}
which rapidly converge for $t_{{\rm cr},a}\ll 1$. Our numerical evaluations show that in a gas consisting of pions of one species at $\lambda \gsim 1$ we have $t_{{\rm cr},a}<1$  for  $0<n\lsim 1.5 n_0$. Range of densities, for which $t_{{\rm cr},a}<1$, is increased, if more pion species are present, e.g., for the gas with two species of equal fractions $t_{{\rm cr},a}<1$  for $0<n\lsim 3 n_0$ and for the isospin-symmetrical gas with three species, $t_{{\rm cr},a}<1$ for densities
$0<n \lsim 5 n_0$ (see discussion in the next Sect.~\ref{sec:fluc}). An increase of $\lambda$ also extends this interval up to higher densities.  Thus, expansion (\ref{ddt-exp}) in $t_{{\rm cr},a}$ is indeed useful.

For completeness we give also expansions for integrals $I_{n}^{a}$ for $T\to T_{\rm cr}^a$:
\begin{align}
I_n^{a}(T\to T_{\rm cr}^a) &=\frac{\,T_{\rm cr}^a}
{2\pi\sqrt{\alpha_a}(T-T_{\rm cr}^{a})}
+\frac{1}{2\pi\sqrt{\alpha_a}}
+ \delta I_{\rm cr, n}^a \nonumber
\\
&+ O((T-T_{\rm cr}^{a})/T_{\rm cr}^{a}) \,.\label{In-atTc}
\end{align}
To derive  first two terms in (\ref{In-atTc}) we used  expansion
at $T$ near $T_{\rm cr}^a$:
\begin{align}
m_a^*(T)-\mu_a(T)
\approx \frac12(\alpha_{a}/\mu_{{\rm cr},a})\,(T-T_{\rm cr}^{a})^2\,,\label{macr}\end{align}
where $\alpha_{a}$ is a coefficient taken at $T=T_{\rm cr}^{a}$.
In Appendix~\ref{app:deriv} we demonstrate calculation of  the coefficient $\alpha_{a}$ on example of the isospin-symmetrical medium.
Finite parts, $\delta I_{\rm cr, n}^a$, can be expressed with the help of Eq.~(\ref{I123-relat}) through the quantities $d_{{\rm cr},a}$, $\tilde{d}_{{\rm cr},a}$ given by (\ref{ddt-exp}) and $\delta I_{\rm cr, 3}^a$ given by
\begin{align}
\delta I_{\rm cr, 3}^a
&=
\frac{t_{{\rm cr},a}^{1/2}}{(2\pi)^{3/2}}\zeta\big({\textstyle\frac12}\big)
+
\frac{3t_{{\rm cr},a}^{3/2}}{8(2\pi)^{3/2}}
\Big[\zeta\big({\textstyle\frac32}\big)
\nonumber\\
&
- \frac{5}{16}\, t_{{\rm cr},a} \zeta\big({\textstyle\frac52}\big)
- \frac{35}{128} t_{{\rm cr},a}^2 \zeta\big({\textstyle\frac72}\big)+...\Big]\,.
\label{dI3-exp}
\end{align}

\begin{figure*}
\centering
\includegraphics[width=15cm]{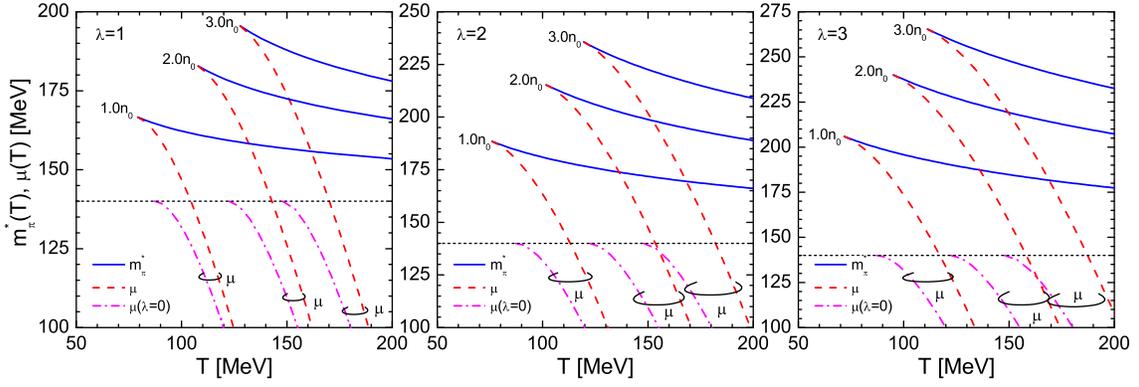}
\caption{The effective pion mass, $m^*$, and the chemical potential as functions  of temperature calculated for three values of the coupling constant $\lambda=1,2,$ and 3, and several values of the pion gas density. Dash-dotted lines show the chemical potentials for the free pion gas ($\lambda=0$).
}
\label{fig:mmu-T}
\end{figure*}

We note that for $T\to T_{\rm cr}^a$ all infinite terms as well as a part of finite terms present in the integrals $I_n^{a}(T\to T_{\rm cr}^a)$ cancel away in the functions $d_a$ and $\tilde{d}_a$, up to the order $t_{{\rm cr},a}^{3/2}$\,, more precisely  cancel away the first two terms in (\ref{In-atTc}) and the first term in (\ref{dI3-exp}). Therefore, it is preferable to  express the quantities remaining finite at $T\to T_{\rm cr}^a$,  through functions $d_{{\rm cr}, a}$ and $\tilde{d}_{{\rm cr}, a}$.

At the end, we should remind  that although formally Eq. (\ref{In-atTc}) is valid for $T\to T_{\rm cr}^a$ for all $a=\pm,0$, in reality, our consideration  should be modified   for  temperatures below $\max T_{\rm cr}^a = T_{\rm cr}^{\tilde{a}}$, since for such temperatures we already need to include the BEC for the species $\tilde{a}$.

\section{Fluctuations in the isospin-symmetrical gas}\label{sec:fluc}

In this section, we consider the isospin-symmetrical pion gas, where $\mu_a=\mu$ and $m^{*2}_a=m^{*2}$ for $a= \pm,0$. The dependences of the effective pion mass $m^{*2}$ and the chemical potential $\mu$ on the temperature $T$ and the particle density $n$ are determined by the set of equations, cf. Eq. (\ref{pi-op}),
\begin{align}
\begin{array}{l}
m^{*2} = m_\pi^2+10\lambda \, m^{*2} d(m^{*2},\mu,T)\,,
\\
n = 3 \, h(m^{*2},\mu,T)\,.
\end{array}\,
\label{eqs-sym}
\end{align}
Solutions of the system of equations (\ref{eqs-sym}) for the effective pion mass and the chemical potential for various temperatures and densities are shown in Fig.~\ref{fig:mmu-T}.
In the  system with the interaction the effective pion mass is larger than the free pion mass. In Fig.~\ref{fig:mmu-T} we see that the effective pion mass and chemical potential  decrease with increase of the temperature, and grow with increase of the density  and the coupling constant $\lambda$.

\begin{figure}
\centering
\includegraphics[width=6cm]{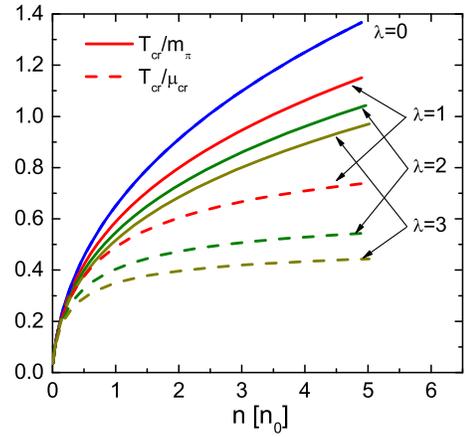}
\caption{Critical temperature of the Bose-Einstein instability in the isospin-symmetrical pion gas in units $m_\pi$, $T_{\rm cr}/m_\pi$, and the ratio $T_{\rm cr}/\mu_{\rm cr}$, $\mu_{\rm cr}=m^*(T_{\rm cr})$, as functions of the particle density (in units $n_0$) for different values $\lambda$.
}
\label{fig:Tmuc}
\end{figure}

As found in~\cite{KV18}, the critical temperature, $T_{\rm cr}$, is a monotonously increasing function of the density and it decreases with an increase of the interaction constant $\lambda$. The ratios  $T_{\rm cr}/m_\pi$,  and  $t_{\rm cr}=T_{\rm cr}/\mu_{\rm cr}$ are shown in Fig.~\ref{fig:Tmuc} by solid and dashed lines, respectively, as functions of a particle density for various values of $\lambda$.    Since $m^*(T)>  m_\pi$ for all temperatures and densities, we have also $t_{\rm cr}<T_{\rm cr}/m_\pi$.
 As it  is seen in Fig.~\ref{fig:Tmuc},  $t_{\rm cr}$  depends more weakly on the density
than $T_{\rm cr}/m_{\pi}$. For $n\gsim (1\mbox{--}2)\,n_0$, $t_{\rm cr}$ begins to flatten out.
Note also that for a given value of $\lambda$,  $t_{\rm cr}$ is limited from above.
Indeed, using Eqs. ~(\ref{eqs-sym}) we can combine an equation relating $t_{\rm cr}$ and $n$:
\begin{align}
&\frac{n}{3m_\pi^3}=\frac{\tilde{h}(t_{\rm cr})}{(1-10\lambda d_{\rm cr})^{3/2}}\,,
\label{constr-eq}
\end{align}
where $\tilde{h}(t_{\rm cr})\equiv h(\mu_{\rm cr}^2,\mu_{\rm cr},T_{\rm cr})/\mu_{\rm cr}^3\,$.
The right hand side of this equation depends only on $t_{\rm cr}$ and  must be positive. Therefore we have the constraint $d_{\rm cr}<1/(10\lambda)$. Hence, since $d_{\rm cr}$ is an increasing function of $t_{\rm cr}$, we have the constraint $t_{\rm cr}< t_{\rm cr}^{\rm (max)}$, where $t_{\rm cr}^{\rm(max)}$ is the solution of equation $d_{\rm cr}(t_{\rm cr})=1/(10\lambda)$. For example, for $\lambda=1$, 2, and 3 we have $t_{\rm cr}^{\rm (max)} =1.023$, 0.6646, and 0.5145, respectively.
Thus Fig.~\ref{fig:Tmuc} allows us to determine applicability range and precision of expansions (\ref{ddt-exp}). As we see, for densities $n<5\, n_0$ the value $t_{\rm cr}$ does not exceed 0.8 for $\lambda=1$ and 0.4 for $\lambda=3$. For these values the expansion (\ref{ddt-exp}) converges very rapidly and first three terms are enough to reproduce the full value with a deviation on the level of 0.3\% for $d_{\rm cr}$ and of 1.4\% for $\tilde{d}_{\rm cr}$.

Now we apply the results derived in the previous section to study fluctuations of various quantities in the isospin-symmetrical pion gas, which properties are described by Eqs. (\ref{eqs-sym}). In this case $I_n^{+}=I_n^{-}=I_n^{0}=I_n$ and relations (\ref{mixed-derivtives}) are essentially simplified.

Consider now fluctuations of the number of pions of a particular species (\ref{varpi-def}). For charged pions we get
\begin{align}
\varpi_{\pm\pm} &
=3\frac{m^{*2}T}{n}\Big(I_1
- \frac{\lambda I_3^2(4+ 5\lambda I_2)}{2(1+\lambda I_2) (2+5\lambda I_2)}\Big)\,,
\label{var-pp}
\end{align}
and thereby
$$\varpi_{\pm\pm}(T\to T_{\rm cr}) \to
\frac{3T_{\rm cr}\mu_{\rm cr}^{2}
}{2n}  I_3 (T\to T_{\rm cr})\to \infty\,.$$
For neutral pions we find
\begin{align}
\varpi_{00} =3\frac{Tm^{*2}}{n}\Big(I_1
-  \frac{\lambda I_3^2(3+5\lambda I_2)}{(1+\lambda I_2)(2+ 5\lambda I_2)}\Big)
 \,.
 \label{var-00}
\end{align}
At $T=T_{\rm cr}$ we obtain that
$$
\varpi_{00}(T_{\rm cr})= \frac{12T_{\rm cr} \mu_{\rm cr}^{2}  }{5\lambda n}
\Big(1+5\lambda \tilde{d}_{\rm cr}\Big)\,,
$$
so the variance of the number of neutral pions remains finite in the critical point. For the cross-variances of charged pions we obtain a negative quantity
\begin{align}
\varpi_{\pm\mp}
=-3\frac{Tm^{*2}}{n}
  \frac{\lambda I_3^2(4+5\lambda I_2)}{2(1+\lambda I_2)(2+5\lambda I_2)}\,.
  \label{var-pm}
\end{align}
For $T\to T_{\rm cr}$ we get $$ \varpi_{\pm\mp}(T\to T_{\rm cr})\to -3\frac{T_{\rm cr}
\mu_{\rm cr}^{2}}{2n} I_3 (T\to T_{\rm cr})\,. $$
At the end, for the cross-variances of the charged and neutral pions we find
\begin{align}
\varpi_{\pm 0}=-3\frac{Tm^{*2}}{n}
 \frac{\lambda I_3^2}{(1+\lambda I_2)(2+5\lambda I_2)} \,.
 \label{var-p0}
\end{align}
At $T=T_{\rm cr}$ this quantity remains finite,
$$\varpi_{\pm 0}(T\to T_{\rm cr})\to -3\frac{T_{\rm cr}\mu_{\rm cr}^{2}}{5\lambda n}  .$$
Thus, we see that all results for variances involving  neutral pions, $\varpi_{00}$ and $\varpi_{\pm 0}$, remain finite at $T_c$, whereas the variances associated  with only charged pions, $\varpi_{\pm\pm}$ and $\varpi_{\pm\mp}$, diverge, whereas the combinations $\varpi_{\pm\pm}+\varpi_{\pm\mp}$
remain finite. Also from Eqs.~(\ref{var-pp}), (\ref{var-00}), (\ref{var-pm}) and (\ref{var-p0} we find useful relation
\begin{align}
\varpi_{00}=\varpi_{\pm\pm}+\varpi_{\pm\mp}-\varpi_{\pm0} \,.
\label{identity}
\end{align}

Now we apply the above formulas for fluctuations of the observables, which can be more directly accessed in experiments. First, we consider the normalized variance of the total number of pions, $N$, which is defined in (\ref{var-N-gen}). For isosispin symmetrical case under consideration in this section  $c_a=1/3$ and therefore
\begin{align}
\varpi_N=\frac13\sum_{a,b} \varpi_{ab}\,.
\label{varN-gen}
\end{align}
Substituting here the results (\ref{var-pp}), (\ref{var-00}), (\ref{var-pm}), and (\ref{var-p0}) we obtain
\begin{align}
\varpi_N=
3\frac{Tm^{*2}}{n}\Big( I_1 - \frac{5 \lambda I_3^2}{2 + 5\lambda I_2} \Big) \,,
\label{var-N}
\end{align}
recovering thereby the expression derived in~\cite{KV18}. Note that Eq. (\ref{var-N})  has  the same form  Eq.~(\ref{waa-single}) but with   $\xi_a\to\xi_N=\frac52$.
Making use  the similarity between Eqs.~(\ref{var-N}) and (\ref{waa-single}) we can write the expression for $\varpi_N (T_{\rm cr})$ substituting $\xi_N$ in Eq.~(\ref{waa-fin}) instead of $\xi_a$, then we have
\begin{align}
\varpi_N (T_{\rm cr})&=\frac{6}{5} \frac{ T_{\rm cr}\mu_{\rm cr}^2}{\lambda n}\Big[ 1  +  10\lambda
\tilde{d}_{\rm cr}  \Big]\,.
\label{varN-Tc}
\end{align}

Now we turn to another quantity, $G$, which characterizes imbalance between charged and neutral pions. Its variance is defined in Eq.~(\ref{var-G-gen}). If all pion species are equally populated, the average of $G$ vanishes, $\langle G\rangle=0$. Then normalized variance of this quantity can be written
through the partial fluctuations (\ref{varpi-def}) as follows
\begin{align}
\varpi_G=
\frac14(\varpi_{++}+2\varpi_{+-}+\varpi_{--})+\varpi_{00}
-\varpi_{+0}-\varpi_{-0}\,.
\label{varG-gen}
\end{align}
With the help of relations (\ref{mixed-derivtives}) and (\ref{identity}) one can show that
\begin{align}
\varpi_G=\frac32\big(\varpi_N-3\varpi_{\pm0}\big)\,.
\end{align}
Since $\varpi_{\pm0}<0$, we immediately conclude that the quantity $\varpi_G$ is greater than $\varpi_N$.
Replacing Eqs.~(\ref{var-pp}), (\ref{var-00}), (\ref{var-pm}), and (\ref{var-p0}) in Eq.~(\ref{varG-gen}) we obtain
\begin{align}
\varpi_G &=
\frac92\frac{Tm^{*2}}{n}\Big[
I_1
- \frac{\lambda I_3^2}{(1+\lambda I_2)}
\Big]
\,.
\label{var-G}
\end{align}
This result is similar to that given by Eq.~(\ref{waa-single}), but now with $\xi_a\to 1$.
For $T\to T_{\rm cr}$ we immediately obtain
\begin{align}
\varpi_G(T_{\rm cr}) &=
\frac92\frac{ T_{\rm cr}\mu_{\rm cr}^{2}}{\lambda n}\Big(1+ 4\lambda\tilde{d}_{\rm cr}
\Big)\,.
\label{varG-Tc}
\end{align}
Thus the variance for $G$ remains finite at $T\to T_{\rm cr}$.

Note that the value of the normalized variance of the imbalance quantity $G$ significantly  differs  from the normalized variance of the total particle number $N$. We emphasize that the result for $\varpi_G$, Eq.~(\ref{var-G}), ought to be used  in analysis of experimental data of~\cite{Ryadovikov12}, where  fluctuations of neutral pions were studied in selected events with a fixed total pion number.

Now consider  fluctuations of the charge,  $Q$, in the system. In the neutral isospin-symmetrical system, the average of this quantity vanishes, $\langle Q\rangle =0$. The normalized  variance of the charge, which we define in (\ref{var-Q-gen}), can be expressed as
\begin{align}
\varpi_Q=\varpi_{++} + \varpi_{--} - 2\varpi_{+-}\,.
\label{varQ-def}
\end{align}
Substituting (\ref{var-pp}) and (\ref{var-pm}) in (\ref{varQ-def}) we find
\begin{align}
\varpi_Q=6\frac{T}{n} \frac{\partial h}{\partial \mu} =6\frac{Tm^{*2}}{n} I_1\,.
\label{var-Q}
\end{align}
This result is similar to that for the ideal pion gas, as all the terms explicitly dependent on $\lambda$ canceled out. Thus, $\varpi_Q$ is divergent at $T\to T_{\rm cr}$.
Applying Eqs.~(\ref{In-atTc}) and (\ref{I123-relat}) we have for  $T\to T_{\rm cr}$:
\begin{align}
\varpi_Q = \frac{3\mu_{\rm cr}^2 T_{\rm cr}}{\pi\sqrt{\alpha}n}\Big[
\frac{T}{T-T_{\rm cr}}+ 2\pi\sqrt{\alpha}\Big(\delta I_{{\rm cr},3} +
2(\tilde{d}_{\rm cr}+d_{\rm cr})
 \Big)\Big]\,,
 \label{var-Q-exp1}
\end{align}
where $\alpha$ is given in Eq.~(\ref{alpha-def}) in Appendix \ref{app:deriv}, cf. also Eq. (\ref{macr}).

One may ask why self-consistent inclusion of the interaction renders the variances $\varpi_N$ and $\varpi_G$ finite but $\varpi_Q$  divergent? To answer this question in Appendix~\ref{app:dFdN} we showed that the variance (\ref{varQ-def}) can be written through the derivatives of the free-energy density $F$ with respect to the charge density $n_Q= n_+-n_-$, for $n_Q\to 0$, cf. Eq.~(\ref{app1:varQ-3}).
The Coulomb contribution to the free-energy density appears because of the formation of a fluctuation characterized by a constant value of charge density $n_{Q}$ in a sphere of radius $R_{\rm fl}$ is equal to
\begin{align}
\delta F_{\rm Coul} = \frac{3}{5} n_Q^2\, V_{\rm fl} \frac{e^2}{R_{\rm fl}} \,,\quad
V_{\rm fl}=\frac{4\pi}{3} R_{\rm fl}^3\,,
\end{align}
where $e^2 \simeq 1/137$.
Replacing this result in Eq. (\ref{app1:varQ-3}) of Appendix~\ref{app:dFdN} we find for $T\to T_{\rm cr}$ that
$$
\varpi_Q (T\to T_{\rm cr})=3\frac{T_{\rm cr}}{n}\Big[\frac{\partial^2 F_{\rm Coul}}{\partial n_{Q}^2}\Big|_{n_Q=0}
\Big]^{-1}=
\frac{15T_{\rm cr}}{8\pi R_{\rm fl}^2 e^2 n}\,.$$
Here we took into account that for $T\to T_{\rm cr}$ there remains only the Coulomb term in the free energy depending on $n_Q$. Notice that $\varpi_N$ doest not depend on the size of the fluctuation,  whereas the Coulomb contribution in $\varpi_Q$ depends on $R_{\rm fl}$ due to a far-distance behavior of the Coulomb force.
Thus we get that $\varpi_Q/\varpi_N\sim \lambda/e^2 \mu^2 R_{\rm fl}^2$.
Since  $e^2\ll \lambda$, for relevant values of $R_{\rm fl}\sim $ several $1/m_\pi$ the resulting quantity $\varpi_Q$ could be considerably larger than $\varpi_N$.
Since in our study in this paper we disregard effects associated with the electromagnetic interaction compared with the effects of the strong interaction, we may employ $w_Q$ as follows from Eqs. (\ref{var-Q}), (\ref{var-Q-exp1}), being divergent for $T\to T_{\rm cr}$.

In terms of the observables, the result, $\varpi_Q\gg \varpi_N$, means that the stronger the multiplicity of an event deviates from the expected mean value the higher will be the probability that the numbers of positive and negative pions are different in this event.

\begin{figure}[t]
\centering
\includegraphics[width=8cm]{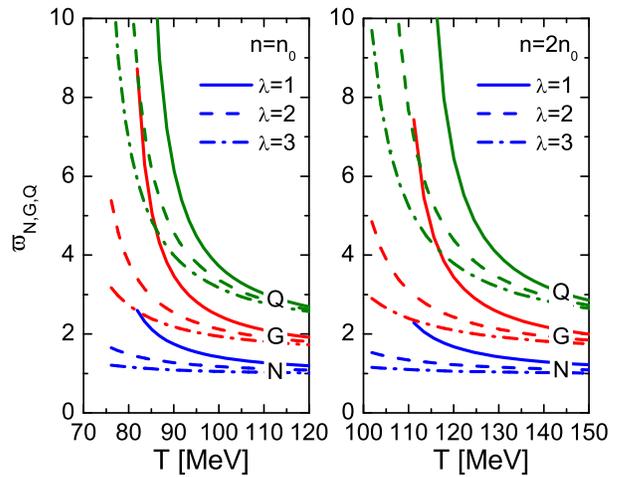}
\caption{Normalized variances of the total number of particles, $\varpi_N$, cf. (\ref{var-N}), the charged versus neutral pion imbalance, $\varpi_G$, cf. (\ref{var-G}), and the charge, $\varpi_Q$, cf. (\ref{var-Q}), in  isospin-symmetrical pion gas as functions of the temperature for $\lambda=1$ (solid lines), $\lambda=2$ (dashed lines), and $\lambda=3$ (dash-dotted lines) and for  $n=n_0$ and $n=2\,n_0$. }
\label{fig:fluct}
\end{figure}

Fig.~\ref{fig:fluct} demonstrates that variances for all three  quantities, $N$, $G$ and $Q$,  increase with a decrease of the temperature, whereby the hierarchy of fluctuation variances is $\varpi_{N}<\varpi_G<\varpi_Q$, and an increase of $\lambda$ leads to a reduction of the variances.

In the given section we have studied behavior of  fluctuations in the isospin-symmetrical pion gas. In the actual  heavy-ion collisions the initial nuclei have a charge. Thereby in a realistic situation  the pion gas has most probably a small positive charge imbalance. In some less probable events the charge imbalance might be also negative. In Sect.~\ref{sec:covariance}  we have derived general expressions valid for the system of arbitrary isospin composition. Now we are at the position to study in a more detail a situation, when at the moment of the chemical freeze-out (when the temperature is high and the chemical potential is tiny) the pion system is formed with an  isospin  imbalance.

\section{Fluctuations in the pion gas with small isospin imbalance}\label{sec:imbalance}

Let the numbers of pions of each species are $n_+$, $n_-$ and $n_0$ and the total density is $n=n_+ + n_- + n_0$. Let the system is  nearly isospin-symmetrical, i.e.,   $|\delta n_a|\ll n/3$, where \begin{align}
\delta n_a = n_a-n/3 \,.
\label{deltana}
\end{align}
We will calculate shifts of critical temperatures for various pion species, $T_{\rm cr}^a$, from the value for the isospin-symmetrical gas to determine the lowest temperature, $\mbox{max}T_{\rm cr}^{a}$, up to which our consideration is valid.

\subsection{Effective mass, chemical potential and critical temperature}\label{subsec:prop}

The condition $|\delta n_a|/n\ll 1$\, implies that the effective masses and the chemical potentials  $\mu_a=\mu+\delta\mu_a$ and $m^{*2}_a=m^2 + \delta m^{*2}_a$, where $\mu$ and $m^{*2}$ satisfy Eqs. (\ref{eqs-sym}), differ only a little from their values in the symmetrical case.  Expanding Eq.~(\ref{na-def}) up to linear order in $\delta n_a$, $\delta m_a^{*2}$ and $\delta\mu_a$ we find the relation
\begin{align}
\delta \mu_a= \Big(\frac{\prt h}{\prt \mu}\Big)^{-1}
\Big[\delta n_a -\frac{\prt h}{\prt m^{*2}}\delta m_a^{*2}\Big]
=\frac{\delta n_a}{m^{*2}I_1} + \frac{I_3}{2I_1} \frac{\delta m_a^{*2}}{m^*} \,.
\label{dmu}
\end{align}
A variation of the effective mass is given by the derivatives of the polarization operator, see Eq.~(\ref{pi-op}), $\delta m_a^{*2}=\sum_b\frac{\partial \Pi_a}{\partial m^{*2}_b}\delta m_b^{*2}+ \sum_b\frac{\partial \Pi_a}{\partial \mu_b}\delta \mu_b$, or explicitly
\begin{align}
\delta m_a^{*2} &= 2\lambda \frac{\partial (dm^{*2})}{\partial m^{*2}}\sum_{b} [\mathcal{K}]_{ab}  \delta m_b^{*2}
\nonumber\\
&+ 2\lambda  \frac{\partial (dm^{*2})}{\partial \mu}
\sum_{b}[\mathcal{K}]_{ab}\delta \mu_b
 \,.
\label{dM}
\end{align}
In Eqs.~(\ref{dmu}) and (\ref{dM}) all derivatives  are taken for $\delta n_a =0$, $\delta m_a^{*2}=0$ and $\delta\mu_a =0$, corresponding to the isospin-symmetrical state.
Substituting (\ref{dmu}) in (\ref{dM}) and solving the system of equations for the effective mass-shifts we find

\begin{align}
\delta m_+^{*2}=\delta m_-^{*2}=\lambda \frac{I_3}{I_1}\frac{(2+5C)(\delta n_+ +\delta n_-) + \delta n_0}{m^*(1+2C)(1+5C)} \,,
\nonumber\\
\delta m_0^{*2}=\lambda \frac{I_3}{I_1}\frac{\delta n_+ +\delta n_- + (3+10C) \delta n_0}{m^*(1+2C)(1+5C)} \,,
\label{eff-mass-exp}
\end{align}
with $C=\frac{\lambda}{2}(I_2-I_3^2/I_1)$.
Substituting (\ref{eff-mass-exp}) in (\ref{dmu}), we recover the changes of the chemical potentials. Quantities $I_{1,3}$ and $C$ in Eqs.~(\ref{dmu}) and (\ref{eff-mass-exp}) are calculated for the isospin-symmetrical matter.

The shift of the critical  temperatures, $\delta T_{\rm cr}^a=T_{\rm cr}^a-T_{\rm cr}$, because of the variation of the particle densities  is determined from the relation $\mu_a(T_{\rm cr}+\delta T_{\rm cr}^a)=m_a^*(T_{\rm cr} +\delta T_{\rm cr}^a)$,  which we rewrite as $\mu(T_{\rm cr}+\delta T_{\rm cr}^a) + \delta\mu_a(T_{\rm cr}) = m^*(T_{\rm cr} +\delta T_{\rm cr}^a) + \delta m^{*}_a(T_{\rm cr})$. For  $|\delta T_{\rm cr}^a|\ll T_{\rm cr}$ in linear approximation we find
\begin{align}
\delta T_{\rm cr}^a=\frac{\delta m_a^{*2}(T_{\rm cr}) - 2m^*(T_{\rm cr})\delta \mu_a(T_{\rm cr})}{2m^*(T_{\rm cr})\Big(\frac{\partial \mu}{\partial T}-\frac{\partial m^*}{\partial T}\Big)\Big|_{T_{\rm cr}}} \,.
\label{Tc-shift}
\end{align}
Partial derivatives appearing in the denominator can be written as
\begin{align}
\Big(\frac{\partial \mu}{\partial T}-\frac{\partial m^*}{\partial T}\Big)=-\frac{n\chi}{T\, m^{*2}I_1(1+ 5 C)}\,,
\label{dmumdt}
\end{align}
where the function $\chi(T)$ is defined in Eq. (\ref{app:chi}) of Appendix \ref{app:deriv} and its limiting value at $T\to T_{\rm cr}$ is given in (\ref{chiTc}).
Knowing the critical temperatures for various pion species we can find the maximal one, $T_{\rm cr}^{\tilde{a}}=\max_a T_{\rm cr}^a$. Our consideration is valid only for temperatures $T\ge T_{\rm cr}^{\tilde{a}}$, since already slightly below the temperature $T_{\rm cr}^{\tilde{a}}$ one has to take into account  presence of the Bose-Einstein condensate of the given pion species $\tilde{a}$. Thus, for $T<\tilde{a}$ the pion excitation spectra for all species must be recalculated. Therefore, the values of other two critical temperatures $T_{\rm cr}^{b}$ for $b\neq \tilde{a}$, being calculated without inclusion of the BEC of the species $\tilde{a}$ prove to be physically irrelevant.

Consider first the system at a fixed density $n$. Then
\begin{align}
\sum_a\delta n_a=0\label{sum}
\end{align}and we can rewrite
\begin{align}
\delta n_\pm &=\frac13\delta n_G\pm\frac12\delta n_Q\,,
\nonumber\\
\delta n_0  &= -\frac23\delta n_G\,,
\label{dnG-dnQ}
\end{align}
where $\delta n_Q$ is a charged density of the system and $\delta n_G$ characterizes the excess of the number of charged pions above the neutral ones. Now two variables  $\delta n_Q$ and $\delta n_G$ can be considered as independent ones,  instead of the two variables chosen from $\delta n_\pm$ and $\delta n_0$, with  the relation (\ref{sum}) between them.
Fluctuations  of the quantities $Q$ and $G$ introduced in Eqs.~(\ref{var-Q-gen}) and (\ref{var-G-gen}), respectively, are characterized by the variances $\varpi_Q$ and $\varpi_G$.

Consider specific variations:

\begin{figure}
\centering
\includegraphics[width=8.5cm]{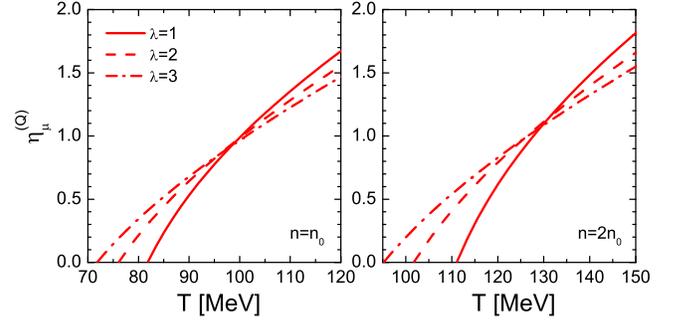}
\caption{
Susceptibility parameters of the chemical potential, $\eta_\mu^{(Q)}$, to variations of the charge density plotted as functions of a temperature for  $\lambda=1$, 2 and 3 and  $n=n_0$ and $2n_0$.  }
\label{fig:eta-Q}
\end{figure}

(i) {\em Variations of $\delta n_Q$ at $\delta n_G=0$.}
Then $\delta n_\pm=\pm\delta n_Q/2$ and $\delta n_0=0$, and from Eqs.~(\ref{dmu})  and (\ref{eff-mass-exp}) we obtain
\begin{align}
\delta m_a^{*2}=\delta \mu_0=0\,,\,\, \,\frac{\delta\mu_\pm}{m^*}=\pm \frac{\delta n_Q}{2n}\eta_\mu^{(Q)}
\,,\,\,\,
\eta_\mu^{(Q)}=\frac{n}{m^{*3} I_1}\,,
\label{etaQ}
\end{align}
where we introduced the susceptibility  $\eta_\mu^{(Q)}$.
Thus, the variation of the  charge of the system, while keeping the total number of particles fixed, does not lead to a change of the effective pion mass, and only chemical potentials of charged pions change to accommodate the difference in $\pi^+$ and $\pi^-$ concentrations.

The quantity $\eta^{(Q)}_\mu$ is shown in Fig.~\ref{fig:eta-Q} as a function of the temperature for two values of the density and three values of the coupling constant. As we see, $\eta^{(Q)}_\mu$ decreases monotonously with the temperature decrease and vanishes at the critical temperature of the BEC.
Note that the lines do not cross at one point, but there are three crossings at values of the temperature $T$ separated by $\simeq \pm2$\,MeV.

The variation of the critical temperature is obtained after substitution of  Eq.~(\ref{etaQ}) in (\ref{Tc-shift}) and using   Eq.~(\ref{dmumdt}),
\begin{align}
\frac{\delta T_{\rm cr}^\pm}{T_{\rm cr}} &=\pm\frac{\delta n_Q}{2n}\frac{\eta_{T_{\rm cr}}^{(Q)}}{\eta_n}\,,\quad \delta T_{\rm cr}^0=0\,,\label{dTc-Q}
\\
\eta_n &= 1-\frac{2\mu_{\rm cr}^3}{n}(\tilde{d}_{\rm cr}+d_{\rm cr}) \,,
\nonumber\\
\eta_{T_{\rm cr}}^{(Q)} &=
1 + \frac{5\lambda}{\chi_{\rm cr}}(\tilde{d}_{\rm cr} + d_{\rm cr})
- 20\lambda  \frac{\mu_{\rm cr}^3}{ n\chi_{\rm cr}} (\tilde{d}_{\rm cr} + d_{\rm cr})^2\,.\nonumber
\end{align}
The factor $\eta_n$ is separated so that for $\lambda=0$ and $\eta_{T_{\rm cr}}^{(Q)}=\eta_\mu^{(Q)}(T=T_{\rm cr})=1$ Eq. (\ref{dTc-Q}) reduces (after the replacement $\mu=m^*\to m_\pi$)  to the one  following directly from the variation of the second Eq. (\ref{eqs-sym}) at $m^*=m_\pi$, $\mu=m_\pi$.

\begin{figure}
\centering
\includegraphics[width=5cm]{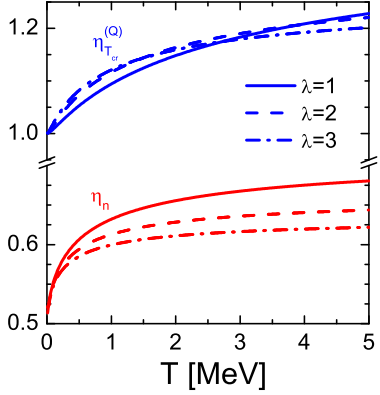}
\caption{
Susceptibility parameters of the critical temperature, $\eta_{T_{\rm cr}}^{(Q)}$ and $\eta_n$, to variations of the charge density plotted as functions of density for three values of the coupling constant $\lambda$. }
\label{fig:eta-TQ}
\end{figure}

The functions $\eta_{T_{\rm cr}}^{(Q)}$ and $\eta_n$ are shown in Fig.~\ref{fig:eta-TQ}.
They are growing functions of $n$ and both are limited from above in view of the constraint $d_{\rm cr}<1/(10\lambda)$ following from the relation (\ref{constr-eq}).
The lower limits of these functions are realized for $n\to 0$ when $d_{\rm cr}\to 0$ and $\tilde{d}_{\rm cr}/d_{\rm cr}\to\frac12$ and we have
\begin{align}
({\mu_{\rm cr}^3}d_{\rm cr}/{n})|_{n\to 0}\to 1/6
\,,
\label{d-n-zero}
\end{align}
that follows from the expression for the critical temperature
$$\big({T_{\rm cr}}/{\mu_{\rm cr}}\big)^{3/2}=\frac{(2\pi)^{3/2}\, n}{\zeta(\frac32)3m^3}$$
 valid in the non-relativistic limit.

\begin{figure}
\centering
\includegraphics[width=8cm]{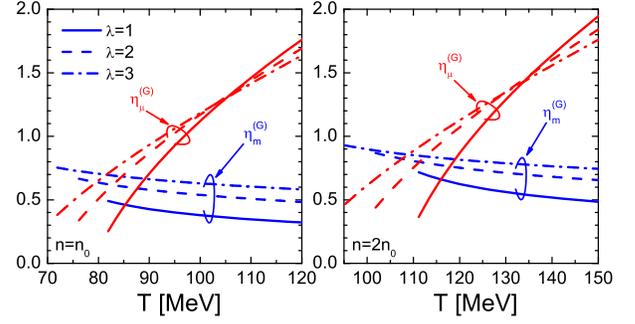}
\caption{
Susceptibility parameters of the effective mass, $\eta_m^{(G)}$, and chemical potential, $\eta_\mu^{(G)}$, to variations of the charged-versus-neutral pion imbalance density plotted as functions of temperature for  $\lambda=1$, 2 and 3 and for  $n=n_0$ and $2n_0$. }
\label{fig:eta-G}
\end{figure}

(ii) {\em Variations of $\delta n_G$ at $\delta n_Q =0$.} Then $\delta n_\pm=\delta n_G/3$ and $\delta n_0=-2\delta n_G/3$. Expressions  (\ref{dmu}) and (\ref{eff-mass-exp}) yield now
\begin{align}
\frac{\delta m_\pm^{*2}}{m^{*2}} & = -\frac12\frac{\delta m_0^{*2}}{m^{*2}}= \frac{\delta n_G}{3 n} \eta_m^{(G)}\,,\,\,\eta_m^{(G)} = \frac{2\lambda I_3 n}{m^{*3} I_1 (1+2C)} \,,
\nonumber\\
\frac{\delta\mu_\pm}{m^*}  & = -\frac12\frac{\delta\mu_0}{m^*} = \frac{\delta n_G}{3 n}\eta_\mu^{(G)}
\,,\eta_\mu^{(G)} = \frac{n}{m^{*3} I_1}
\Big[1+\frac{\lambda I_3^2/I_1}{1+2C}\Big]\,.
\label{nG-res}
\end{align}
The susceptibilities  $\eta_\mu^{(G)}$ and $\eta_m^{(G)}$ are plotted in Fig.~\ref{fig:eta-G} as functions of the temperature for $n=n_0$ and $2n_0$ and for $\lambda=1$, 2 and 3.  We see that the susceptibility parameter $\eta_m^{(G)}$ decreases very weakly with a temperature increase and increases with an increase of $\lambda$ and $n$. On the other hand, the quantity $\eta_\mu^{(G)}$ is  rapidly and monotonously increasing function of $T$. At $T=T_{\rm cr}$ we have
$\eta_m^{(G)}(T_{\rm cr}) = 2\eta_\mu^{(G)}(T_{\rm cr})$. Notice also that, as in Fig.~\ref{fig:eta-Q}, the lines for $\eta_{T_{\rm cr}}^{(G)}$ do not cross in one point.

The imbalanced system with $\delta n_G\neq 0$ changes dynamically its composition in  reactions $\pi^+ +\pi^-\leftrightarrow 2\pi^0$. These reactions are controlled by the difference of chemical potentials $\Delta\mu=
\mu_+ + \mu_- - 2\mu_0=\delta\mu_+ + \delta\mu_- - 2\delta\mu_0$. As follows from (\ref{nG-res}), $\Delta\mu=\gamma\, \delta n_G$, where $\gamma=\frac43\frac{m^*}{n}\eta_\mu^{(G)}>0$. Hence, if  $\delta n_G>0$ (this means an excess of charged pions) then $\Delta\mu>0$ and the reaction balance is shifted to the conversion of charged pions into the neutral ones. Oppositely, if $\delta n_G<0$ and there are more neutral pions than charged ones, the neutral pions are  converted into the charged ones, since $\Delta\mu<0$. Thus, the interacting isospin-symmetrical pion gas is stable with respect to deviations from the equilibrium between charged and neutral pions, i.e. fluctuations in the  quantity $G$ do not grow spontaneously.

\begin{figure}
\centering
\includegraphics[width=5cm]{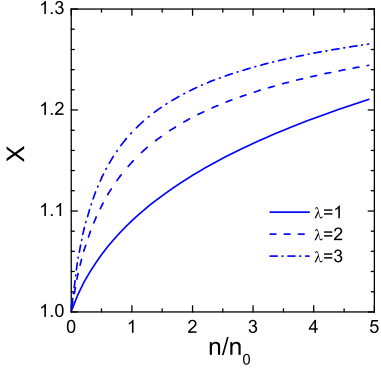}
\caption{Ratio $X=\eta_{T_{\rm cr}}^{(Q)}/\eta_{T_{\rm cr}}^{(G)}$ determining shifts of  critical temperatures of the Bose-Einstein instability because of variations of the charge density $\delta n_Q$, see Eq.~(\ref{dTc-Q}), and the variation of the charged-vs.-neutral pion densities, $\delta n_G$, see Eq.(\ref{dTcr-G}), around the isospin-symmetrical state with the density $n$. Solid, dashed and dash-dotted lines correspond to $\lambda =1$, 2 and 3, respectively.}
\label{fig:X-n}
\end{figure}

Shifts of the critical temperatures induced by the variations of $\delta n_G$
are obtained by substituting Eq.~(\ref{nG-res}) in (\ref{Tc-shift}),
\begin{align}
\frac{\delta T_{\rm cr}^\pm}{T_{\rm cr}}& = -\frac12\frac{\delta T_{\rm cr}^0}{T_{\rm cr}} = \frac{\delta n_G}{3n}
\frac{\eta_{T_{\rm cr}}^{(G)}}{\eta_n}\,,
\nonumber\\
 \eta_{T_{\rm cr}}^{(G)}&=\eta_{T_{\rm cr}}^{(Q)}/X \,,
\label{dTcr-G}
\end{align}
where
\begin{align}
X =
\frac{1 + 4 \lambda \tilde{d}_{\rm cr}}{1+ 2\lambda(\tilde{d}_{\rm cr}-d_{\rm cr})}\,.
\label{X-def}
\end{align}
This quantity is plotted in Fig.~\ref{fig:X-n} as a function of the density for three values of the coupling constant $\lambda$. We see that the ratio is a monotonically increasing function of $n$ and it increases with an increase of $\lambda$. Taking into account the constraint $d_{\rm cr}<1/(10\lambda)$ following from (\ref{constr-eq}) and that $\frac12<{\tilde{d}_{\rm cr}}/{d_{\rm cr}}<1$ as follows from definitions (\ref{d-def}) and (\ref{def-d-dt}) we can limit $X$ from above as
\begin{align}
X< 1 + 4 \lambda d_{\rm cr} < X_{\rm max}=\frac{14}{9}\,.
\label{X-max}
\end{align}

Now consider a general case when both $\delta n_Q$ and $\delta n_G$ can be nonzero.
Variations can be constructed as a linear superposition of the above results obtained in limit cases $\delta n_G =0$ and $\delta n_Q =0$. In the general case, it proves to be more convenient to  express $\delta n_Q$ and $\delta n_G$ through $\delta n_\pm$  as
$\delta n_Q=\delta n_+-\delta n_-$ and
$\delta n_G=\frac12(\delta n_+ +\delta n_-)-\delta n_0=\frac32(\delta n_+ +\delta n_-)$, where in the last expression we used that $\sum_a \delta n_a=0$.
As a result, the effective masses and chemical potentials  (up to terms linear in $\delta n_\pm$) are expressed as:
\begin{align}
\delta m_\pm^{*2} &= -\frac12 \delta m_0^{*2} =  \frac{m^{*2}}{2n} \eta_m^{(G)}
( \delta n_+ +\delta n_-)\,,
\label{dm-gen}
\end{align}
and
\begin{align}
\delta\mu_\pm  &=\frac{m^*}{2n}\Big[(\eta_\mu^{(Q)}+\eta_\mu^{(G)}) \delta n_\pm
-(\eta_\mu^{(Q)}-\eta_\mu^{(G)}) \delta n_\mp \Big] \,,
\nonumber\\
\delta\mu_0  &=-\frac{m^*}{n}\eta_\mu^{(G)}(\delta n_+ +\delta n_-)\,.
\label{dmu-gen}
\end{align}
Correspondingly, shifts of the critical temperatures of the BEC for various pion species are
\begin{align}
\delta T_{\rm cr}^\pm &= \frac{m^*}{2n\eta_n}\Big[
 ( \eta_{T_{\rm cr}}^{(Q)}  + \eta_{T_{\rm cr}}^{(G)}) \delta n_\pm
-( \eta_{T_{\rm cr}}^{(Q)}  - \eta_{T_{\rm cr}}^{(G)}) \delta n_\mp
\Big]\,,
\nonumber\\
\delta T_{\rm cr}^0 &=-\frac{m^*}{n\eta_n}\eta_{T_{\rm cr}}^{(G)}(\delta n_+ +\delta n_-)
\,.
\label{dT-dn}
\end{align}

\begin{figure*}
\centering
\includegraphics[width=5.9cm]{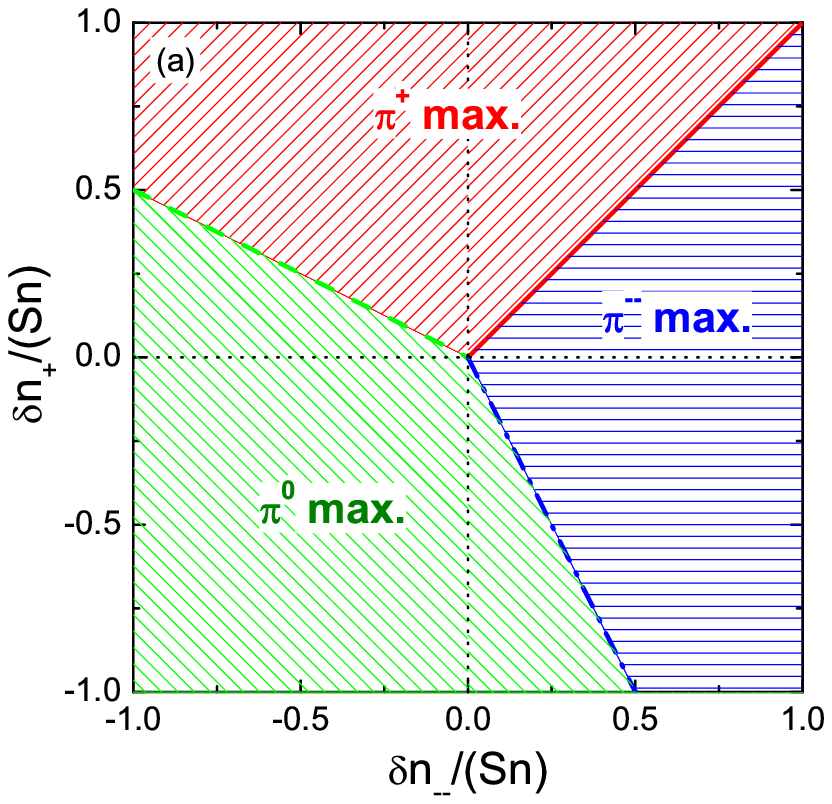}
\includegraphics[width=5.9cm]{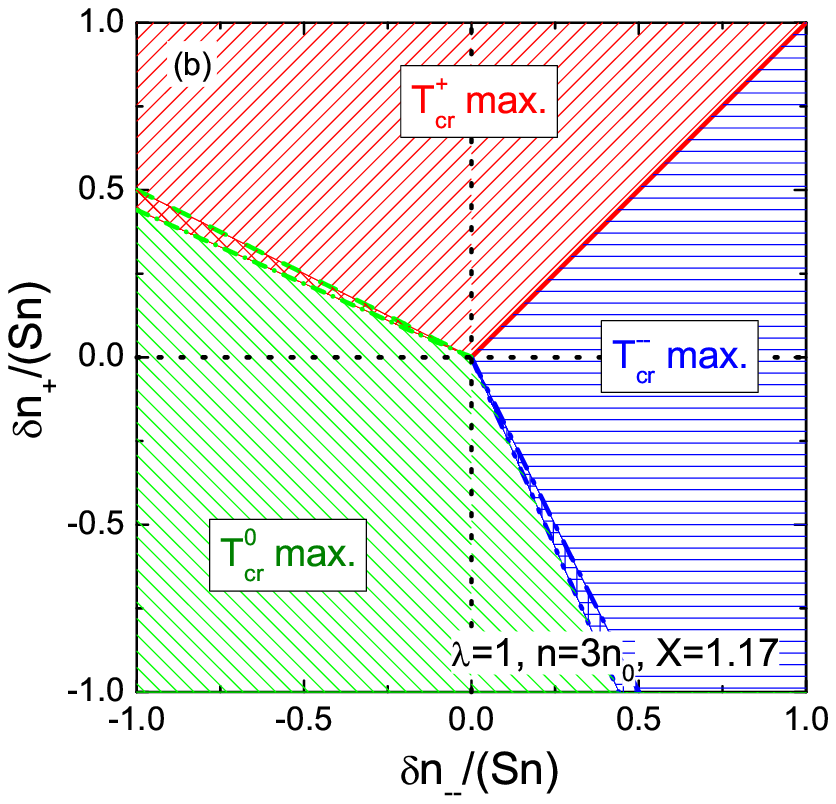}
\includegraphics[width=5.9cm]{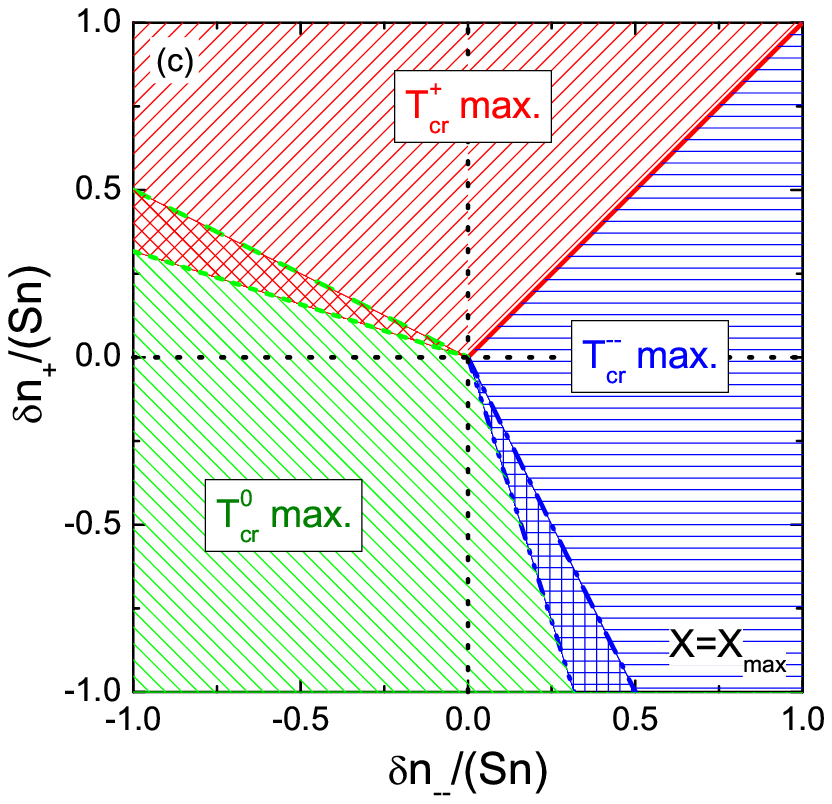}
\caption{The $(\delta n_-,\delta n_+)$ plane showing a state with an arbitrary small isospin imbalance at a condition $\sum_a\delta n_a=0$. Since the results for the ratios $\delta n_\pm/n$ are valid only for small density variations we introduce a scaling factor $S\ll 1$.
Panel~(a): Regions with maximal density of pion species $a$. Solid, dashed and dash-dotted lines are the zero lines for inequalities (\ref{nz-1}), (\ref{nz-2}) and (\ref{nz-3}), respectively.
Panel~(b): Regions where the maximal critical temperatures are realized for pions of a specific species. Short-dashed and dash-dot-dotted lines are the zero lines for inequalities (\ref{nzT-2}) and (\ref{nzT-3}), calculated for $\lambda=1$ and $n=3\, n_0$, that corresponds to $X=1.17$. Solid, dashed and dash-dotted lines are the same as on panel~(a). Doubly-hatched regions correspond to the case when the neutral pions are most abundant in the system but the maximal critical temperature is $T_{\rm cr}^{+}$ in the upper region and $T_{\rm cr}^{-}$ in the lower region.  On panel~(c) it is shown the same as in panel~(b), but for  $X=X_{\rm max}$ given in Eq.~(\ref{X-max}).  }
\label{fig:pm-plane}
\end{figure*}

Now we may address the key question of this section, if  a species $\tilde{a}$, for which the density is higher than for other species, has the largest critical temperature $T_{\rm cr}^{\tilde{a}}$. An arbitrary deviation from isospin-symmetrical state can be characterized by two densities $\delta n_+$ and $\delta n_-$ with the variation of the neutral pion density given by $\delta n_0=-(\delta n_+ +\delta n_-)$ following Eq. (\ref{sum}). Then we can  fix  regions in  2-dimensional $(\delta n_-,\delta n_+)$ plane, where the variations $\delta n_{\pm,0}$ and the values of the critical temperatures $T_{\rm cr}^{\pm,0}$ are maximal.

Let us study  regions characterized by different relations between densities of the species.
The regions are determined by three inequalities comparing the density variations $\delta n_{+,-,0}$. Using  the relation $\delta n_0=-(\delta n_+ + \delta n_-)$ we can reduce inequalities among three $\delta n_a$ to inequalities between $\delta n_+$ and $\delta n_-$
and obtain
\begin{align}
&\delta n_+ - \delta n_- \ge 0\,,
\label{nz-1}\\
&\delta n_+ -\delta n_0 \ge 0 \rightarrow \delta n_+\ge -\frac12 \delta n_-\,,
\label{nz-2}\\
&\delta n_- -\delta n_0\ge 0 \rightarrow \delta n_+ \ge -2\delta n_-\,.
\label{nz-3}
\end{align}
From the analysis of these inequalities we find three regions on the $(\delta n_-,\delta n_+)$ plane,  where $\pi^+$, either $\pi^-$ or $\pi^0$  are the most abundant species.  These regions are shown in Fig.~\ref{fig:pm-plane}a by different hatching.
For $\delta n_->0$ the solid line in Fig.~\ref{fig:pm-plane}a,  $\delta n_+ = \delta n_-$, divides the regions with maximal concentrations of $\pi^+$ (above the line) and $\pi^-$ (below). For $\delta n_-<0$ the dashed line, $\delta n_+ = -\frac12 \delta n_-$, divides regions of $\pi^+$ and $\pi^0$ dominance (above and below the line, respectively).
Also, the dash-dotted line, $\delta n_+ =-2\delta n_-$, separates regions of $\pi^-$ and $\pi^0$ dominance.

To find regions of the maximal $\delta T_{\rm cr}^a$,  we consider three differences of the critical temperature shifts,
\begin{align}
&\delta T_{\rm cr}^+ - \delta T_{\rm cr}^- =
   2\frac{m^*\eta_{T_{\rm cr}}^{(Q)}}{2n\eta_n} \,( \delta n_+ -\delta n_-)\,,
\nonumber\\
&\delta T_{\rm cr}^+ -\delta T_{\rm cr}^0  =
 \frac{m^*\eta_{T_{\rm cr}}^{(G)}}{2n\eta_n}\big[ ( X + 3)\, \delta n_+ - ( X  - 3)\, \delta n_-\big]\,,
\nonumber\\
&\delta T_{\rm cr}^- - \delta T_{\rm cr}^0 =\frac{m^*\eta_{T_{\rm cr}}^{(G)}}{2n\eta_n}\big[  ( X + 3)\, \delta n_- - ( X  - 3)\, \delta n_+\big]\,,
\nonumber
\end{align}
where we used (\ref{dT-dn}) and  (\ref{X-def}). These relations hold
at least in the range of the applicability of the linear approximation that we use in this section.

We can reduce three inequalities among $\delta T_{\rm cr}^a$ to inequalities between $\delta n_+$ and $\delta n_-$,
\begin{align}
\delta T_{\rm cr}^+ - \delta T_{\rm cr}^-\ge 0 &\rightarrow \delta n_+\ge \delta n_-\,,
\label{nzT-1}\\
\delta T_{\rm cr}^+ - \delta T_{\rm cr}^0\ge 0 &\rightarrow \delta n_+\ge -\Big(\frac12 -
\frac{3(X-1)}{2(3+X)}\Big)\delta n_-\,,
\label{nzT-2}\\
\delta T_{\rm cr}^- - \delta T_{\rm cr}^0\ge 0 &\rightarrow \delta n_+\ge -\Big(2 +
\frac{3(X-1)}{(3-X)}\Big)\delta n_-\,,
\label{nzT-3}
\end{align}
where we employed that $X<X_{\rm max}<3$.
We brought here inequalities (\ref{nzT-2}) and (\ref{nzT-3}) in the form closely resembling  inequalities (\ref{nz-2}) and (\ref{nz-3}), respectively. Therefore, we can directly see how the border lines between regions with maximal critical temperature for a certain pion species may differ from the regions, where the given species is most abundant.

For $\lambda=0$ we have $X=1$ and inequalities in~(\ref{nzT-1}), (\ref{nzT-2}), and (\ref{nzT-3}) become identical to inequalities in~(\ref{nz-1}), (\ref{nz-2}), and (\ref{nz-3}). This means that  \emph{in the case of an ideal pion gas the critical temperature $T_{\rm cr}^{\tilde{a}}$ is maximal for the most abundant pion species $\tilde{a}$.}
For $\lambda \neq 0$ we have $X>1$, since $0<\tilde{d}<d$ following Eqs. (\ref{d-def}), (\ref{def-d-dt}), see also Fig. \ref{fig:X-n}. We see that the slope of the border line [Eq.~(\ref{nzT-2})] between the regions with maximal $T_{\rm cr}^{+}$ and $T_{\rm cr}^{0}$  decreases, and for $\delta n_-<0$ this border lies below the border [Eq.~(\ref{nz-2})], which separates regions of dominance of $\pi^+$ and $\pi^0$ meson, respectively, cf. dashed and short-dashed lines in Fig.~\ref{fig:pm-plane}b. On the other hand, the slope of border line [Eq.~(\ref{nzT-3})] separating regions with maximal  $T_{\rm cr}^{-}$ and $T_{\rm cr}^{0}$,  becomes steeper and for $\delta n_->0$ this line falls below the line [Eq.~(\ref{nz-3})], which separates the regions of the $\pi^-$ and $\pi^0$ dominance, cf. dash-dotted and dash-dot-dotted lines in Fig.~\ref{fig:pm-plane}b.  Thus, there appear \emph{two regions, where although the most abundant species is $\pi^0$, the maximal critical temperature is realized for $\pi^+$ mesons in one region and for $\pi^-$ mesons in the other one.} Both regions are marked by double hatching.  Figure~\ref{fig:pm-plane}b is calculated for $X=1.17$ corresponding to the density $n=3n_0$ and $\lambda=1$.
With an increase of $X$ the anomalous regions grow and the case of $X=X_{\rm max}=14/9$ is shown in Fig.~\ref{fig:pm-plane}c.

 {(iii)~\em{Isospin imbalance with a variation of the density.}}
Let us show how the relations derived above at the fulfilled condition (\ref{sum}) are applied in the case when this condition is not satisfied.
Assume we have an isospin-symmetrical system with a density $n$, i.e. at $n_+=n_-=n_0=n/3$. Let us  change the densities of $\pi^+$, $\pi^-$ and $\pi^0$ by small quantities $\Delta n_+$, $\Delta n_-$ and $\Delta n_0$, respectively, with $\sum_a\Delta n_a \neq 0$.
The total density is now $n'=n+\sum_a \Delta n_a$. In order the quantity $n'$ would be the density of the isospin-symmetrical system characterized by the densities of each pion species  $n'/3$, the deviations from this equilibrium density should satisfy equations
\begin{align}
\delta n_+ &= \frac{n}{3} + \Delta n_+ - \frac{n'}{3}
 = \frac23 \Delta n_+ -\frac13( \Delta n_- + \Delta n_0) ,
\nonumber\\
\delta n_- &=\frac{n}{3} + \Delta n_-  - \frac{n'}{3}
= \frac23 \Delta n_- -\frac13( \Delta n_+ + \Delta n_0) ,
\nonumber\\
\delta n_0 &=\frac{n}{3} + \Delta n_0  - \frac{n'}{3}
=\frac23 \Delta n_0 -\frac13( \Delta n_+ + \Delta n_-).
\label{deltn-shifted}
\end{align}
In terms of $\delta n_Q$ and $\delta n_G$ we have, respectively
\begin{align}
\delta n_Q=\Delta n_+ -\Delta n_- \,,\,\, \delta n_G =\frac{\Delta n_+ +\Delta n_-}{2}-
\Delta n_0\,.
\end{align}
We see that the densities (\ref{deltn-shifted}) satisfy now the condition $\sum_a \delta n_a=0$ and
 the expressions derived above, (\ref{dm-gen}), (\ref{dmu-gen}), and (\ref{dT-dn}), are valid after the replacements $n\to n'$ with $m^*$, $\mu$  and all $\eta$'s  now evaluated at the density $n'$.

For completeness let us now  recalculate effective masses, chemical potentials and the critical temperature for new densities in the isospin-symmetrical case. We can still apply expressions (\ref{dmu}), (\ref{eff-mass-exp}) and (\ref{Tc-shift}) for a small variation of the total particle density without any change of the isospin composition $\delta n_a =\delta n/3 =
\frac13\sum_a\Delta n_a$. Then we find
\begin{align}
\frac{\delta m_a^{*2}}{m^{*2}} &= \frac{\delta n}{3n}\eta_m^{(N)}\,,
\,\,
\eta_m^{(N)}=\frac{5\lambda I_3 n}{m^{*3} I_1(1 + 5C)} \,,
\nonumber\\
\frac{\delta\mu_a}{m^*} &= \frac{\delta n}{3 n}\eta_\mu^{(N)}\,,
\,\,
\eta_\mu^{(N)}=\frac{n\big(1+\frac52\lambda I_2\big)}{m^{*3} I_1(1+5C)}\,.
\label{etas-N}
\end{align}
Certainly, the same relations, which we derived here from expansions of Eq.~(\ref{na-def}), could be obtained directly from the variations performed in Eqs. (\ref{eqs-sym}).

\begin{figure}
\centering
\includegraphics[width=8cm]{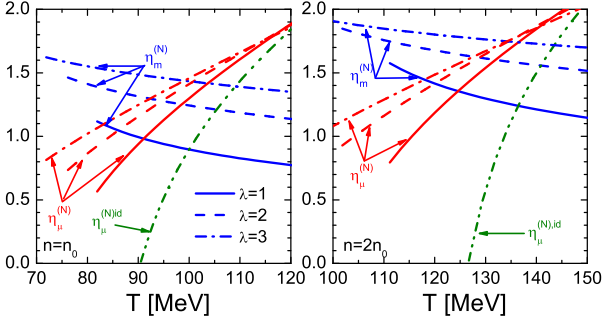}
\caption{
Susceptibility parameters of the effective mass, $\eta_m^{(N)}$, and chemical potential, $\eta_\mu^{(N)}$, to variations of the particle density plotted as functions of temperature for  $\lambda=1$, 2 and 3 and  $n=n_0$ and $2n_0$. The dash-double-dotted lines depict the susceptibility parameter $\eta_\mu^{(N),\rm id}$ for the ideal pion gas, see Eq.~(\ref{eta-mu0}). }
\label{fig:eta-N}
\end{figure}

Quantities $\eta_m^{(N)}$ and $\eta_\mu^{(N)}$ are shown in Fig.~\ref{fig:eta-N} as functions of a  temperature. In general $\eta_m^{(N)}$ decreases rather weakly with increase of $T$.
The quantity $\eta_\mu^{(N)}$ demonstrates much stronger dependence on $T$, than $\eta_m^{(N)}$, increasing by a factor $3-4$ with an increase of $T$ from $T\simeq T_{\rm cr}$ to $\sim 1.5 T_{\rm cr}$. It is instructive to compare the $\eta_\mu^{(N)}$ parameter with the corresponding parameter for the ideal pion gas
\begin{align}
\eta_\mu^{(N), \rm id}=\frac{n}{m_\pi^3 I_1}\,,
\label{eta-mu0}
\end{align}
which is shown in Fig.~\ref{fig:eta-N} by the dash-dotted line.  We see the qualitative difference in the behaviour of this parameter, when the temperature approaches $T_{\rm cr}$ from above and the interaction is switched on. Since $I_1$ increases strongly for $T\to T_{\rm cr}$, the quantity $\eta_\mu^{(N),\rm id}$ also decreases strongly and tends to zero in contrast to $\eta_\mu^{(N)}$ in Eq.~(\ref{etas-N}), where the divergency of $I_1$ in  the denominator is compensated by the divergency of $I_2$ in the numerator.
We also have $\eta_m^{(N)}(T_{\rm cr})=2\eta_\mu^{(N)}(T_{\rm cr})$, and therefore
\begin{align}
\Big(\frac{\partial m^*}{\partial n} - \frac{\partial \mu}{\partial n}\Big)\Big|_{T\to T_{\rm cr}}\to 0\,.
\end{align}

The shift of the critical temperature, which in the given case is the same for all pion species, can be presented with the help of Eq.~(\ref{Tc-shift}) as
\begin{align}
\frac{\delta T_{\rm cr}^a}{T_{\rm cr}} &= \frac{\delta n}{3n}\frac{\eta_{T_{\rm cr}}^{(N)}}{\eta_n}
\,,\,\,  \eta_n=1-\frac{2\mu_{\rm cr}}{n}(\tilde{d}_{\rm cr}+d_{\rm cr}) \,,
\nonumber\\
 \eta_{T_{\rm cr}}^{(N)} &= 1 - 10\lambda\frac{(\tilde{d}_{\rm cr} + d_{\rm cr})^2}{\mu_{\rm cr} n\chi_{\rm cr}}
 \,.
 \label{etaTc-N}
\end{align}

Now we  present the results for shifts of effective masses, chemical potentials and critical temperature for the density variations (\ref{deltn-shifted}):

for mass shifts
\begin{align}
\delta m^{*2}_\pm &= \frac{m^{*2}}{3n}
\Big[
\frac{\eta_m^{(G)} +2 \eta_m^{(N)}}{2} (\Delta n_+ +\Delta n_-)
\nonumber\\
&+
(\eta_m^{(N)} - \eta_m^{(G)})\Delta n_0
\Big],
\nonumber\\
\delta m^{*2}_0 &= \frac{m^{*2}}{3n}
\Big[(\eta_m^{(N)} - \eta_m^{(G)})(\Delta n_+ +\Delta n_-)
\nonumber\\
&+ (2\eta_m^{(G)} + \eta_m^{(N)})\Delta n_0
\Big],
\label{dm-general}
\end{align}
for the chemical potentials
\begin{align}
\delta\mu_\pm &=\frac{m^*}{2n}\Big[
\big(\frac{\eta_\mu^{(G)} +2 \eta_\mu^{(N)}}{3} \pm \eta_\mu^{(Q)}\big)\Delta n_+
\nonumber\\
&\big(\frac{\eta_\mu^{(G)} +2 \eta_\mu^{(N)}}{3} \mp \eta_\mu^{(Q)}\big)\Delta n_-
+\frac{\eta_\mu^{(N)} - \eta_\mu^{(G)}}{3} \Delta n_0\Big],
\nonumber\\
\delta \mu_0 &= \frac{m^*}{3n}\Big[
(\eta_\mu^{(N)} - \eta_\mu^{(G)})(\Delta n_+ +\Delta n_-)
\nonumber\\
&+
(2 \eta_\mu^{(G)} + \eta_\mu^{(N)})\Delta n_0 \Big],
\label{dmu-general}
\end{align}
and for the critical temperatures
\begin{align}
\delta T_{\rm cr}^\pm &=\frac{m^*}{2n\eta_n}\Big[
\big(\frac{\eta_{T_{\rm cr}}^{(G)} +2 \eta_{T_{\rm cr}}^{(N)}}{3} \pm \eta_{T_{\rm cr}}^{(Q)}\big)\Delta n_+
\nonumber\\
&\big(\frac{\eta_{T_{\rm cr}}^{(G)} +2 \eta_{T_{\rm cr}}^{(N)}}{3} \mp \eta_{T_{\rm cr}}^{(Q)}\big)\Delta n_-
+\frac{\eta_{T_{\rm cr}}^{(N)} - \eta_{T_{\rm cr}}^{(G)}}{3} \Delta n_0\Big] ,
\nonumber\\
\delta T_{\rm cr}^0 &= \frac{m^*}{3n\eta_n}\Big[
(\eta_{T_{\rm cr}}^{(N)} - \eta_{T_{\rm cr}}^{(G)})(\Delta n_+ +\Delta n_-)
\nonumber\\
&+
(2 \eta_{T_{\rm cr}}^{(G)} + \eta_{T_{\rm cr}}^{(N)})\Delta n_0 \Big]\,.
\label{Tc-general}
\end{align}

\subsection{Variances and cross-variances of the particle number}\label{subsec:fluct-imbal}

Here we apply the relations derived above to analyze the particle number fluctuations.
We assume that the relation  $\sum_a\delta n_a=0$ is fulfilled\,. We calculate the variances in special cases considered above.

(i) {\em Variations of $\delta n_Q$ at $\delta n_G=0$.}\\
Consider first a slightly charged gas at $\delta n_G =0$.
Then pion densities are $n_\pm= n/3 \pm \delta n_Q/2$\,, $n_0=n/3$, and $\delta n_0 =0$\,.
In this case $\delta n_+ = -\delta n_-$, and, as one can see, in Fig.~\ref{fig:pm-plane}\,(b,c)  the line $\delta n_+=-\delta n_-$ passes in the 2nd and 4th quadrants through the regions corresponding  to the maximal critical temperature $T_{\rm cr}^+$, if $\delta n_Q=-\delta n_->0$ and $T_{\rm cr}^-$, if $\delta n_Q<0$.

To be specific let first $\delta n_Q>0$.
We are interested to find characteristics of fluctuations for   $T>T_{\rm cr}^{+}$, especially when $T$ approaches $T_{\rm cr}^+$  from above. In this case, only quantities $I_n^{+}(T\to T_{\rm cr}^+)$ diverge as in (\ref{I3-limit}) and others, $I_n^{-}(T\to T_{\rm cr}^+)$ and $I_n^{0}(T\to T_{\rm cr}^+)$, remain finite. Then, setting in relations (\ref{mixed-derivtives}) $I_n^{+}\to \infty$, taking into account Eq.~(\ref{fin-comb-d})  and keeping terms $I_n^{-,0}(T_{\rm cr}^+)$, we obtain
\begin{align}
\varpi_{++}(T_{\rm cr}^{+}) &=\frac{\mu_{\rm cr,+}^{2}T_{\rm cr}^{+}}{\lambda n_+}
\Big[1+ 4\lambda \tilde{d}_{+}(T_{\rm cr}^{+})
\nonumber\\
&+\frac{\lambda I_2^{0}(T_{\rm cr}^{+})} { 4 + 5\lambda  I_2^{0}(T_{\rm cr}^{+}) } +  \lambda I_2^{-}(T_{\rm cr}^{+})
\Big]
\,,
\nonumber\\
\varpi_{--}(T_{\rm cr}^{+}) &= \frac{T_{\rm cr}^{+}m^{*2}_-(T_{\rm cr}^+)}{n_-} I_1^{-}(T_{\rm cr}^{+})\,,
\nonumber\\
\varpi_{+-}(T_{\rm cr}^{+}) &=
\varpi_{-+}(T_{\rm cr}^{+}) =
 -\frac{T_{\rm cr}^{+} \mu_{{\rm cr}, +} m^{*}_-(T_{\rm cr}^+)}{\sqrt{n_-n_+}}I_3^{-}(T_{\rm cr}^{+}) \,,
\nonumber\\
\varpi_{00}(T_{\rm cr}^{+}) &= \frac{T_{\rm cr}^{+}m^{*2}_0(T_{\rm cr}^+)}{n_0}\Bigg[
 I_1^{(0)}(T_{\rm cr}^{+}) -
\frac{5\lambda \big[I_3^{(0)}(T_{\rm cr}^{+})\big]^2 }{4 + 5\lambda I_2^{(0)}(T_{\rm cr}^{+})}
\Bigg] \,,
\nonumber\\
\varpi_{0+}(T_{\rm cr}^{+}) &=
\varpi_{+0}(T_{\rm cr}^{+}) = -\frac{2T_{\rm cr}^{+}}{\sqrt{n_+n_0}}\frac{ \mu_{{\rm cr}, +}
m^{*}_0(T_{\rm cr}^+)I_3^{0}(T_{\rm cr}^{+})}
{4 + 5\lambda I_2^{0}(T_{\rm cr}^{+})}\,,
\nonumber\\
\varpi_{0-}(T_{\rm cr}^{+}) &= \varpi_{-0}(T_{\rm cr}^{+}) = 0\,,
\label{mixed-deriv-nQ}
\end{align}
where $\mu_{{\rm cr}, +} =m_+^*(T_{\rm cr}^+)$\,.
We have to emphasize that these expressions
can be used for any isospin composition of the system (without invoking smallness of $\delta n_Q$ and the expansions derived in Sect.~\ref{subsec:prop}) with the only constraint  $T>T_{\rm cr}^+$, where $T_{\rm cr}^+$ is  the maximal critical temperature. We note that
for a non-vanishing value of $\delta n_Q$ all partial variances in (\ref{mixed-deriv-nQ}) take finite values at $T_{\rm cr}^+$ (if $\delta n_Q>0$). In contrast, in the isospin-symmetrical case the variances $\varpi_{\pm\pm}$ and $\varpi_{\pm\mp}$ in  Eqs.~(\ref{var-pp}) and (\ref{var-pm}) diverge when the temperature  tends to $T_{\rm cr}$.

Now let us analyze relations (\ref{mixed-deriv-nQ}) for  $\delta n_Q\ll n$.  The result proves to be dependent on the order, whether   first $T\to T_{\rm cr}^{+}$ for fixed $\delta n_Q$ and then $\delta n_Q\to 0$ or $\delta n_Q\to 0$ at fixed $T$ and then $T\to T_{\rm cr}^{+}$.
In the former case expressions in (\ref{mixed-deriv-nQ}) are not valid but we may use the original relations (\ref{var-pp}), (\ref{var-00}), (\ref{var-pm}), (\ref{var-p0}). After letting $\delta n_Q\to 0$ we  reduce temperature down to  $T_{\rm cr}$ and recover  the results for the isospin-symmetrical gas.

If we put first $T\to T_{\rm cr}^{+}$ at fixed $\delta n_Q >0$ to evaluate changes of variances at decreasing values of $\delta n_Q $, we may
use results of Appendix~\ref{app:Iexpan} for expansions of $I^-(T_{\rm cr}^+)$ and
$I^0(T_{\rm cr}^+)$ given in Eqs.~(\ref{Imin-exp}) and (\ref{I0-exp}), respectively,
expressions for the effective masses $ m^*_{-,0}(T_{\rm cr}^{+})$  given by Eqs.~(\ref{mpTcr-m}),  expressions for the chemical potential $\mu_{\rm cr,+}$ given by Eqs.~(\ref{mpTcr-mu}) and (\ref{I1exp}), and the expression for the critical temperature
$T_{\rm cr}^+$ from Eq.~(\ref{Tcrp-Tc}). Then, keeping terms $\propto 1/\delta n_Q$ and those not depending on $\delta n_Q$ we obtain
\begin{align}
\varpi_{++}(T_{\rm cr}^{+}) &= \phantom{-}
\frac{3\mu_{\rm cr}^{2}T_{\rm cr}}{n} \Big(\beta^{(Q)} \frac{n}{\delta n_Q} + \frac{6} {5\lambda}
+ b_{++}^{(Q)} \Big) +O(\delta n_Q)
\,,
\nonumber\\
\varpi_{--}(T_{\rm cr}^{+}) &= \phantom{-}
\frac{3\mu_{\rm cr}^2T_{\rm cr}}{n} \Big(\beta^{(Q)}\frac{n}{\delta n_Q} + b_{--}^{(Q)}\Big) +O(\delta n_Q)\,,
\nonumber\\
\varpi_{\pm\mp}(T_{\rm cr}^{+}) &=
- \frac{3 \mu_{\rm cr}^2 T_{\rm cr}}{n} \Big(\beta^{(Q)}\frac{n}{\delta n_Q} + b_{+-}^{(Q)}
\Big)+O(\delta n_Q) \,,
\nonumber\\
\varpi_{00}(T_{\rm cr}^{+}) &=\phantom{-}
\frac{3\mu_{\rm cr}^2 T_{\rm cr}}{n}\Big( \frac{4}{5\lambda} + b_{00}^{(Q)}\Big)
+O(\delta n_Q) \,,
\nonumber\\
\varpi_{0+}(T_{\rm cr}^{+}) &= \varpi_{+0}(T_{\rm cr}^{+})=
-\frac{3\mu_{\rm cr}^2 T_{\rm cr}}{n}\frac{2}{5\lambda}+O(\delta n_Q) \,,
\nonumber\\
\varpi_{0-}(T_{\rm cr}^{+}) &= \varpi_{-0}(T_{\rm cr}^{+})=
\phantom{-}0
\,,
\label{mixed-deriv-nQ-exp}
\end{align}
where
\begin{align}
\beta^{(Q)} = \sqrt{\frac{\mu_{\rm cr}T_{\rm cr}^2}{
(2\pi)^{3}\alpha^{1/2}_{\rm cr}n}
\frac{\eta_n}{\eta_{T_c}^{(Q)}}}
\,,
\label{betQ-def}
\end{align}
and the background terms are equal to
\begin{align}
b_{00}^{(Q)} &=b_{I}=\delta I_{{\rm cr},1} + \delta I_{{\rm cr},2} -2\, \delta I_{{\rm cr},3}=  4\tilde{d}_{\rm cr}
\,,
\nonumber\\
b_{+-}^{(Q)} &= b_{T_{\rm cr}} +  \delta I_{{\rm cr},3} \,,\,\,\label{bkg-terms}
\\
b_{T_{\rm cr}} &=
\beta^{(Q)}
\frac{\eta_{T_c}^{(Q)}}{2\eta_n}
\Big(\frac52 +\frac54\beta\frac{T_{\rm cr}}{\mu_{\rm cr}} +
\pi\alpha^{1/2}_{\rm cr}\delta I_{{\rm cr},1}\Big)
\,,
\nonumber\\
b_{++}^{(Q)} &= b_{I} -  \frac{3}{2}\beta^{(Q)} + b_{T_{\rm cr}} + \delta I_{{\rm cr},2}\,,
\nonumber\\
b_{--}^{(Q)} &=  \frac{3}{2}\beta^{(Q)} + b_{T_{\rm cr}} + \delta I_{{\rm cr},1} \,.\nonumber
\end{align}
There are three sources for background terms: variation of the density for $ \frac{3}{2}\beta^{(Q)}$; variation of the critical temperature for $b_{T_{\rm cr}}$; and the finite part of the integrals (\ref{I-def}) for $\delta I_n$ and $b_I$.
All quantities on the right-hand sides in Eqs.~(\ref{mixed-deriv-nQ-exp}) are calculated for the isospin-symmetrical pion gas at the critical temperature $T_{\rm cr}$ (although we continue to consider $T\ge T_{\rm cr}^{+}>T_{\rm cr}$).

In (\ref{mixed-deriv-nQ-exp}) we separated explicitly potentially large terms controlling the magnitude of variances  $\propto 1/\delta n_Q$.
Also explicitly are separated terms $\propto 1/\lambda$ to recover the limit of the ideal gas. Note that the pole terms,  $\propto 1/\delta n_Q$ and $\propto 1/\lambda$, cancel out, e.g., in the combinations
$\varpi_{++}-\varpi_{00}-\varpi_{0+}+\varpi_{\pm\mp}$ and
$\varpi_{--} + \varpi_{\pm\mp}$\,.

The obtained values for the cross-variances (\ref{mixed-deriv-nQ-exp})  differ essentially from the corresponding  expressions (\ref{var-pp}), (\ref{var-00}), (\ref{var-pm}), (\ref{var-p0}) for the isospin-symmetrical case. The expressions for the isospin-symmetrical case are not recovered in the limit $\delta n_Q \to 0$.

Substituting expressions (\ref{mixed-deriv-nQ-exp}) and (\ref{bkg-terms})
in the definitions of the variances $\varpi_N$ and $\varpi_G$, Eq.~(\ref{var-N-gen}) and (\ref{var-G-gen}), we obtain
\begin{align}
\varpi_N(T_{\rm cr}^+) =\frac{\mu_{\rm cr}^2T_{\rm cr} }{n}
\Big(\frac{6}{5\lambda}+12 \tilde{d}_{\rm cr} +
O(\delta n_Q)\Big) \,,
\nonumber\\
\varpi_G(T_{\rm cr}^+) =\frac{9\mu_{\rm cr}^2T_{\rm cr} }{2\lambda n}
\Big(1+4\lambda \tilde{d}_{\rm cr} +
O(\delta n_Q)\Big) \,.
\label{varNG-dnQ-exp}
\end{align}
In the limit $\delta n_Q\to 0$ these expressions reduce to those found for the isospin-symmetrical case at $T=T_{\rm cr}$, (\ref{varN-Tc}) and (\ref{varG-Tc}). The differences may appear only at the first order  in $\delta n_Q$. Oppositely, the variance of the total charge in the system, which behaves as $1/(T-T_{\rm cr})$ for $T\to T_{\rm cr}$  in the isospin-symmetrical system, see Eq.~(\ref{var-Q}), proves to be finite
for $T\to T_{\rm cr}^+$  in a slightly asymmetrical system,
\begin{align}
\varpi_Q(T_{\rm cr}^+) &= \frac{3\mu_{\rm cr}^2T_{\rm cr} }{n}
\Big( 4\beta^{(Q)}\frac{ n}{\delta n_Q}
+\frac{6}{5\lambda}+ 2b_{I}
\nonumber\\
&
+ 3 \delta I_{{\rm cr},3}
 +4 b_{T_{\rm cr}}\Big)
 +O(\delta n_Q)
\label{varQ-dnQ-exp}
\end{align}
for small but finite values of $\delta n_Q$. Interestingly, the critical value of the variance (\ref{varQ-dnQ-exp}) depends now explicitly on the value of the self-interaction constant  $\lambda$, that was not the case in the purely isospin-symmetrical case, see Eqs.~(\ref{var-Q}) and (\ref{var-Q-exp1}).

So far we assumed $\delta n_Q>0$. If the system is slightly negatively charged, i.e.,  $\delta n_Q<0$, the maximal critical temperature is $T_{\rm cr}^-$ and Eqs.~(\ref{mixed-deriv-nQ-exp}), (\ref{varNG-dnQ-exp}), and (\ref{varQ-dnQ-exp})
can be used after the replacements $T_{\rm cr}^+\to T_{\rm cr}^-$ and $\delta n_Q\to -\delta n_Q$. 

(ii) {\em Variations of $\delta n_G$ at $\delta n_Q =0$.}\\
Consider now another case of isospin  asymmetry, when the number of charged and neutral pions, $(n_+ + n_-)/2 - n_0$, differs by $\delta n_G$,
\begin{align}
n_+=n_-=\frac{n}{3}+\frac{\delta n_G}{3}\,,\quad n_0=\frac{n}{3}-   \frac{2\delta n_G}{3}\,.
\label{n-caseG}
\end{align}
As we show below, the values of variances  are essentially different in dependence on the sign of $\delta n_G$. Therefore, we consider  separately the cases $\delta n_G<0$ and $\delta n_G>0$.

First, let us consider the case $\delta n_G<0$.
As we can see in Fig.~\ref{fig:pm-plane}, along the line $\delta n_Q=0$ the maximal critical temperature is realized for neutral pions, $T_{\rm cr}^0$. Then the quantities $I_{1,2,3}^{(0)}(T\to T_c^{(0)}+0)$ diverge, whereas $I_{1,2,3}^{(\pm)}(T_c^{(0)})$ are finite for finite values of $|\delta n_G|$.  Then in the limit $T\to T_{\rm cr}^0$, expressions (\ref{mixed-derivtives}) take the following forms:
\begin{align}
\varpi_{\pm\pm}(T_{\rm cr}^{0}) &= \frac{T_{\rm cr}^{0}m_+^{*2}(T_{\rm cr}^{0})}{n_{+}}\Big(
I_1^{+}(T_{\rm cr}^{0})
- \frac{5\lambda  \big[I_3^{+}(T_{\rm cr}^{0})\big]^2 }
{6+10\lambda I_2^{+}(T_{\rm cr}^{0})  }\Big) \,,
\nonumber\\
 \varpi_{\pm\mp}(T_{\rm cr}^{0}) &= -\frac{T_{\rm cr}^{0}m_+^{*2}(T_{\rm cr}^{0})}{n_{+}}
\frac{5\lambda [I_3^{+}(T_{\rm cr}^{0})]^2}
{6+10\lambda I_2^{+}(T_{\rm cr}^{0}) } \,,
\nonumber\\
\varpi_{00}(T_{\rm cr}^{0}) &=
\frac{2\mu_{{\rm cr},0}^2 T_{\rm cr}^0}{3\lambda n_{0}} \Big(
1 +
6\lambda \tilde{d}_{0}(T_{\rm cr}^0)  +
\frac{\lambda I_2^{+}(T_{\rm cr}^{0}) }{3 + 5\lambda I_2^{+}(T_{\rm cr}^{0})}
\Big) \,,
\nonumber\\
\varpi_{\pm,0}(T_{\rm cr}^{0}) &= \varpi_{0,\pm}(T_{\rm cr}^{0}) =-\frac{T_{\rm cr}^{0}m_+^{*}(T_{\rm cr}^{0})}{\sqrt{n_{+} n_{0}}}
\frac{ 2\mu_{{\rm cr},0} I_3^{+}(T_{\rm cr}^{0}) }
{ 6+10\lambda I_2^{+}(T_{\rm cr}^{0})}\,.
\label{mixed-deriv-nGm}
\end{align}
Here $\mu_{{\rm cr}, 0} =m_0^*(T_{\rm cr}^0)$. Take into account that for $\delta n_Q=0$ we have $\mu_+=\mu_-$ and $m^*_+=m^*_-$, and, therefore,  put $I_n^{(+)}=I_n^{(-)}$.
These expressions do not rely on the smallness of $|\delta n_G|$ and can be used for any neutral system with an access of $\pi^0$ mesons, when the maximal critical temperature is $T_{\rm cr}^0$.
For a non-vanishing value of $\delta n_G$ all partial variances in (\ref{mixed-deriv-nGm}) take finite values at  $T_{\rm cr}^0$.

To employ expressions (\ref{mixed-deriv-nGm}) at $0<-\delta n_G\ll n$  we use  expansions of $I_n^{+}(T_{\rm cr}^0)$ given by Eq.~(\ref{IpTcr0})  in Appendix~\ref{app:dnG-1} and  expansions for the critical temperature (\ref{Tc0epx-1}),
the effective mass (\ref{mpTc0}) and the chemical potential (\ref{mu0Tc0}).
Substituting them in Eqs.~(\ref{mixed-deriv-nGm}), we obtain
\begin{align}
\varpi_{\pm\pm}(T_{\rm cr}^0) &=-\frac{3\mu_{\rm cr}^2 T_{\rm cr}}{n}
\Big(\frac{\beta^{(G)}}{2}
\frac{n}{\delta n_G}  -\frac{3}{10\lambda} - b_{++}^{(G)} \Big)
+O(\delta n_G)\,,
\nonumber\\
\varpi_{\pm\mp}(T_{\rm cr}^0) &=\frac{3\mu_{\rm cr}^2 T_{\rm cr}}{n}
\Big(\frac{\beta^{(G)}}{2} \frac{n}{\delta n_G}
 +\frac{3}{10\lambda} + b_{+-}^{(G)} \Big)+O(\delta n_G)\,,
\nonumber\\
\varpi_{00}(T_{\rm cr}^0) &=\frac{3\mu_{\rm cr}^2 T_{\rm cr}}{n}
\Big(\frac{4}{5\lambda} + b_{00}^{(G)} \Big)+O(\delta n_G)\,,
\label{mixed-deriv-nGm-exp}\\
\varpi_{\pm0}(T_{\rm cr}^0) &= \varpi_{0\pm}(T_{\rm cr}^0) =-\frac{3\mu_{\rm cr}^2 T_{\rm cr}}{n}\frac{1}{5\lambda}+O(\delta n_G)\,,
\nonumber\end{align}
where
\begin{align}
\beta^{(G)} = \beta^{(Q)} \frac{\sqrt{3}}{2}\frac{\eta^{(Q)}_{T_{\rm cr}}}
{\eta^{(G)}_{T_{\rm cr}}}\,,
\label{betG-def}
\end{align}
and the background terms are
\begin{align}
b_{++}^{(G)} &= \frac12 b_I +\frac{b_{T_{\rm cr}}}{\sqrt{3}} +\frac12\delta I_{\rm cr, 1}
+\frac34 \beta^{(G)} -b_m\,,
\nonumber\\
b_{+-}^{(G)} &= \frac12 b_I - \frac{b_{T_{\rm cr}}}{\sqrt{3}} - \frac12\delta I_{\rm cr,1}
- \frac34 \beta^{(G)} + b_m\,,
\nonumber\\
b_{00}^{(G)} &= b_I\,,\quad
b_{m}= \frac{\beta^{(G)} \lambda n}{4\mu_{\rm cr}^{3}(1+2C)}\,.
\label{bck-G}
\end{align}
The quantities $b_I$ and $b_{T_{\rm cr}}$ are the same as in Eq.~(\ref{bkg-terms}).
We see the qualitative difference in the structure of limiting variances for the temperature approaching the temperature of the Bose-Einstein instability  for variations $n_Q$ and $n_G$, cf. Eqs.~(\ref{mixed-deriv-nQ-exp}) and (\ref{mixed-deriv-nGm-exp}).
The variances $\varpi_{--}$ and $\varpi_{\pm\mp}$ as functions of  $\delta n_Q$  do not contain terms $\propto 1/\lambda$ in contrast to the same variances as functions of  $\delta n_G$. Also $\varpi_{0-}$ and $\varpi_{-0}$ vanish as functions of $\delta n_Q$  but remain finite as functions of  $\delta n_G$. Nevertheless, $\varpi_{0+}$ and $\varpi_{+0}$ are equal for both cases,  since $b_{00}^{(Q)}=b_{00}^{(G)}$.
In spite of these differences, if we combine variances (\ref{mixed-deriv-nGm-exp})  in the variances $\varpi_N$ and $\varpi_G$ defined by Eqs.~(\ref{varN-gen}) and (\ref{varG-gen}), we obtain
\begin{align}
\varpi_N(T_{\rm cr}^0) &= \frac{\mu_{\rm cr}^2T_{\rm cr} }{n}
\Big(\frac{6}{5\lambda}+12\lambda\tilde{d}_{\rm cr} +
O(\delta n_G)\Big) \,,
\nonumber\\
\varpi_G(T_{\rm cr}^0) &=
\frac{9\mu_{\rm cr}^2T_{\rm cr} }{2\lambda n}
\Big(1+4\lambda \tilde{d}_{\rm cr} +
O(\delta n_G)\Big) \,.
\label{varNG-dnG-exp}
\end{align}
We see that the leading terms are the same as for the isospin symmetric system (\ref{varN-Tc}) and (\ref{varG-Tc}) and also for the case of $\delta n_Q$ variations (\ref{varNG-dnQ-exp}).
On the other hand the expression for the variance $\varpi_Q$ is quite different,
\begin{align}
\varpi_Q(T_{\rm cr}^0) &= \frac{6\mu_{\rm cr}^2T_{\rm cr} }{n}
\Big(-\beta^{(G)} \frac{n}{\delta n_G} + 2 b_{++}^{(G)} - b_I\Big)\,.
\label{varQ-dnG-exp}
\end{align}
This expression does not  contain terms $\propto 1/\lambda$ in difference with the result (\ref{varQ-dnQ-exp}).

We turn now to the case $\delta n_G>0$. Then $T_{\rm cr}^{+}=T_{\rm cr}^{-}$ is the largest critical temperature, and $I_{1,2,3}^{\pm}$ diverge at $T\to T_{\rm cr}^{+}+0$. Expressions~(\ref{mixed-derivtives}) yield in this case:
\begin{align}
\varpi_{\pm\pm}(T_{\rm cr}^+) &= \frac{T_{\rm cr}^+\mu_{\rm cr,+}^2}{2n_+}I_1^{+}(T\to T_{\rm cr}^+)
\nonumber\\
&+ \frac{\mu_{{\rm cr},+}^2 T_{\rm cr}^+}{4\lambda n_+}\Big(1 +8\lambda
\tilde{d}_\pm(T_{\rm cr}^+)
+
\frac{ \lambda I_2^{0}(T_{\rm cr}^+)}{4 + 5\lambda I_2^{0}(T_{\rm cr}^+)} \Big)\,,\,\,\,
\nonumber\\
 \varpi_{\pm\mp}(T_{\rm cr}^+) &=
\varpi_{\pm\pm}(T_{\rm cr}^+) - \frac{T_{\rm cr}^+\mu_{\rm cr,+}^2}{n_+}I_1^{+}(T\to T_{\rm cr}^+)\,,
\nonumber\\
\varpi_{00}(T_{\rm cr}^+) &= \frac{T_{\rm cr}^+m_0^2(T_{\rm cr}^+)}{n_{0}}\Big(
I_1^{0}(T_{\rm cr}^+)
-
\frac{5\lambda \big[I_3^{0}(T_{\rm cr}^+)\big]^2 }{ 4 + 5\lambda I_2^{0}(T_{\rm cr}^+) }
\Big)\,,
\nonumber\\
\varpi_{\pm,0}(T_{\rm cr}^+) &= \varpi_{0,\pm}(T_{\rm cr}^+)=-\frac{T_{\rm cr}^+m_0(T_{\rm cr}^+)}{\sqrt{n_{+} n_{0}}}
\frac{\mu_{{\rm cr},+}  I_3^{0}(T_{\rm cr}^+)} {4 + 5\lambda I_2^{0}(T_{\rm cr}^+)}\,.
\label{mixed-deriv-nGp}
\end{align}
Here, $I_1^{+}(T\to T_{\rm cr}^+)$ is a divergent term given by expansion (\ref{In-atTc}).
We see that the variances (\ref{mixed-deriv-nGp}) obtained for $\delta n_G>0$ differ from the variances (\ref{mixed-deriv-nQ}) and (\ref{mixed-deriv-nGm}) for $\delta n_Q$ and $\delta n_G<0$ variations.
Particularly, in the present case the variances for charged pions, $\varpi_{\pm\pm}$ and $\varpi_{\pm\mp}$, are divergent. However, the sums $\varpi_{\pm\pm}+\varpi_{\pm\mp}$ remain finite. As Eqs.~(\ref{mixed-deriv-nQ}) and (\ref{mixed-deriv-nGm}), these expressions follow directly from (\ref{mixed-derivtives}) and can be used for arbitrary isospin composition, provided the charge of the system in zero and the maximal critical temperature is realized for $\pi^\pm$ mesons.

To exploit relations (\ref{mixed-deriv-nGp}) in the limit of a small isospin imbalance $0<\delta n_G\ll n$  we need the expansion in $\delta n_G$ for the critical temperature (\ref{Tcrp-exp}), the effective mass (\ref{m0Tcp}) and the chemical potential (\ref{mu0Tcp}). Substituting these relations in Eq.~(\ref{mixed-deriv-nGp}) together with the expansion for  $I_n^{0}(T_{\rm cr}^+)$ given by  Eq.~(\ref{I0Tcp}) we obtain
\begin{align}
\varpi_{\pm\pm}(T_{\rm cr}^+) &= \phantom{-}\frac12\varpi_{\infty}
+ \frac{3\mu_{\rm cr}^2 T_{\rm cr}}{n}\Big(\frac{3}{10\lambda} + \frac12 b_{00}^{(G)} \Big)+
O(\delta n_G)\,,
\nonumber\\
 \varpi_{\pm\mp}(T_{\rm cr}^+) &= -\frac12\varpi_{\infty}
+ \frac{3\mu_{\rm cr}^2 T_{\rm cr}}{n}\Big(\frac{3}{10\lambda} + \frac12 b_{00}^{(G)} \Big)+
O(\delta n_G)\,,
\nonumber\\
\varpi_{00}(T_{\rm cr}^+) &=\frac{3\mu_{\rm cr}^2 T_{\rm cr}}{n}\Big(\frac{4}{5\lambda} + b_{00}^{(G)} \Big)+
O(\delta n_G)\,,
\nonumber\\
\varpi_{\pm0}(T_{\rm cr}^+) &= \varpi_{0\pm}(T_{\rm cr}^+)=-\frac{3\mu_{\rm cr}^2 T_{\rm cr}}{n}\frac{1}{5\lambda}+
O(\delta n_G) \,,
\label{mixed-deriv-nGp-exp}
\end{align}
where we denoted the divergent term as
\begin{align}
\varpi_{\infty} &=\frac{3\mu_{\rm cr}^2 T_{\rm cr}}{n}\Big(\frac{T_{\rm cr}}{2\pi\sqrt{\alpha}(T-T_{\rm cr}^{+})}
+\frac{1}{2\pi\sqrt{\alpha}}
+ \delta I_{\rm cr, 1} \Big)\,.
\nonumber
\end{align}
We see that these results are quite different from the corresponding results (\ref{mixed-deriv-nGm-exp}) obtained for $\delta n_G<0$. Particularly, expressions~(\ref{mixed-deriv-nGp-exp}) do not contain terms $\propto 1/\delta n_G$. Nevertheless, if we substitute expressions~(\ref{mixed-deriv-nGp-exp}) in the definitions~(\ref{varN-gen}) and (\ref{varG-gen}), we obtain
\begin{align}
\varpi_N(T_{\rm cr}^+) &= \frac{\mu_{\rm cr}^2T_{\rm cr} }{n}
\Big(\frac{6}{5\lambda}+12\lambda\tilde{d}_{\rm cr} +
O(\delta n_G)\Big) \,,
\nonumber\\
\varpi_G(T_{\rm cr}^+) &=
\frac{9\mu_{\rm cr}^2T_{\rm cr} }{2\lambda n}
\Big(1+4\lambda \tilde{d}_{\rm cr} +
O(\delta n_G)\Big) \,,
\label{varNG-dnGp-exp}
\end{align}
i.e., the leading terms are exactly the same as in Eqs.~(\ref{varNG-dnGp-exp}) and (\ref{varNG-dnQ-exp}) up to  linear terms in $\delta n_G$ and $\delta n_Q$  and agree with the results for the isospin symmetrical system.
For the variance of the charge of the system defined by Eq.~(\ref{varQ-def}) we find from (\ref{mixed-deriv-nGp-exp})
\begin{align}
\varpi_Q(T_{\rm cr}^+)=2\varpi_\infty
\end{align}
that agrees with the result (\ref{var-Q-exp1}) for the isospin symmetrical case.

The main result of this section is that the partial variances of the numbers of various pion species, $\varpi_{ab}$, are quite sensitive to a small isospin imbalance of the system. So, the variances of charged pions, $\varpi_{\pm\pm}$ and $\varpi_{\pm\mp}$, and the variance of the total charge of the system, $\varpi_Q$, turn out to be finite at the critical point of the BEC for finite values of $\delta n_Q\neq 0$ or $\delta n_G<0$, whereas they are divergent in the isospin symmetrical case and if
$\delta n_G$ is finite but positive. On the other hand, the variances $\varpi_N$ and $\varpi_G$ are weakly sensitive to the isospin imbalance and up to terms $O(\delta n_{Q,G})$ coincide with those  for the isospin symmetrical case.

\section{Conclusion}\label{sec:conclusion}

Experimental evidence for  formation of the baryon-poor medium at midrapidity at SPS, RHIC and LHC energies~\cite{Afanasiev2002,Alt2008,Nayak:2012np,Abelev:2012wca,Adamczyk:2017iwn}
calls for investigations of properties of a dense and hot purely pion gas. In this paper, we studied fluctuations in the self-interacting pion gas, which could be formed at an intermediate or latest stage at the heavy-ion collisions of ultrarelativistic energies. Characteristics of fluctuations measured at different experimental conditions may give an information on different moments of the fireball evolution, like the chemical freeze-out, at the total density and the temperature $(n_{\rm chem},T_{\rm chem})$, and the kinetic one, at $(n_{\rm kin},T_{\rm kin})$. Right after the chemical freeze-out the pion annihilation and creation processes cease and the total number of pions $N$ almost does not change. Thus, if we consider an ideal detector with full $4\pi$ geometry, a variance of the total pion number reflects the state of the system at the chemical freeze-out. The same quantity, taken for $T=T_{\rm kin}$ and the volume $V(T_{\rm kin})$, characterizes fluctuations of the volume of the pion fireball at the kinetic freeze-out at measurements done in the $4\pi$ geometry.

Although in the time interval between chemical and kinetic freeze-outs the total number of pions remains fixed, an exchange of particles between pion species continues owing to the $2\leftrightarrow 2$ reactions. Thus, if pions are measured in experiments with incomplete geometry and/or in a restricted momentum range, then the elastic pion-pion reactions and processes  $\pi^0\pi^0\leftrightarrow\pi^+\pi^-$ change populations of pions of different isospin species and in different momentum bins. Therefore, there exists a kind of thermodynamic reservoir for the subsystem of pions, which reach the detector later, and the grand-canonical formulation can be relevant in such a situation. If one measures correlations between pions emitted at different angles and in various momentum bins, one may get an information about the state of the pion fireball at the kinetic freeze-out.

The temperature decreases when the pion system evolves from the chemical to the kinetic freeze-out. Chemical potentials $\mu_a$ of pions for all three species $a$ grow towards their effective masses $m^*_a$ and the system may reach the critical point of the Bose-Einstein condensation, first for one of the species and with a further decrease of the temperature for other pion species, cf.~\cite{Voskresensky:1994uz}. If a significant growth of fluctuation characteristics were observed, it could be associated with a closeness to the pion Bose-Einstein condensation either at the chemical freeze-out or the thermal freeze-out, depending on the specifics of the measurement.

In the given paper, continuing our recent studies \cite{BKV18,KV18} we consider behavior of the strongly interacting pion gas with a dynamically fixed number of particles within  the self-consistent Hartree approximation in the $\lambda\phi^4$ model. Within the grand-canonical approach for the pion system of arbitrary particle composition we calculated normalized cross-variances of the numbers of pions of various species, $\varpi_{ab}$, $a,b=\pm,0$ (defined in Eq.~(\ref{cross-var})) and variances of the total number, $N=N_{\pi^+} + N_{\pi^-} +N_{\pi^0}$, the charge number, $Q=N_{\pi^+}-N_{\pi^-}$, and the imbalance between   charged  and neutral pions, $G=(N_{\pi^+} + N_{\pi^-})/2- N_{\pi^0}$ (defined in Eqs.~(\ref{var-N-gen}), (\ref{var-Q-gen}), (\ref{var-G-gen})).

Then we focused on the description of the isospin-symmetrical system with equal densities of $\pi^+$, $\pi^-$ and $\pi^0$, being considered above the critical temperature of the Bose-Einstein condensation, $T_{\rm cr}$, which value  in this case is one and the same for $\pi^+$, $\pi^-$ and $\pi^0$. We found that various variances show different behaviors when the temperature approaches $T_{\rm cr}$: the quantities describing only the charged particles, $\varpi_{\pm\pm}$ and $\varpi_{\pm\mp}$, diverge at $T_{\rm cr}$, whereas variances involving neutral pions, $\varpi_{\pm0}$ and $\varpi_{00}$, remain finite. Also the particular combination of variances, $\varpi_{\pm\pm}+\varpi_{\pm\mp}$, does not diverge at $T\to T_{\rm cr}$. Then we analyzed variances of the total particle number, the charge and the difference between charged and neutral numbers $\varpi_{N,Q,G}$ (see Eqs.~(\ref{var-N}), (\ref{var-G}), and (\ref{var-Q}), respectively). All these quantities increase with a decrease in the temperature keeping a hierarchy $\varpi_N<\varpi_G<\varpi_Q$, see Fig.~\ref{fig:fluct}. When the system approaches the critical temperature, variances $\varpi_{NG}$ and $\varpi_{G}$ remain finite because of the self-consistent account of a pion interaction, whereas $\varpi_{Q}$ diverges. The results for $\varpi_G$ ought to be used in the analysis of the experimental data, where fluctuations of neutral pions are studied in selected events with a fixed total pion number. To understand why the self-consistent inclusion of the interaction renders the variances $\varpi_N$ and $\varpi_G$ finite but $\varpi_Q$  divergent we showed that the variance of the charge can be written through the derivatives of the free-energy density $F$ with respect to the charge density $n_Q= n_+-n_-$, for $n_Q\to 0$, when only the Coulomb term remains in the free energy depending on $n_Q$. It proved to be that the variance of the total pion number $\varpi_N$ does not depend on the size of the fluctuation region whereas the Coulomb contribution in $\varpi_Q$ depends on the size of the fluctuation, $R_{\rm fl}$, because of the long-range Coulomb force. Thus we get that $\varpi_Q/\varpi_N\sim \lambda/e^2 \mu^2 R_{\rm fl}^2$. Since  $e^2=1/137\ll \lambda$, for relevant values of $R_{\rm fl}\sim $ several $1/m_\pi$ the resulting quantity $\varpi_Q$ could be considerably larger than $\varpi_N$. Thus, disregarding effects associated with the electromagnetic interaction compared with the effects of the strong interaction we may employ $w_Q$, being divergent for $T\to T_{\rm cr}$.

In the actual heavy-ion collisions initial nuclei have a charge and may have unequal numbers of protons and neutrons, i.e. an isospin imbalance. Therefore, in a realistic situation, the created pion-enriched system could have some  charge imbalance and an imbalance between the numbers of charged and neutral pions. To address this situation we studied a pion system formed with a small isospin imbalance in more details.
First we considered variations for the system at a fixed total density $n$ allowing for a small variation of the charge density, $\delta n_Q$, and the isospin density, $\delta n_G=(n_+ +n_-)/2- \delta n_0$. We considered specific cases: (i) a variation of $\delta n_Q$ at $\delta n_G=0$ and (ii) a variation of $\delta n_G$ at $\delta n_Q=0$.
The variation of $\delta n_Q$ only, case (i), does not change the effective pion mass, whereas
the variation of $\delta n_G$, case (ii), results in a change of the pion mass: an increase of $\delta n_G$ leads to an increase of masses of charged pions and a reduction of the neutral pion mass, (\ref{nG-res}).
Variations of chemical potentials of pions were parameterized in both cases as $\delta \mu_a/m^* = (\delta n_a/n)\,\eta_\mu^{(N,Q,G)}$, where $\delta n_a$ is a variation of the density of pions of type $a=+,-,0$. The susceptibility parameters $\eta_\mu^{(Q,G)}$ decrease with a temperature decrease, and $\eta_\mu^{(Q)}$ vanishes at $T_{\rm cr}$, whereas $\eta_\mu^{(G)}$ remains finite, see Figs.~\ref{fig:eta-Q} and \ref{fig:eta-G}, respectively.
Then we calculated shifts of critical temperatures of Bose-Einstein instabilities (determined by the equation $m_a^*(T_{\rm cr}^a)=\mu_a(T_{\rm cr}^a)$) with respect to the critical temperature of the Bose-Einstein instability, $T_{\rm cr}$ for an isospin symmetrical pion gas, with the density $n$.
In both cases defined above, the shift $ \delta T_{\rm cr}^a= T_{\rm cr}^a- T_{\rm cr}$ can be written as
$\delta T_{{\rm cr},a}/T_{\rm cr}=(\delta n_a/n)\, \eta_{T_{\rm cr}}^{(G,Q)}$, where  susceptibilities $\eta_{T_{\rm cr}}^{(N,Q)}$ are illustrated in Figs.~\ref{fig:eta-TQ} and \ref{fig:X-n}.
Using this result and exploiting  established relations between $\eta_{T_{\rm cr}}^{(G)}$  and $\eta_{T_{\rm cr}}^{(Q)}$, Eq.~(\ref{X-def}), we were able to determine what pion species would have a largest $T_{\rm cr}^a$ in the case of an arbitrary isospin imbalance. For an ideal gas the answer is obvious: the most abundant species will have the largest $T_{\rm cr}^a$. For an interacting system we have found  two regions, where  neutral pions are most abundant but the maximal critical temperature is realized either for positive pions
or for negative pions, see Fig.~\ref{fig:pm-plane}.

Having determined the maximal among three values of the critical temperatures, $T_{\rm cr}^{\tilde{a}}=\max_a T_{\rm cr}^a$, we can specify the temperature above which our consideration is valid, since already slightly below the temperature $T_{\rm cr}^{\tilde{a}}$ one has to take into account  presence of the Bose-Einstein condensate of the given pionic species $\tilde{a}$ that we did not do.

As next, we addressed the question of how the pion properties will change, if in an isospin-symmetrical  system with density $n$  the densities of various pion species are changed by some small amounts, $\delta n_a$. The total density of the isospin-symmetrical system changes in this case, and we calculated susceptibilities of the effective mass, the chemical potential and the critical temperature to small variations of the density, $\eta_{m,\mu,T_{\rm cr}}^{(N)}$, see Fig.~\ref{fig:eta-N} and Eq.~(\ref{etaTc-N}). Then, Eqs.~(\ref{dm-general}), (\ref{dmu-general}), and
(\ref{Tc-general}) provide responses to the changes of $m_a^*$, $\mu_a$ and $T_{\rm cr}^a$ in general case.

Finally, we studied how the pion number variances, $\varpi_{ab}$ behave when the temperature approaches $T_{\rm cr}^{\tilde{a}}$ from above, i.e., when the system is still in a non-condensed phase, for the system with an isospin imbalance. Equations~(\ref{mixed-deriv-nQ}), (\ref{mixed-deriv-nGm}) and (\ref{mixed-deriv-nGp}) provide the results for the cases (i) and (ii), specified above, without invoking smallness of the isospin imbalance. It was found that the variances are quite sensitive to isospin imbalance. For example the variances of charged pions, $\varpi_{\pm\pm}(T_{\rm cr}^{\tilde{a}})$, $\varpi_{\pm\mp}(T_{\rm cr}^{\tilde{a}})$, and  the variance of the total charge of the system, $\varpi_Q(T_{\rm cr}^{\tilde{a}})$, turn out to be finite, if $\delta n_Q\neq 0$ or $\delta n_G <0$, whereas they are divergent (provided the Coulomb interaction is not taken into account in calculation of $\varpi_Q$) in the isospin-symmetrical medium and, if $\delta n_G$ is finite but positive. Interestingly, the variances $\varpi_N$ and $\varpi_G$ are only weakly sensitive to the isospin imbalance and are equal to those for the isospin-symmetrical case up to terms $O(\delta n_{Q,G})$.

\acknowledgments
The work was supported in part by
Slovak grant VEGA--1/0348/18, by German-Slovak collaboration grant in framework of DAAD PPP project and by THOR the COST Action CA15213. E.E.K. acknowledges the support by the Plenipotentiary of the Slovak Government at JINR, Dubna. The work of D.N.V. was  supported by the Ministry of
Science and High Education of the Russian Federation within the state assignment,
project No 3.6062.2017/6.7.

\appendix

\section{First and second $T$-derivatives of $m^*-\mu$  in isospin-symmetrical system}\label{app:deriv}

The dependence of $m^*$ and $\mu$ on the temperature and the density in the isospin-symmetrical gas is determined by  set of equations (\ref{eqs-sym}).
Differentiating them with respect to $T$ at fixed $n$ we obtain the set of equations
\begin{align}
\Big(2m^*-\frac{\partial \Pi}{\partial m^*}\Big)\frac{\partial m^*}{\partial T}\Big|_n-
\frac{\partial \Pi}{\partial \mu}\frac{\partial \mu}{\partial T}\Big|_n &=\phantom{-}\frac{\partial\Pi}{\partial T} \,,
\nonumber\\
\frac{\partial h}{\partial m^*}\frac{\partial m^*}{\partial T}\Big|_n +
\frac{\partial h}{\partial \mu}\frac{\partial \mu}{\partial T}\Big|_n&=-\frac{\partial h}{\partial T} \,,
\label{eq-deriv-1}
\end{align}
 where quantities $h$ and $$\Pi=10\lambda m^{*2} d(m^{*2},\mu,T)$$ are treated as functions of variables $m^{*2}$, $\mu$ and $T$ and, the partial derivatives with respect to one of these variables are taken at the fixed other two.  The solution of Eqs. (\ref{eq-deriv-1}) is as follows:
\begin{align}
\frac{\partial\mu}{\partial T}\Big|_n &=
\frac{
\Big(\frac{\partial \Pi}{\partial m^{*2}} -1
\Big)\frac{\partial h}{\partial T}
- \frac{\partial h}{\partial m^{*2}}
\frac{\partial\Pi}{\partial T}
}
{\Big(1-\frac{\partial \Pi}{\partial m^{*2}}\Big)\frac{\partial h}{\partial \mu}
+\frac{\partial \Pi}{\partial \mu}\frac{\partial h}{\partial m^{*2}}} \,,
\label{dmudT}\\
\frac{\partial m^{*}}{\partial T}\Big|_n &=
\frac{ \frac{1}{2m^*}\frac{\partial h}{\partial \mu}
\frac{\partial\Pi}{\partial T}
-\frac{1}{2m^*}\frac{\partial \Pi}{\partial \mu}\frac{\partial h}{\partial T}
}
{\Big(1-\frac{\partial \Pi}{\partial m^{*2}}\Big)\frac{\partial h}{\partial \mu}
+\frac{\partial \Pi}{\partial \mu}\frac{\partial h}{\partial m^{*2}}} \,.
\label{dmdT}
\end{align}
Derivatives of $h$ and $\Pi$ with respect to $T$ can be expressed through the quantities $I_n$, or through $I_3$, $d$ and $\tilde{d}$, with the help of Eq.~(\ref{I123-relat}) as
\begin{align}
\frac{\partial h}{\partial T} &= \frac{n - \mu m^{*2} I_1 +m^{*3} I_3}{T}\nonumber\\
&=\frac{m^*-\mu}{T} m^{*2} I_3 +\frac{n-2\mu m^{*2}(\tilde{d}+d)}{T}
\,,
\nonumber\\
\frac{\partial \Pi}{\partial T} &= 5\lambda \frac{m^{*2} I_1-\mu m^{*}I_3}{T}
\nonumber\\
&=5\lambda m^{*2}\frac{m^*-\mu}{T} I_3+\frac{10\lambda m^{*2}}{T}(\tilde{d}+d)
\, ,
\label{deriv-hPi-t}
\end{align}
and for the derivatives with respect to $m^{*2}$ and $\mu$ we have
\begin{align}
\frac{\partial \Pi}{\partial m^{*2}} &= -\frac52\lambda  I_2=-\frac52
\lambda\big( {I_3} +2\,(\tilde{d} - d)\big)
\,,\,\,\nonumber\\
\frac{\partial \Pi}{\partial \mu} &= 5\lambda m^* I_3 \,,
\quad
\frac{\partial h}{\partial m^{*2}} = -\frac{m^*}{2} I_3
\,,\,\,\nonumber\\
\frac{\partial h}{\partial \mu} &= m^{*2}I_1=m^{*2}\big( I_3 + 2\,(\tilde{d} + d)\big)\,.
\label{deriv-hPi-mum}
\end{align}
Now, using these expressions we can simplify the denominators in Eqs.~(\ref{dmudT}) and (\ref{dmdT}) as
\begin{align}
&\Big(1-\frac{\partial \Pi}{\partial m^{*2}}\Big)\frac{\partial h}{\partial \mu}
+\frac{\partial \Pi}{\partial \mu}\frac{\partial h}{\partial m^{*2}}
=
\frac{\partial h}{\partial \mu}+5C m^{*2} I_1
\nonumber\\
&=m^{*2} \big( I_3 + 2\,(\tilde{d} + d)\big)(1+5C)\,,
\label{app:2}
\end{align}
where the quantity
\begin{align}
C=\frac{\lambda}{2}\left(I_2-\frac{I_3^2}{I_1}\right)=2\lambda \frac{\tilde{d}}{d } \frac{d\,\,I_3 +(\tilde{d}^2-d^2)} {I_3 +2(\tilde{d}+d)}\,,
\label{app:C}
\end{align}
wherefrom for $T\to T_{\rm cr}$ using (\ref{fin-comb-d}) we get
\begin{align}
C(T_{\rm cr})=2\lambda {\tilde{d}_{\rm cr}}
\,.
\label{CatTc}
\end{align}
From Eq.~(\ref{deriv-hPi-mum}) we find another useful relations for the derivatives of  $\Pi$ and $h$:
\begin{align}
&\frac{\partial\Pi}{\partial m^*}+\frac{\partial\Pi}{\partial \mu} =
-10\lambda\,{m^{*}}{(\tilde{d} - d)}\,,
\nonumber\\
&\frac{\partial h}{\partial m^*}+\frac{\partial h}{\partial \mu} =2{m^{*2}}(\tilde{d}+d)\,.
\label{Dp-act}
\end{align}
These expressions are finite at $T=T_{\rm cr}$.
Oppositely, differences of these derivatives are divergent at $T\to T_{\rm cr}$ and the leading terms
are
\begin{align}
&\Big(\frac{\partial\Pi}{\partial \mu} - \frac{\partial\Pi}{\partial m^*}\Big)\Big|_{T\to T_{\rm cr}} \to
10\lambda \mu_{\rm cr} I_3(T\to T_{\rm cr})\,,\
\nonumber\\
&\Big(\frac{\partial h}{\partial \mu} - \frac{\partial h}{\partial m^*}\Big)\Big|_{T\to T_{\rm cr}} \to 2 \mu_{\rm cr}^2 I_3 (T\to T_{\rm cr})\,.
\label{Dm-act}
\end{align}

Now we can express temperature derivatives of $m^*$ and $\mu$ as
\begin{align}
\frac{\partial\mu}{\partial T}\Big|_n &= \frac{-n\chi_\mu(T)}{TI_1(1+5C)}
\,,\,\,
\frac{\partial m^{*}}{\partial T}\Big|_n = \frac{-n\chi_m(T)}{T m^{*2}I_1(1+5C)} \,,
\label{dmudT-2}
\end{align}
where
\begin{align}
\chi_\mu(T) &= \frac52\lambda{I_3} \Big(1-2(\mu+m^*)\frac{m^{*2}(\tilde{d}+d)}{n}\Big) \label{chi-mu}\\
& +\Big(1  +5\,\lambda{(\tilde{d} - d)}\Big)
\nonumber\\
& \times\Big(
\frac{m^{*2}(m^*-\mu) I_3}{n} + 1 - \frac{2\mu m^{*2}}{n}(\tilde{d}+d)\Big)\,,
\nonumber \\
\chi_m(T) &= \frac52\lambda {I_3}\Big(1-\frac{4m^{*3}}{n}{(\tilde{d}+d})\Big)\,.
\label{chi-m}
\end{align}
At $T\to T_{\rm cr}$ we have $\chi_m=\chi_\mu\propto I_3$
since $(m-\mu)I_3\to 0$ in view of (\ref{I3-limit}). Therefore
 both derivatives (\ref{dmudT-2}) equal to each other and are finite,
\begin{align}
\beta&=\frac{\partial\mu}{\partial T}\Big|_{n, T_{\rm cr}} =  \frac{\partial m^{*}}{\partial T}\Big|_{n, T_{\rm cr}}\nonumber\\
&= -\frac{5\lambda n}{2T_{\rm cr} \mu_{\rm cr}^2}
\frac{1-\frac{4\mu_{\rm cr}^3}{n}(\tilde{d}_{\rm cr} + d_{\rm cr})}
{1+10\lambda \tilde{d}_{\rm cr}}\,.
\label{a1-def}
\end{align}
We see that at $T\to T_{\rm cr}$ the divergency  in denominator ($\propto I_1$) is canceled by the divergency in numerator ($\propto I_3$).

Note that for the ideal gas, when $\lambda=0$, both derivatives (\ref{a1-def}) vanish.   Thus, the finiteness of the derivatives (\ref{dmudT-2}) is another manifestation of the effect of the self-consistent account of the interaction.
Using Eq.~(\ref{ddt-exp}) we can write the expansion of the coefficient $\beta$  for $\mu_{\rm cr}\ll T_{\rm cr}$\, as
\begin{align}
\beta &= -\frac{5\lambda n}{2T_{\rm cr}\mu_{\rm cr}^2}\Big\{
1-\frac{\zeta(\frac32)t_{\rm cr}^{3/2}}{2(2\pi)^{3/2}}\frac{m^{*3}}{n}
\Big(6 +\frac{15}{4}\frac{\zeta(\frac52)t_{\rm cr}}{\zeta(\frac32)}
\nonumber\\
&+5\lambda\frac{n}{\mu_{\rm cr}^3}\Big[1+\frac{9}{8} \frac{\zeta(\frac52)t_{\rm cr}}{\zeta(\frac32)}
-\frac{3\zeta({\textstyle\frac32})t_{\rm cr}^{3/2}}{(2\pi)^{3/2}}\Big(\frac{\mu_{\rm cr}^3}{n} + \frac56\lambda\Big)
\Big]\Big)\Big\}
\nonumber\\
&+O(t_c^{7/2})\,.
\end{align}

\begin{figure}
\centering
\includegraphics[width=5cm]{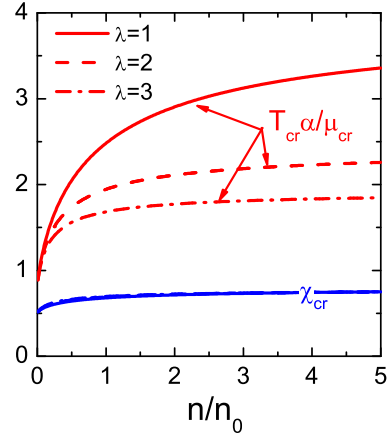}
\caption{{Quantities $\chi_{\rm cr}$ and $\frac{T_{\rm cr}\alpha}{\mu_{\rm cr}}$ given in Eqs.~(\ref{chiTc}) and (\ref{alpha-def}), respectively, as functions of a density $n$ for three values of the coupling parameter $\lambda$.}}
\label{fig:alf-chi}
\end{figure}

We turn now to the combinations of partial derivatives appearing in Eq.~(\ref{Tc-shift}). From Eq.~(\ref{dmudT-2}) we can construct
\begin{align}
&\frac{\partial(\mu-m)}{\partial T}\Big|_n =-\frac{n\chi(T)}{T m^{*2}I_1(1+5C)}\,,
\label{deriv-1}
\end{align}
where the function $\chi(T)$ is
\begin{align}
&\chi(T) = \chi_\mu(T) - \chi_m(T)
\nonumber\\
&= \Big(1+5\lambda (\tilde{d}-d)\Big)
\frac{m^{*2}(m^*-\mu) I_3}{n}
 \label{app:chi}\\
&+\Big(1+10\lambda \tilde{d}\Big) \Big(1-\frac{2\mu m^{*2}}{n}(\tilde{d}+d)\Big)
 +10\lambda\frac{m^{*2}(\tilde{d}+d)^2}{n} \,.\nonumber
\end{align}
We observe that the  terms in $\chi_\mu$ and $\chi_m$ divergent at $T\to T_{\rm cr}$
cancel each other exactly and, therefore, in the limit $T\to T_{\rm cr}$ the function $\chi$ reduces to the finite quantity
\begin{align}
\chi_{\rm cr} =1+{5\lambda}(\tilde{d}_{\rm cr}-d_{\rm cr})
-\frac{2\mu_{\rm cr}^3}{n}(\tilde{d}_{\rm cr}+d_{\rm cr})(1-  10\lambda d_{\rm cr})\,.
\label{chiTc}
\end{align}
The lower limit of $\chi_{\rm cr}$ is realized for $n\to 0$ when $d_{\rm cr}\to 0$ and $\tilde{d}_{\rm cr}/d_{\rm cr}\to\frac12$ and using Eq.~(\ref{d-n-zero}) we have
$\chi_{\rm cr}(n\to 0) \to \frac12$\,. The quantity $\chi_{\rm cr}$ is illustrated in Fig.~\ref{fig:alf-chi}. As we see, it exhibits a rather weak dependence on the coupling constant $\lambda$ and on the pion density $n$.

Since quantity $C$ remains finite at $T=T_{\rm cr}$ and $I_1$ diverges at $T\to T_{\rm cr}$
the derivative ${\partial(\mu-m^*)}/{\partial T}$  at $T=T_{\rm cr}$ vanishes.  The same can be seen  also directly from (\ref{a1-def}). Thus, if we want to estimate the dependence of $\mu-m^*$ on the temperature for $T$ close to $T_{\rm cr}$ we have to calculate the second derivatives.

After differentiating Eq.~(\ref{eq-deriv-1}) second time with respect to $T$, we obtain
\begin{align}
\Big(2m^*-\frac{\partial \Pi}{\partial m^*}\Big)\frac{\partial^2 m^*}{\partial T^2}\Big|_n-
\frac{\partial \Pi}{\partial \mu}\frac{\partial^2 \mu}{\partial T^2}\Big|_n &=\phantom{-}\frac{\partial^2\Pi}{\partial T^2} + A\,,
\nonumber\\
\frac{\partial h}{\partial m^*}\frac{\partial^2 m^*}{\partial T^2}\Big|_n +
\frac{\partial h}{\partial \mu}\frac{\partial^2 \mu}{\partial T^2}\Big|_n&=-\frac{\partial^2 h}{\partial T^2}-m^{*}B\,,
\label{eq-deriv-2}
\end{align}
where
\begin{align}
A &= \hat{D}_T\frac{\partial \Pi}{\partial T}+\hat{D}_T^2(\Pi-m^{*2})\,,
\nonumber\\
m^{*} B &= \hat{D}_T\frac{\partial h}{\partial T}  + \hat{D}_T^2 h
\label{A-B-def}
\end{align}
with $\hat{D}_T$ standing for the differential operator
\begin{align}
\hat{D}_T &=
\frac{\partial (\mu-m^*)}{\partial T}\hat{D}_-
+\frac{\partial (\mu+m^*)}{\partial T}\hat{D}_+\,,
\nonumber\\
&\quad \hat{D}_{\pm} = \frac12\Big(\frac{\partial }{\partial \mu}\pm \frac{\partial}{\partial m^*}\Big)
\,.
\end{align}
From Eq.~(\ref{deriv-hPi-t}) second derivatives with respect to $T$ can be written as
\begin{align}
\frac{\partial^2\Pi}{\partial T^2} &= -\frac{1}{T}\frac{\partial\Pi}{\partial T}
+\frac{5\lambda}{T}\Big(m^{*2}\frac{\partial I_1}{\partial T} - \mu m^{*}\frac{\partial I_3}{\partial T}\Big) \,,
\nonumber\\
\frac{\partial^2h}{\partial T^2} &= -\frac{1}{T}\frac{\partial h}{\partial T}
+\frac{m^{*2}}{T}\Big(m^{*}\frac{\partial I_3}{\partial T} - \mu \frac{\partial I_1}{\partial T}\Big) \,.
\label{dhPidT}
\end{align}
The solution of the system (\ref{eq-deriv-2}) for the second derivatives,  $\partial^2 m^*/\partial T^2$ and
$\partial^2\mu/\partial T^2$, can be cast in the same form as for the first derivatives, Eqs.~(\ref{dmudT}) and (\ref{dmdT}) with the replacement
of $\partial h/\partial T$ and $\partial\Pi/\partial T$ through $\frac{\partial^2 h}{\partial T^2} + m^{*}B$ and $\frac{\partial^2\Pi}{\partial T^2} + A$, repsectively. Now using   (\ref{app:2}), (\ref{Dp-act}), and (\ref{Dm-act}) we  obtain
\begin{align}
\frac{\partial^2(m^*-\mu)}{\partial T^2}\Big|_n &=
\frac{1+5\lambda(\tilde{d}-d)} {m^* I_1(1+5C)}
\Big(\frac{1}{m^{*}}\frac{\partial^2 h}{\partial T^2}+B\Big)
\nonumber\\
&+
\frac{(\tilde{d}+d)}{ m^* I_1(1+5C)}
\Big(\frac{\partial^2\Pi}{\partial T^2}+A\Big)\,.
\label{deriv-2-2}
\end{align}
The full evaluation of this expression is very cumbersome because of the necessity to calculate additional derivatives in Eqs.~(\ref{A-B-def}) and (\ref{dhPidT}). However, the task is simplified, if we are interested in the value of this quantity at $T\to T_{\rm cr}$. In this case the quantity (\ref{deriv-2-2}) does not tend to zero only,
if the divergency of $I_1(T\to T_{\rm cr})$ in the denominator is compensated by another divergent term. Since quantities $d$ and $\tilde{d}$ are finite at $T_{\rm cr}$ the divergence has to be in the second derivatives with respect to the temperature and/or in the quantities $A$ and $B$.  To isolate these terms we can use the following considerations.
First, we note that the quantities $\Pi$, $m^*$ and $h$, which derivatives enter in Eqs.~(\ref{A-B-def}) and (\ref{dhPidT}) are finite at $T_{\rm cr}$. The partial derivative with respect to the temperature cannot increase the degree of the divergence of the integrals, since they lead to the multiplication of an integrand by the quantity $(\om-\mu)\, f(\om-\mu)$, where $f$ is the Bose-Einstein distribution, being regular in the limit $k\to0$ and $\mu\to m^*$.
Similarly, the differential operator $\hat{D}_+$ does not lead to an enhancement of the divergence power of integrals, e.g., see Eq~(\ref{Dp-act}), since the differentiation of the distribution function $f$ is accompanied by the factor $\frac12(m^*-\om)/\om$ killing  additional divergencies at $k\to 0$.
Thus, only terms with the $\hat{D}_-$ operator can be potentially divergent and, therefore, survive in (\ref{deriv-2-2}) at $T_{\rm cr}$. An additional analysis shows that the terms linear in  $\hat{D}_-$ produce at the end terms proportional to $\frac{I_3}{I_1}\frac{\partial (\mu-m^*)}{\partial T}\propto
I_3/I_1^2$, which  vanish at $T\to T_{\rm cr}$. Concluding, we see that in the limit $T\to T_{\rm cr}$ we may keep in Eq.~(\ref{deriv-2-2}) only the terms quadratic in $\hat{D}_-$. As the result we find
\begin{align}
&\frac{\partial^2(m^*-\mu)}{\partial T^2}\Big|_{n, T\to T_{\rm cr}} \to \frac{1}{m^*}
\Big[\frac{\partial (\mu-m^*)}{\partial T}\Big]^2
\nonumber\\
&\times
\left(
\frac{
1+5\lambda(\tilde{d}-d)
} {m^* I_1(1+5C)}\hat{D}_-^2 h
+
\frac{(\tilde{d}+d)}{ I_1(1+5C)}
\hat{D}_-^2\Pi
\right)_{T_{\rm cr}}
\,,
\label{deriv-2-3}
\end{align}
or after account for Eq.~(\ref{Dm-act}) we have
\begin{align}
\frac{\partial^2(m^*-\mu)}{\partial T^2}\Big|_{n,T_{\rm cr}} &=
\frac{n^2\chi_{\rm cr}^2\,}{T_{\rm cr}^2 \mu_{\rm cr}^4
\big(1+10\lambda \tilde{d}\big)^2}
\Big[\frac{\hat{D}_-I_3}{I_1^3}\Big]\Big|_{T\to T_{\rm cr}}
\,,
\label{deriv-2-4}
\end{align}
where
\begin{align}
\Big[\frac{\hat{D}_-I_3}{I_1^3}\Big]\Big|_{T\to T_{\rm cr}}=\frac{4\pi \mu_{\rm cr}}{\, T_{\rm cr}^2}\,.
\end{align}
Combining all terms we  write
the expansion of $m^*-\mu$ for  $T$ close to $ T_{\rm cr}$
as
\begin{align}
m^*-\mu\approx \frac{\alpha}{2\mu_{\rm cr}} (T-T_{\rm cr})^2 \,,
\label{m-mu-exp}
\end{align}
with
\begin{align}
\alpha=
\frac{4\pi n^2\chi_{\rm cr}^2\,}{\mu_{\rm cr}^{2} T_{\rm cr}^4
\big(1+10\lambda \tilde{d}\big)^2}
\,.
\label{alpha-def}
\end{align}
The dependence of the quantity $T_{\rm cr}\,\alpha/\mu_{\rm cr}$ on a density is shown in Fig.~\ref{fig:alf-chi}. We see that this product is finite for $n\to 0$.

\section{$w_Q$ in  isospin-symmetrical system}\label{app:dFdN}

In Sect.~\ref{sec:fluc} we have obtained that the variance of  the total charge, $\varpi_Q$, diverges at $T\to T_{\rm cr}$ in the isospin-symmetrical system even with taking into account of a strong pion-pion interaction within $\lambda\phi^4$ model.  In this Appendix we  study, under which conditions this divergency can be eliminated. For this  we rewrite the variances derived in Sect.~\ref{sec:covariance}, as the derivatives of the densities with respect to  chemical potentials through the derivatives of the chemical potentials with respect to densities. Below we do not indicate explicitly the dependence on the number of $\pi^0$ mesons, assuming that it is fixed.

The chemical potentials $\mu_a$ and the particle densities are connected by Eq.~(\ref{na-def}). We consider this relation as an equation for $n_a(\mu_+,\mu_-)$. Then we can calculate derivatives in (\ref{n-deriv}). On the other hand, the system of equations (\ref{na-def}) implicitly defines functions $\mu_a=\mu_a(n_+,n_-)$.
The relation among the partial derivatives of direct and inverse function implies in this case
\begin{align}
\delta_{ab}=\sum_{c=\pm}\frac{\partial n_a}{\partial \mu_c}\frac{\partial \mu_c}{\partial n_b}\,,\quad a,b=\pm\,.
\end{align}
Terms with $\pi^0$ vanish here, as we assume that $n_{\pi^0}$ is fixed. Then the letter relation can be written as  inversion of a $2\times 2$ matrix,
\begin{align}
\frac{\partial n_a}{\partial \mu_b}=[M^{-1}]_{ab},\quad
M=\left[
\begin{array}{cc}
\frac{\partial \mu_+}{\partial n_+} &
\frac{\partial \mu_+}{\partial n_-}\\
\frac{\partial \mu_-}{\partial n_+} &
\frac{\partial \mu_-}{\partial n_-}
\end{array}\right]\,,
\end{align}
or explicitly as
\begin{align}
\frac{\partial n_\pm}{\partial\mu_\pm} &=\frac{1}{E}\frac{\partial\mu_\mp}{\partial n_\mp}
\,,\quad
\frac{\partial n_\pm}{\partial\mu_\mp}=-\frac{1}{E}\frac{\partial \mu_\pm}{\partial n_\mp}\,,
\nonumber\\
E &=\frac{\partial \mu_+}{\partial n_+}\frac{\partial \mu_-}{\partial n_-}
-\frac{\partial \mu_+}{\partial n_-}\frac{\partial \mu_-}{\partial n_+}\,.
\label{app:D-def}
\end{align}
Using these relations the charge variance (\ref{varQ-def}) can be written as
\begin{align}
\varpi_Q &=
3\frac{T}{n}\frac{1}{E}\Big[\frac{\partial \mu_-}{\partial n_-}
+\frac{\partial \mu_+}{\partial n_+}
+\frac{\partial \mu_+}{\partial n_-}
+\frac{\partial \mu_-}{\partial n_+}
\Big]
\nonumber\\
&=3\frac{T}{n}\frac{1}{E}\Big(\frac{\partial}{\partial n_+} +\frac{\partial}{\partial n_-}\Big)( \mu_+ + \mu_- )\,.
\label{app1:varQ}
\end{align}
Now we can introduce densities $n_{\rm ch}=n_++n_-$ and $n_Q=n_+-n_-$ and the corresponding derivatives
\begin{align}
\frac{\partial}{\partial n_\pm}=\frac{\partial}{\partial n_{\rm ch}} \pm\frac{\partial}{\partial n_Q}\,.
\end{align}
So, the quantity $E$ in (\ref{app:D-def}) can be rewritten as
\begin{align}
E
 &= \frac{\partial (\bar\mu_+ - \bar\mu_-)}{\partial n_Q}
\frac{\partial (\bar\mu_++\bar\mu_-)}{\partial n_{\rm ch}}
-\frac{\partial (\bar\mu_+ - \bar\mu_-)}{\partial n_{\rm ch}}\frac{\partial (\bar\mu_+ + \bar\mu_-)}{\partial n_Q} \,.
\label{app1:D}
\end{align}
Since we consider fluctuations in the isospin-symmetrical system we need these derivatives  for $n_Q=0$ and $n_{\rm ch}=\frac23 n$. Since $\mu_+ - \mu_-$ and $\mu_+ + \mu_-$ are odd and even functions of $n_Q$, respectively, the last term in (\ref{app1:D}) vanishes.\footnote{
Symmetry properties of the functions $g_s(n_{\rm ch},n_Q)= \mu_+ + \mu_-$ and $g_a(n_{\rm ch},n_Q)= \mu_+ - \mu_-$ can be verified, if we formally rename positive and negative pions, i.e. interchange all minuses and pluses (``$+$''$\leftrightarrow$``$-$''), that leads to the relations
\begin{align}
g_s(n_{\rm ch},\pm n_Q) &= (\mu_- + \mu_+)= \phantom{-}g_s(n_{\rm ch}, n_Q)\,,
\nonumber\\
g_a(n_{\rm ch},\pm n_Q)&= (\mu_- - \mu_+)= -g_a(n_{\rm ch}, n_Q)\,.
\nonumber
 \end{align}
. }
 Then from Eqs.~(\ref{app1:varQ}) and (\ref{app1:D}) we find a simple relation
\begin{align}
\varpi_Q& =6\frac{T}{n}\Big[\frac{\partial (\bar\mu_+-\bar\mu_-)}{\partial n_Q}
\Big|_{n_Q=0}\Big]^{-1}\,.
\label{app1:varQ-2}
\end{align}
Alternatively we can rewrite it through the free energy density $F(n_+,n_-,T)$, when
the chemical potentials as functions of densities can be obtained as partial derivatives of  $\mu_\pm=\frac{\partial F}{\partial n_\pm}$ at fixed $T$ and $V$. Then we have
$\mu_+ - \mu_-=2\frac{\partial F}{\partial n_{\rm Q}}$ and therefore Eq.~(\ref{app1:varQ-2}) takes the form
\begin{align}
\varpi_Q& =3\frac{T}{n}\Big[\frac{\partial^2 F}{\partial n_{Q}^2}\Big|_{n_Q=0}
\Big]^{-1}\,.
\label{app1:varQ-3}
\end{align}

\section{$I_n (T\to T_{\rm cr}^{\tilde{a}})$ at non-zero isospin imbalance}\label{app:Iexpan}

In this Appendix we consider pion gas with a small isospin imbalance, $\delta n_Q\neq 0$ or $\delta n_G\neq 0$, so the  critical temperature of the BEC is largest for the  species $\tilde{a}$, i.e., $T_{\rm cr}^{\tilde{a}}=\max_a\{ T_{\rm cr}^{a}\}$. We are interested in quantities $I_n^b(T\to T_{\rm cr}^{\tilde{a}})$ for $b\neq \tilde{a}$ in the case of very small imbalance $|\delta n_{Q,G}|\ll n$, i.e. when $|T_{\rm cr}^{b,\tilde{a}}-T_{\rm cr}|\ll T_{\rm cr}$, where $T_{\rm cr}$ is the critical temperature for the isospin-symmetrical system with the density $n$. The derived results are needed for expansion of Eqs.~(\ref{mixed-deriv-nQ}), (\ref{mixed-deriv-nGm}), and (\ref{mixed-deriv-nGp}).
To find leading and next-to-leading terms in the $\delta n_{Q,G}$ expansion of $I_n^{b}(T_{\rm cr}^{\tilde{a}})$ for small $\delta n_{Q,G}$, we first separate the divergent part (\ref{I3-limit}), determined by the small momenta in the integrals (\ref{I-def}). Then we have
\begin{align}
I_n^{b}(T_{\rm cr}^{\tilde{a}}) \approx
\frac{T_{\rm cr}^{\tilde{a}}}{2^{3/2}\pi
\sqrt{m_b^*(T_{\rm cr}^{\tilde{a}})[m_b^*(T_{\rm cr}^{\tilde{a}})-\mu_b(T_{\rm cr}^{\tilde{a}})}]} + \delta I_{{\rm cr}, n}\,,
\label{In-atT+reg}
\end{align}
where in the regular term $\delta I_{{\rm cr}, n}$ we can take all quantities in the
isospin symmetrical limit.

\subsection{ Variation $\delta n_Q$  at $\delta n_G=0$}

Here we consider variations  $\delta n_Q\neq 0$ at $\delta n_G=0$, studied  in Sect. \ref{subsec:fluct-imbal}, point (i). The condition $\delta n_G=0$ implies $\delta n_+ = -\delta n_-$ and along this line, as we see in Fig.~\ref{fig:pm-plane} (b,c), the maximal critical temperature is realized for the most abundant charged species, i.e. $T_{\rm cr}^{+} \lessgtr T_{\rm cr}^{-}$ provided  $\delta n_+ \lessgtr \delta n_-$\,, respectively.
To be specific  consider $\delta n_Q>0$, then the critical temperature of positively charged pions, $T_{\rm cr}^{+}$, is the highest one.

Our aim here is to find quantities $I_n^{0,-}( T_{\rm cr}^{+})$ for $\delta n_Q\ll n$ and, therefore, for $\delta T_{\rm cr}^{+}=T_{\rm cr}^{+} - T_{\rm cr}\ll T_{\rm cr}$, where $T_{\rm cr}$ is the  critical temperature of the isospin-symmetrical pion gas with density $n$.
To use Eq.~(\ref{In-atT+reg}) we have to evaluate the differences $m_a^*(T_{\rm cr}^{+}) - \mu_a(T_{\rm cr}^{+})$  for $a=0,-$. Taking into account  (\ref{etaQ}), we find
\begin{align}
&m_{-}^*(T_{\rm cr}^{+}) - \mu_{-}(T_{\rm cr}^{+})=m_{+}^*(T_{\rm cr}^{+})-\mu_{+}(T_{\rm cr}^{+})
\nonumber\\
&\quad +\mu_{+}(T_{\rm cr}^{+}) - \mu_{-}(T_{\rm cr}^{+})
\approx 2\delta \mu_{+}(T_{\rm cr}^{+}) \approx \frac{\delta n_Q}{\mu_{\rm cr}^2 I_1(T_{\rm cr}^{+})},
\nonumber\\
&m_{0}^*(T_{\rm cr}^{+}) - \mu_{0}(T_{\rm cr}^{+}) = m_{+}^*(T_{\rm cr}^{+})-\mu_{+}(T_{\rm cr}^{+})
\nonumber\\
&\quad +\mu_{+}(T_{\rm cr}^{+}) - \mu_{0}(T_{\rm cr}^{+})
\approx
\delta \mu_{+}(T_{\rm cr}^{+}) \approx \frac{\delta n_Q}{2\mu_{\rm cr}^2 I_1(T_{\rm cr}^{+})},
\label{mminmu}
\end{align}
where we used that the effective masses do not depend on $\delta n_Q$, i.e.
$m^*_{-,0}(T_{\rm cr}^{+})=m^*_{+}(T_{\rm cr}^{+})$ and corrections to the chemical potentials are given by Eq.~(\ref{etaQ}). Also we can write the expansion for the effective mass and the chemical potential
\begin{align}
m^*_{-}(T_{\rm cr}^{+}) &\approx m^*_{0}(T_{\rm cr}^{+}) \approx m^*_{+}(T_{\rm cr}^{+}) = \mu_+(T_{\rm cr}^{+})\,,
\label{mpTcr-m}\\
\mu_+(T_{\rm cr}^{+}) & \approx
\mu(T_{\rm cr}^{+}) + \frac{\delta n_Q}{2\mu_{\rm cr}^2I_1(T_{\rm cr}^{+})}
\nonumber\\
&\approx \mu_{\rm cr}+\beta \, T_{\rm cr}\frac{\delta n_Q}{2n}\frac{\eta_{T_{\rm cr}}^{(Q)}}{\eta_n} +
\frac{\delta n_Q}{2\mu_{\rm cr}^2I_1(T_{\rm cr}^{+})} \,,
\label{mpTcr-mu}
\end{align}
where we used Eqs.~(\ref{a1-def}) and
that the critical temperature $T_{\rm cr}^+$ is shifted with respect to $T_{\rm cr}$ according to Eq.~(\ref{dTc-Q}),
\begin{align}
T_{\rm cr}^+\approx T_{\rm cr}\Big(1+\frac{\delta n_Q}{2n}\frac{\eta_{T_{\rm cr}}^{(Q)}}{\eta_n}\Big)
\,.
\label{Tcrp-Tc}
\end{align}
In Eqs.~(\ref{mminmu}) and (\ref{mpTcr-mu}) the quantity $I_1(T_{\rm cr}^{+})$ is calculated with effective masses and chemical potentials computed for the isospin-symmetrical pion gas but at the temperature $T_{\rm cr}^+$, i.e. $m^*(T_{\rm cr}^{+})$ and $\mu(T_{\rm cr}^{+})$, respectively.

To evaluate $I_1(T_{\rm cr}^{+})$ we  use  (\ref{In-atT+reg}), where we replace  $m_a^*$ and $\mu_a$ to $m^*$ and $\mu$, respectively and
take into account  that
\begin{align}
m^*(T_{\rm cr}^+)\approx \mu_{\rm cr} +\beta\, T_{\rm cr} \frac{\delta n_Q}{2n}\frac{\eta_{T_{\rm cr}}^{(Q)}}{\eta_n}
\label{mTcr}
\end{align}
according to  Eqs.~(\ref{a1-def}) and (\ref{dTc-Q}).
The mass difference in the numerator of the singular term in $I_1(T_{\rm cr}^{+})$ can be rewritten as
\begin{align}
&m^*(T_{\rm cr}^{+})-\mu(T_{\rm cr}^{+})=m^*(T_{\rm cr}+\delta T_{\rm cr}^{+})-\mu(T_{\rm cr}+\delta T_{\rm cr}^{+})
\nonumber\\
&\quad
\approx \frac12\frac{\alpha}{\mu_{\rm cr}}\,[\delta T_{\rm cr}^{+}]^2
\approx\frac{\alpha}{2\mu_{\rm cr}}\,
\frac{(\delta n_Q)^2}{4n^2}\frac{\eta_{T_{\rm cr}}^{(Q)2}}{\eta_n^2} T_{\rm cr}^2\,,
\label{Dmsmu}
\end{align}
where we used Eq.~(\ref{m-mu-exp}) with the coefficient $\alpha$ given in Eq.~(\ref{alpha-def}). Thus, using Eqs.~(\ref{Tcrp-Tc}), (\ref{mTcr}), and (\ref{Dmsmu}) we obtain
\begin{align}
I_1(T_{\rm cr}^{+}) &\approx
\frac{1}{\pi\alpha^{1/2}_{\rm cr}}
\Big(\frac{n}{\delta n_Q}\frac{\eta_n}{\eta_{T_c}^{(Q)}}
+\frac12-\frac{\beta}{4}\frac{T_{\rm cr}}{\mu_{\rm cr}}\Big) +
 \delta I_{{\rm cr},1}\,.
\label{I1exp}
\end{align}
Now taking  $I_n^{a}(T_{\rm cr}^{+})$ from (\ref{In-atT+reg}) we can write for $a=``-"$:
\begin{align}
I_n^{-}(T_c^{+}) &\approx
 \beta^{(Q)}
\frac{n}{\delta n_Q}
\Big(
1
+\frac{\delta n_Q}{2n}\frac{\eta_{T_c}^{(Q)}}{\eta_n}
 \nonumber\\
&\times \Big(\frac32-\frac34\beta\frac{T_{\rm cr}}{\mu_{\rm cr}} +
\pi\alpha^{1/2}_{\rm cr}\delta I_{{\rm cr},1}\Big)
\Big)
+ \delta I_{{\rm cr},n}\,,
\label{Imin-exp}
\end{align}
where
$$\beta^{(Q)} = \sqrt{\frac{\mu_{\rm cr}T_{\rm cr}^2}{
(2\pi)^{3}\alpha^{1/2}_{\rm cr}
n}
\frac{\eta_n}{\eta_{T_c}^{(Q)}}
}\,.$$
 Analogously, for neutral pions we get
\begin{align}
I_n^{(0)}(T_c^{(+)})\approx \sqrt{2} [I_n^{(-)}(T_c^{(+)}) -\delta I_{{\rm cr},n}] +\delta I_{{\rm cr},n}\,.
\label{I0-exp}
\end{align}

Some comments about parameters of our expansions are in order. Expressions for shifts of pion effective masses and chemical potentials obtained in Sect.~\ref{subsec:prop} are derived as expansions in $\delta n_a$ up to linear terms $O(\delta n_a)$, see Eqs.~(\ref{dmu}) and (\ref{eff-mass-exp}). The results are valid for any temperature $T\ge \max_a \{T_{\rm cr}^a\}$. The difference between the effective mass and the chemical potential, e.g. in Eq.~(\ref{mminmu}), is $\propto\delta n_Q/I_1(T)$. Formally for arbitrary temperatures $T>T_{\rm cr}^+$ this result is  of the order $O(\delta n_Q)$. However,  for $T\to T_{\rm cr}^+=T_{\rm cr}+O(\delta n_Q)$ the quantity $I_1$ behaves at the leading order like $I(T_{\rm cr}^+)\propto 1/\delta n_Q$, see Eq.~(\ref{I1exp}). Therefore, the expansion~(\ref{mminmu}) is effectively of the order $O(\delta n_Q^2)$. The same expansion order is explicitly seen in the difference  $m^*-\mu$ at $T_{\rm cr}^+$ for the isospin symmetric medium, Eq.~(\ref{Dmsmu}), which is based on the expansion (\ref{m-mu-exp})  independently on the $\delta n_Q$ and that $\delta T_{\rm cr}^a\propto \delta n_Q $.
The final expressions of this section (\ref{I1exp}), (\ref{Imin-exp}), and (\ref{I0-exp}) hold up to terms linear in $\delta n_Q$.

\subsection{ Variation $\delta n_G$ at $\delta n_Q =0$ }\label{app:dnG-1}

{\it (a)} Let now $\delta n_Q=0$ and $\delta n_G<0$. In this case,  the maximal is the critical temperature of the BEC for neutral pions, $T_{\rm cr}^{0}$. In order to expand relations  (\ref{mixed-deriv-nGm}) in small quantity $-\delta n_G\ll n$ we need the corresponding expansions of $I_n^{+}(T_{\rm cr}^0)$. We can use a relation analogous to (\ref{In-atT+reg})
and expand the mass-chemical potential difference in the denominator with the help of Eq.~(\ref{nG-res}) as follows
\begin{align}
&m_{+}^*(T_{\rm cr}^{0}) - \mu_{+}(T_{\rm cr}^{0}) \approx
m_0^*(T_{\rm cr}^{0})+3\delta m^*_+(T_{\rm cr}^{0})
\nonumber\\
&- \mu_0(T_{\rm cr}^{0}) - 3\delta\mu_+(T_{\rm cr}^{0})=
3\big(\delta m^*_+(T_{\rm cr}^{0}) -\delta\mu_+(T_{\rm cr}^{0})\big)
\nonumber\\
&
= m^{*}(T_{\rm cr}^{0})\frac{\delta n_G}{ n} \big(\frac12\eta_m^{(G)}(T_{\rm cr}^{0}) -\eta_\mu^{(G)}(T_{\rm cr}^{0})\big)\,.
\label{Dmmu-G}
\end{align}
We observe that although each of the quantities $\eta_m^{(G)}(T_{\rm cr}^{0})$ and $\eta_\mu^{(G)}(T_{\rm cr}^{0})$ remains finite, if $T_{\rm cr}^0\to T_{\rm cr}$, see Eq.~(\ref{nG-res}), their difference in (\ref{Dmmu-G}) vanishes and hence integrals $I_n^+(T_{\rm cr}^0)$ are enhanced, when $\delta n_G\to 0$ and $T_{\rm cr}^0\to T_{\rm cr}$. Being interested only in terms $\sim O(\delta n_G)$ in the small-$\delta n_G$ expansion, we can write
\begin{align}
m_{+}^*(T_{\rm cr}^{0}) - \mu_{+}(T_{\rm cr}^{0})
&\approx  -\frac{\delta n_G}{m^{*2} I_1(T_{\rm cr}^{0}) }
\frac{1 +\lambda (I_2 -I_3)}{1+2C}\Big|_{T_{\rm cr}}
\nonumber\\
&= -\frac{\delta n_G}{\mu_{\rm cr}^2 I_1(T_{\rm cr}^{0})}\frac{\eta_{T_{\rm cr}}^{(G)}}{\eta_{T_{\rm cr}}^{(Q)}}\,,
\label{Dmpmup}
\end{align}
where in the last equation we used Eqs.~(\ref{I123-relat}), (\ref{dTcr-G}), (\ref{X-def}),
and (\ref{CatTc}). Next-to-leading terms in these expansions are of the order $O([\delta n_G/ I_1(T_{\rm cr}^{0})]^2)$.
For the integral $I_1(T_{\rm cr}^{0})$ we can use the expansion (\ref{In-atT+reg}), where we replace  $m_a^*$ and $\mu_a$ by $m^*$ and $\mu$, respectively. According to Eq.~(\ref{dTcr-G}) we have
\begin{align}
T_{\rm cr}^0=T_{\rm cr}\Big(1-\frac23\frac{\delta n_G}{n}
\frac{\eta_{T_{\rm cr}}^{(G)}}{\eta_n}\Big) + O\big((\delta n_G)^2\big)
\,.
\label{Tc0epx-1}
\end{align}
The expression for the $m-\mu$ difference can be written in analogy to Eq.~(\ref{Dmsmu}) using Eqs.~(\ref{m-mu-exp}) and (\ref{Tc0epx-1}):
\begin{align}
&m^*(T_{\rm cr}^{0})-\mu(T_{\rm cr}^{0})
\approx \frac{\alpha_{\rm cr}}{2\mu_{\rm cr}}\,\frac{4}9
\frac{(\delta n_G)^2}{n^2}\frac{\eta_{T_{\rm cr}}^{(G)2}}{\eta_n^2} T_{\rm cr}^2
\,.
\label{m-mu-T0}
\end{align}
Since, as we show below, $I_1(T_{\rm cr}^0)\propto 1/\delta n_G$, the expansion (\ref{Dmpmup}) is of the same quadratic order in $\delta n_G$ as expansion (\ref{m-mu-T0}).
Additionally, to get the expansion $I_1(T_{\rm cr}^{0})$ we need the expansion for the effective mass,
\begin{align}
m^*(T_{\rm cr}^0) & = m^*(T_{\rm cr})+\beta\delta T_{\rm cr}^0
+ O\big((\delta T_{\rm cr}^0)^2\big)
\nonumber\\
&= \mu_{\rm cr}
-\beta\, T_{\rm cr} \frac23\frac{\delta n_G}{n}\frac{\eta_{T_{\rm cr}}^{(Q)}}{\eta_n}
+O\big((\delta n_G)^2\big)
\label{mTcro}
\end{align}
where we used (\ref{a1-def}) and (\ref{Tc0epx-1}).

Thus, we obtain
\begin{align}
I_1(T_{\rm cr}^{0}) & =
-\frac{1}{\pi\alpha^{1/2}_{\rm cr}}
\Big(\frac{3n}{4\delta n_G}\frac{\eta_n}{\eta_{T_{\rm cr}}^{(G)}}
-\frac12
+\frac{\beta}4 \frac{T_{\rm cr}}{\mu_{\rm cr}}
\Big)
\nonumber\\
&+ \delta I_{{\rm cr},1} + O(\delta n_G)\,.
\label{I1exp-G}
\end{align}
To derive the expansion of $I_n^{+}(T_{\rm cr}^{0})$ we also need the expansion for the effective mass $m_+^*(T_{\rm cr}^0)$, which we obtain using  relations (\ref{nG-res}),
\begin{align}
&m_{+}^*(T_{\rm cr}^{0})
\approx  m_0^*(T_{\rm cr}^{0})+3\delta m^*_+(T_{\rm cr}^{0})
\nonumber\\
&=\mu(T_{\rm cr}^{0})-2\delta\mu_+(T_{\rm cr}^{0}) + 3\delta m^*_+(T_{\rm cr}^{0})
\nonumber\\
&\approx\mu_{\rm cr}+\beta \delta T_{\rm cr}^0-2\delta\mu_+(T_{\rm cr}^{0}) + 3\delta m^*_+(T_{\rm cr}^{0})
\nonumber\\
& \approx
\mu_{\rm cr}\Big(1-\beta\frac23\frac{\delta n_G}{n}\frac{\eta_{T_{\rm cr}}^{(G)}}{\eta_n}\frac{T_{\rm cr}}{\mu_{\rm cr}}
\nonumber\\
& + \frac{\delta n_G}{3n}\Big(\frac32\eta_m^{(G)}(T_{\rm cr})-2\eta_\mu^{(G)}(T_{\rm cr})\Big)
\Big)
\label{mpTc0}\\
&\approx
\mu_{\rm cr}\Big(1-\beta\frac23\frac{\delta n_G}{n}\frac{\eta_{T_{\rm cr}}^{(G)}}{\eta_n}\frac{T_{\rm cr}}{\mu_{\rm cr}}  + \frac{\lambda\delta n_G}{3\mu_{\rm cr}^{3}(1+2C)}
\Big) + O\big((\delta n_G)^2\big).\nonumber
\end{align}
The expansion for the chemical potential $\mu_{\rm cr,0}$ is obtained using Eqs.~(\ref{nG-res}) and (\ref{Tc0epx-1}),
\begin{align}
&\mu_{\rm cr,0} = \mu_0(T_{\rm cr}^0) = \mu(T_{\rm cr}^0)+\delta \mu_0(T_{\rm cr}^0)
\label{mu0Tc0}\\
& \simeq \mu_{\rm cr}+\beta \delta T_{\rm cr}^0
- \mu_{\rm cr}\frac23 \frac{\delta n_G}{n}\eta_\mu^{(G)}(T_{\rm cr})
 +O\big((\delta T_{\rm cr}^0)^2,(\delta n_G)^2\big)
\nonumber\\
& \approx
\mu_{\rm cr}\Big(1  -\beta\frac23\frac{\delta n_G}{n}\frac{\eta_{T_{\rm cr}}^{(G)}}{\eta_n}\frac{T_{\rm cr}}{\mu_{\rm cr}}
-\frac{2\lambda\delta n_G}{3\mu_{\rm cr}^{3}(1+4\lambda \tilde{d}_{\rm cr})}
\Big)+ O\big((\delta n_G)^2\big)\,.\nonumber
\end{align}

Now substituting Eqs.~(\ref{Dmpmup}), (\ref{Tc0epx-1}), and (\ref{mpTc0}) in Eq.~(\ref{In-atT+reg}) we can expand
\begin{align}
&I_n^{+}(T_{\rm cr}^0) \approx
-\beta^{(G)}\frac{n}{\delta n_G}
\Big[1 -\frac{\lambda\delta n_G}{6\mu_{\rm cr}^3(1+4\lambda\tilde{d}_{\rm cr})}
\nonumber\\
&- \frac{2\delta n_G}{3n}\frac{\eta_{T_{\rm cr}}^{(G)}}{\eta_n}
 \Big(\frac32 -\frac{3}{4}\beta\frac{T_{\rm cr}}{\mu_{\rm cr}} +
\pi\alpha^{1/2}_{\rm cr}\delta I_{\rm cr 1}(T_{\rm cr})
\Big) \Big]
\nonumber\\
&+ \delta I_{{\rm cr},n} + O(\delta n_G) \,,
\label{IpTcr0}
\end{align}
where
\begin{align}
\beta^{(G)} = \beta^{(Q)} \frac{\sqrt{3}}{2}\frac{\eta^{(Q)}_{T_{\rm cr}}}
{\eta^{(G)}_{T_{\rm cr}}}\,.
\end{align}

{\it (b)} Now we consider the case $\delta n_G>0$. The maximal critical temperature is now $T_{\rm cr}^{+}$. To expand relations (\ref{mixed-deriv-nGp}) we have to find expansions of $I_n^{0}(T_{\rm cr}^+)$ for $\delta n_G\ll n$. For this we need the expansion for the critical temperature,
\begin{align}
T_{\rm cr}^+=T_{\rm cr}\Big(1+ \frac13\frac{\delta n_G}{n}
\frac{\eta_{T_{\rm cr}}^{(G)}}{\eta_n} \Big) +O\big((\delta n_G)^2\big)\,,
\label{Tcrp-exp}
\end{align}
and for the mass difference
\begin{align}
m_{0}^*(T_{\rm cr}^{+}) - \mu_{0}(T_{\rm cr}^{+})
&\approx 3\big(\delta \mu_+(T_{\rm cr}^+)-
\delta m_+^* (T_{\rm cr}^+)\big)
\nonumber\\
&=   \frac{\delta n_G}{\mu_{\rm cr}^2 I_1(T_{\rm cr}^{+})}\frac{\eta_{T_{\rm cr}}^{(G)}}{\eta_{T_{\rm cr}}^{(Q)}}\,,
\label{dm0dmu0}
\end{align}
which we derived in a similar way as in Eqs.~(\ref{Dmmu-G}) and (\ref{Dmpmup}).
To expand $I_1(T_{\rm cr}^{+})$ we need the difference
\begin{align}
&m^*(T_{\rm cr}^{+})-\mu(T_{\rm cr}^{+})\approx\frac{\alpha}{2\mu_{\rm cr}}\,
\frac{(\delta n_G)^2}{9n^2}\frac{\eta_{T_{\rm cr}}^{(G)2}}{\eta_n^2} T_{\rm cr}^2\,,
\label{Dmsmu-1}
\end{align}
obtained using Eqs.~(\ref{m-mu-exp}) and (\ref{Tcrp-exp}),
and the relation for the effective mass,
\begin{align}
m^*(T_{\rm cr}^+)\approx \mu_{\rm cr}
+\beta\, T_{\rm cr} \frac13\frac{\delta n_G}{n}\frac{\eta_{T_{\rm cr}}^{(Q)}}{\eta_n}
+O\big((\delta n_G)^2\big)\,,
\label{mTcrp}
\end{align}
where we used Eq.~(\ref{a1-def})  and (\ref{Tcrp-exp}).
As we have argued above for the cases described by  Eqs.~(\ref{mminmu}) and (\ref{Dmsmu}) and Eqs.~(\ref{Dmpmup}) and (\ref{m-mu-T0}), the differences between an effective mass and chemical potentials in Eqs.~(\ref{dm0dmu0}) and (\ref{Dmsmu-1}) prove to be  of the order $(\delta n_G)^2$, that is seen after  taking into account that $I_1(T_{\rm cr}^{+})\propto 1/\delta n_G$.

Now substituting Eqs.~(\ref{Dmsmu-1}) and (\ref{Tcrp-exp}) in Eq.~(\ref{In-atT+reg}) with the pion mass and the chemical potential taken as in the isospin symmetrical matter, we obtain
\begin{align}
I_1(T_{\rm cr}^{+}) &\approx
\frac{1}{\pi\alpha^{1/2}_{\rm cr}}
\Big(\frac{3n}{2\delta n_G}\frac{\eta_n}{\eta_{T_{\rm cr}}^{(G)}}
+\frac12 -\frac{\beta}{4}\frac{T_{\rm cr}}{\mu_{\rm cr}}\Big)
\nonumber\\
&+  \delta I_{\rm cr,1} +O(\delta n_G) \,.
\label{I1Tcrp-1}
\end{align}
Now using this result we can evaluate Eq.~(\ref{dm0dmu0}) and substitute it in Eq.~(\ref{In-atT+reg}) together with the effective mass
\begin{align}
m_{0}^*(T_{\rm cr}^{+})
& =
\mu_{\rm cr}\Big(1 +\beta\frac13\frac{\delta n_G}{n}\frac{\eta_{T_{\rm cr}}^{(G)}}{\eta_n}\frac{T_{\rm cr}}{\mu_{\rm cr}}
\nonumber\\
&- \frac{\delta n_G}{3n}\Big(\frac32\eta_m^{(G)}(T_{\rm cr})-\eta_\mu^{(G)}(T_{\rm cr})\Big)
\Big) +O\big((\delta n_G)^2\big)
\nonumber\\
& =
\mu_{\rm cr}\Big(1+\beta\frac13\frac{\delta n_G}{n}\frac{\eta_{T_{\rm cr}}^{(G)}}{\eta_n}\frac{T_{\rm cr}}{\mu_{\rm cr}} - \frac{\lambda\delta n_G}{3\mu_{\rm cr}^{3}(1+2C)}
\Big)
\nonumber\\
& +O\big((\delta n_G)^2\big)\,,
\label{m0Tcp}
\end{align}
and the chemical potential
\begin{align}
\mu_{\rm cr,+}
 &=
\mu_{\rm cr}\Big(1+\beta\frac13\frac{\delta n_G}{n}\frac{\eta_{T_{\rm cr}}^{(G)}}{\eta_n}\frac{T_{\rm cr}}{\mu_{\rm cr}} + \frac{\lambda\delta n_G }{3\mu_{\rm cr}^{3}(1+2C)}
\Big)
\nonumber\\
& +O\big((\delta n_G)^2\big)\,,
\label{mu0Tcp}
\end{align}
obtained in the same way as Eqs.~(\ref{mu0Tc0}) and (\ref{mu0Tc0}),
and the critical temperature (\ref{Tcrp-exp}). Finally  we obtain
\begin{align}
I_n^{0}(T_{\rm cr}^+) & \approx
\sqrt{2} \beta^{(G)}\frac{n}{\delta n_G}
\Big[1 + \frac{\lambda \delta n_G}{6\mu_{\rm cr}^{3}(1+2C)}
\nonumber\\
&+ \frac{\delta n_G}{3n}\frac{\eta_{T_{\rm cr}}^{(G)}}{\eta_n}
 \Big(\frac32 -\frac34\beta \frac{T_{\rm cr}}{\mu_{cr}} +
\pi\alpha^{1/2}_{\rm cr} \delta I_{\rm cr,1}
\Big) \Big]
\nonumber\\
&+ \delta I_{{\rm cr},n}  +O\big(\delta n_G\big) \,.
\label{I0Tcp}
\end{align}


\begin{thebibliography}{99}

\bibitem{Montvay-Zimanyi-79}
I.~Montvay and J.~Zimanyi, Hadron chemistry in heavy ion collisions, Nucl. Phys. A {\bf 316}, 490 (1979).

\bibitem{ZFJ}
J.~Zimanyi, G.~Fai, B.~Jakobsson, Bose-Einstein condensation of pions in energetic heavy-ion collisions?, Phys. Rev. Lett. {\bf 43}, 1705 (1979).

\bibitem{Migdal:1990vm}
A.B.~Migdal, E.E.~Saperstein, M.A.~Troitsky and D.N.~Voskresensky,
Pion degrees of freedom in nuclear matter,
Phys.\ Rept.\  {\bf 192}, 179 (1990).

\bibitem{Voskresensky:1993ud}
D.N.~Voskresensky, Many particle effects in nucleus-nucleus collisions, Nucl.\ Phys.\ A {\bf 555}, 293 (1993).

\bibitem{Reisdorf:2006ie}
W.~Reisdorf {\it et al.} [FOPI Collaboration], Systematics of pion emission in heavy ion collisions in the 1 $ A$- GeV regime,
  Nucl.\ Phys.\ A {\bf 781}, 459 (2007).


\bibitem{Afanasiev2002}
S.V.~Afanasiev et al. [NA49 Collab.], Energy dependence of pion and kaon production in central Pb+Pb collisions,
Phys.\ Rev.\ C {\bf 66}, 054902 (2002).

\bibitem{Alt2008}
C.~Alt et al. [NA49 Collab.], Pion and kaon production in central Pb+Pb collisions at 20$A$ and 30$A$\,GeV: Evidence for the onset of deconfinement, Pys.\ Rev.\ C {\bf 77}, 024903 (2008).

\bibitem{Nayak:2012np}
T.K.~Nayak, Heavy ions: results from the large hadron collider, Pramana {\bf 79}, 719 (2012).

\bibitem{Abelev:2012wca}
B.~Abelev {\it et al.} [ALICE Collab.], Pion, kaon, and proton production in central Pb--Pb collisions at $\sqrt{s_{NN}} = 2.76$ TeV,
Phys.\ Rev.\ Lett.\  {\bf 109}, 252301 (2012).

\bibitem{Adamczyk:2017iwn}
L.~Adamczyk {\it et al.} [STAR Collab.], Bulk properties of the medium produced in relativistic heavy-ion collisions from the beam energy scan program, Phys. Rev. C {\bf 96}, 044904 (2017).

\bibitem{Kataja:1990tp}
M.~Kataja and P.V.~Ruuskanen, Nonzero chemical potential and the shape of the $p_T$ distribution of hadrons in heavy-ion collisions,
Phys.\ Lett.\ B {\bf 243}, 181 (1990).

\bibitem{mishust}
I.N.~Mishustin, L.N.~Satarov, J.~Maruhn, H.~St\"ocker, and W.~Greiner,
Pion production and Bose-enhancement effects in relativistic heavy-ion collisions,
Phys.\ Lett.\ B {\bf 276}, 403 (1992).

\bibitem{Lee-Heinz-Schnee-90}
K.S.~Lee, U.~Heinz, and E.~Schnedermann, Search for collective transverse flow using particle transverse momentum spectra in relativistic heavy-ion collisions,
Z. Phys. C {\bf 48}, 525 (1990).

\bibitem{Schnedermann:1993ws}
E.~Schnedermann, J.~Sollfrank, and U.W.~Heinz,
Thermal phenomenology of hadrons from 200-A/GeV S+S collisions,
Phys.\ Rev.\ C {\bf 48}, 2462 (1993).

\bibitem{Ferenc}
D.~Ferenc, U.~Heinz, B.~Tom\'a\v{s}ik, U.A.~Wiedemann, and J.G.~Cramer,
Universal pion freeze-out phase-space density,
Phys. Lett. B {\bf 457}, 347 (1999).

\bibitem{Tomasik-Heinz02}
B.~Tom\'a\v{s}ik and U.~Heinz, Flow effects on the freeze-out phase-space density in heavy-ion collisions, Phys. Rev. C {\bf 65}, 031902(R) (2002).

\bibitem{Bertsch94}
G.F.~Bertsch, Meson phase-space density in heavy-ion collisions from interferometry,
Phys. Rev. Lett. {\bf 72}, 2349 (1994); [Erratum Phys. Rev. Lett. {\bf 77}, 789 (1996)].

\bibitem{Goity:1989gs}
J.L.~Goity and H.~Leutwyler, On the mean free path of pions in hot matter,
Phys.\ Lett.\ B {\bf 228}, 517 (1989).

\bibitem{Gerber:1990yb}
P.~Gerber, H.~Leutwyler, and J.L.~Goity, Kinetics of an expanding pion gas,
Phys.\ Lett.\ B {\bf 246}, 513 (1990).

\bibitem{Voskresensky:1994uz}
D.N.~Voskresensky, On the possibility of Bose-condensation of pions in ultrarelativistic collisions of nuclei,
J.\ Exp.\ Theor.\ Phys.\  {\bf 78}, 793 (1994)
[Zh.\ Eksp.\ Teor.\ Fiz.\  {\bf 105}, 1473 (1994)].

\bibitem{KV95}
E.E.~Kolomeitsev and D.N.~Voskresensky, Bose-Einstein condensation of pions in ultrarelativistic nucleus-nucleus collisions and spectra of kaons,
Phys. Atom. Nucl. {\bf 58}, 2082 (1995).

\bibitem{Kolomeitsev:1996tv}
E.E.~Kolomeitsev, B.~K\"ampfer, and D.N.~Voskresensky,
Hot and dense pion gas with finite chemical potential,
Acta Phys.\ Polonica B {\bf 27}, 3263 (1996).


\bibitem{Hung:1997du}
C.M.~Hung and E.V.~Shuryak,
Equation of state, radial flow and freezeout in high-energy heavy ion collisions,
Phys.\ Rev.\ C {\bf 57}, 1891 (1998).

\bibitem{Stachel}
J.~Stachel, A.~Andronic, P.~Braun-Munzinger, and K.~Redlich,
Confronting LHC data with the statistical hadronization model,
J. Phys. Conf. Ser. {\bf 509}, 012019 (2014).

\bibitem{Petran:2013lja}
M.~Petr\'{a}n, J.~Letessier, V.~Petr\'{a}\v{c}ek, and J.~Rafelski,
Hadron production and quark-gluon plasma hadronization in Pb-Pb collisions at $\sqrt{s_{NN}}=2.76$ TeV,
Phys.\ Rev.\ C {\bf 88}, 034907 (2013).

\bibitem{Teaney:2002aj}
D.~Teaney, Chemical freezeout in heavy ion collisions,  nucl-th/0204023.

\bibitem{Pratt1999}
S.~Pratt and K.~Haglin, Hadronic phase space density and chiral symmetry restoration in relativistic heavy ion collisions,
Phys.\ Rev.\ C {\bf 59}, 3304 (1999).

\bibitem{Melo:2015wpa}
I.~Melo and B.~Tomasik, Reconstructing the final state of Pb+Pb collisions at $\sqrt{s_{NN}}=2.76$ TeV,
J.\ Phys.\ G {\bf 43}, 015102 (2016).

\bibitem{Prorok:2015vxa}
D.~Prorok, Single freeze-out, statistics and pion, kaon and proton production in central Pb-Pb collisions at $\sqrt{s_{NN}} = 2.76$ TeV,
J.\ Phys.\ G {\bf 43}, 055101 (2016).

\bibitem{Song1997}
C.~Song and V.~Koch, Chemical relaxation time of pions in hot hadronic matter,
Phys.\ Rev.\ C {\bf 55}, 3026 (1997).

\bibitem{Prakash1993}
M.~Prakash, M.~Prakash, R.~Venugopalan, and G.~Welke,
Non-equilibrium properties of hadronic mixtures,
Phys.\ Rept.\ {\bf 227}, 321 (1993).

\bibitem{GreinerGong}
C.~Greiner, C.~Gong, and  B.~M\"uller,
Pion condensation in relativistic heavy ion collisions,
Phys.\ Lett.\ B {\bf 316}, 226 (1993).

\bibitem{Voskresensky:1995tx}
D.N.~Voskresensky, D.~Blaschke, G.~R\"opke, and H.~Schulz,
Nonequilibrium approach to dense hadronic matter,
Int.\ J.\ Mod.\ Phys.\ E {\bf 4}, 1 (1995).

\bibitem{Voskresensky:1996ur}
D.N.~Voskresensky,
Kinetic description of a pion gas in ultrarelativistic collisions of nuclei: Turbulence and Bose condensation,
Phys.\ Atom.\ Nucl.\  {\bf 59}, 2015 (1996)
[Yad.\ Fiz.\  {\bf 59}, 2090 (1996)].

\bibitem{ornik}
U.~Ornik, M.~Pl\"umer, and D.~Strottmann, Bose condensation through resonance decay,
Phys.\ Lett.\ {\bf B314}, 401 (1993).

\bibitem{Voskresensky:2004ux}
D.N.~Voskresensky,
Hadron liquid with a small baryon chemical potential at finite temperature,
Nucl.\ Phys.\ A {\bf 744}, 378 (2004).

\bibitem{CC94}
T.~Cs\"org\H{o} and L.P.~Csernai,
Quark-gluon plasma freeze-out from a supercooled state?
Phys. Lett. B {\bf 333}, 494 (1994).

\bibitem{Blaizot:2011xf}
J.P.~Blaizot, F.~Gelis, J.F.~Liao, L.~McLerran, and R.~Venugopalan,
Bose-Einstein condensation and thermalization of the quark-gluon plasma,
Nucl.\ Phys.\ A {\bf 873}, 68 (2012).

\bibitem{Xu:2014ega}
Z.~Xu, K.~Zhou, P.~Zhuang, and C.~Greiner,
Thermalization of gluons with Bose-Einstein condensation,
Phys.\ Rev.\ Lett.\  {\bf 114}, 182301 (2015).

\bibitem{Kochelev16}
N.~Kochelev,
Ultralight glueballs in quark-gluon plasma,
Phys.\ Part.\ Nucl.\ Lett.\  {\bf 13}, 149 (2016).

\bibitem{Peshier2016}
A.~Peshier and D.~Giovannoni,
The cool potential of gluons,
J.\ Phys.\ Conf.\ Ser.\  {\bf 668}, 012076 (2016).

\bibitem{Tanji:2017suk}
N.~Tanji and R.~Venugopalan,
Effective kinetic description of the expanding overoccupied Glasma,
Phys.\ Rev.\ D {\bf 95}, 094009 (2017).

\bibitem{Begun:2014rsa}
V.~Begun, W.~Florkowski, and M.~Rybczynski, Transverse-momentum spectra of strange particles produced in Pb+Pb collisions at $\sqrt{s_{\rm NN}}=2.76$\,TeV in the chemical non-equilibrium model,
Phys.\ Rev.\ C {\bf 90}, 054912 (2014).


\bibitem{Abelev-coherent}
B. Abelev et al. [ALICE Collaboration],
Two- and three-pion quantum statistics correlations in Pb-Pb collisions at
$\sqrt{s_{NN}} = 2.76$\,TeV at the CERN Large Hadron Collider,
Phys. Rev. C {\bf 89}, 024911 (2014).

\bibitem{Adam-coherent}
J. Adam et al. [ALICE Collaboration], Multipion Bose-Einstein correlations in $pp$, $p$-Pb, and Pb-Pb collisions at energies available at the CERN Large Hadron Collider
Phys. Rev. C {\bf 93}, 054908 (2016).

\bibitem{Akkelin-02}
S.V.~Akkelin, R.~Lednicky, and Yu.M.~Sinyukov, Correlation search for coherent pion emission in heavy ion collisions, Phys. Rev. C {\bf 65}, 064904 (2002).

\bibitem{Wong-Zhang07}
C.Y. Wong and W.N. Zhang, Chaoticity parameter $\lambda$ in Hanbury-Brown–Twiss interferometry, Phys. Rev. C {\bf 76}, 034905 (2007).

\bibitem{Begun:2015ifa}
V.~Begun and W.~Florkowski,
Bose-Einstein condensation of pions in heavy-ion collisions at the CERN Large Hadron Collider (LHC) energies,
Phys.\ Rev.\ C {\bf 91}, 054909 (2015).

\bibitem{Shuryak:2014zxa}
E.~Shuryak,
Strongly coupled quark-gluon plasma in heavy-ion collisions,
Rev.\ Mod.\ Phys.\  {\bf 89}, 035001 (2017).

\bibitem{LLP8}
L.D.~Landau, E.M.~Lifshitz, and L.P.~Pitaevskii, {\it El\-ectro\-dynamics of Continuous Media, Vol. 8} (Pergamon Press,  Oxford, 1984).

\bibitem{Maslov:2019dep}
K.A.~Maslov and D.N.~Voskresensky,
RMF models with $\sigma$-scaled hadron masses and couplings for description of heavy-ion collisions below 2$A$GeV,
Eur.\ Phys.\ J.\ A {\bf 55}, 100 (2019).

\bibitem{Begun:2006gj}
V.V.~Begun and M.I.~Gorenstein,
Bose-Einstein condensation of pions in high multiplicity events,
Phys.\ Lett.\ B {\bf 653}, 190 (2007).

\bibitem{Begun:2008hq}
V.V.~Begun and M.I.~Gorenstein,
Bose-Einstein condensation in the relativistic pion gas: thermodynamic limit and finite size effects,
Phys.\ Rev.\ C {\bf 77}, 064903 (2008).

\bibitem{Kokoulina:2011ed}
E.~Kokoulina, Neutral pion fluctuations in $pp$ collisions at 50 GeV by SVD-2,
Prog.\ Theor.\ Phys.\ Suppl.\  {\bf 193}, 306 (2012).

\bibitem{Ryadovikov12}
V.N.~Ryadovikov, Neutral-pion fluctuations at high multiplicity in $pp$ interactions
at 50 GeV, Phys. At. Nucl. {\bf 75}, 989 (2012).

\bibitem{Asakawa-Kitazawa2015}
M.~Asakawa and M. Kitazawa, Fluctuations of conserved charges in relativistic heavy ion collisions: An introduction, arxiv:1512.05038.

\bibitem{Jeon:2003gk}  S.~Jeon and V.~Koch,
Event by event fluctuations, {\it Quark gluon plasma},
edited by R.C.~Hwa, R.C. et al., 430-490 (2003)
  [hep-ph/0304012].

\bibitem{Heiselberg:2000ti}
H.~Heiselberg and A.~D.~Jackson,
Anomalous multiplicity fluctuations from phase transitions in heavy ion collisions,
Phys.\ Rev.\ C {\bf 63}, 064904 (2001).


\bibitem{KV18}
E.E.~Kolomeitsev and D.N.~Voskresensky, Fluctuations in non-ideal pion gas with dynamically fixed particle number, Nucl. Phys. A {\bf 973}, 89 (2018).

\bibitem{BKV18}
  E.~E.~Kolomeitsev, M.~E.~Borisov and D.~N.~Voskresensky,
  ``Particle number fluctuations in a non-ideal pion gas,''
  EPJ Web Conf.\  {\bf 182}, 02066 (2018).



\bibitem{Begun-Gor-2008-FS}
V.V.~Begun and M.I.~Gorenstein, Bose-Einstein condensation in the relativistic pion gas: Thermodynamic limit and finite size effects,
Phys. Rev. C {\bf 77}, 064903 (2008).

\bibitem{Weinberg68}
S.~Weinberg, Nonlinear realization of chiral symmetry,
Phys. Rev. {\bf 166}, 1568 (1968).

\bibitem{Sawyer:1989nu}
R.F.~Sawyer, Effects of nuclear forces on neutrino opacities in hot nuclear matter,
Phys.\ Rev.\ C {\bf 40}, 865 (1989).

\bibitem{Roepke:2017bad}
G.~R\"opke, D. N.~Voskresensky, I. A.~Kryukov, and D.~Blaschke,
Fermi liquid, clustering, and structure factor in dilute warm nuclear matter,Nucl.\ Phys.\ A {\bf 970}, 224 (2018).

\bibitem{terHaar52}
D. ter Haar, The perfect Bose-Einstein gas in the theory of the quantum-mechanical grand canonical ensembles, Proc. R. Soc. Lond. A {\bf 212}, 552 (1952).

\bibitem{Fierz56}
M.~Fierz, \"Uber die statistischen Schwankungen in einem
kondensierenden System, Helvetica Physica Acta {\bf 29}, 47 (1956).

\bibitem{ZUK77}
R.M.~Ziff, G.E.~Uhlenbeck, and M.~Kac, The ideal Bose-Einstein gas, revisited,
Phys. Rept. {\bf 32}, 169 (1977).



\end{thebibliography}
\end{document}